\documentclass[reqno,12pt]{article}
\usepackage{amsfonts,amssymb,amsmath,amscd,bm,bbold,delarray,subfigure}
\usepackage{hyperref,cite}
\usepackage[dvips]{color}
\usepackage[dvips]{graphicx}

\DeclareMathOperator{\ad}{ad}

\DeclareMathOperator{\im}{im}
\DeclareMathOperator{\id}{id}
\DeclareMathOperator{\Hom}{Hom}
\DeclareMathOperator{\Iso}{Iso}
\DeclareMathOperator{\Fun}{Fun}
\DeclareMathOperator{\Map}{Map}

\DeclareMathOperator{\End}{End}
\DeclareMathOperator{\Aut}{Aut}
\DeclareMathOperator{\Der}{Der}

\DeclareMathOperator{\rank}{rank}
\DeclareMathOperator{\lspan}{span}

\numberwithin{equation}{section}

\newcommand{\sss}{{\hbox{$\sum$}}}

\font\sansserif=cmss12
\font\scriptsansserif=cmss12 at 7 truept
\font\scriptscriptsansserif=cmss10 at 5 truept
\font\euler=eusm10 at 12 truept
\font\scripteuler=eusm7
\font\scriptscripteuler=eusm5 
\def\eul{\fam=12}
\newcommand{\matheul}[1]{{{\eul #1}}}

\newcommand{\mathbfs}[1]{{\boldsymbol{#1}}}

\begin{document}

\hrule\vskip.4cm
\hbox to 14.5 truecm{December 2009\hfil DFUB 12/09}
\hbox to 14.5 truecm{Version 1  \hfil } 
\vskip.4cm
\hrule
\vskip.7cm
\begin{large}
\centerline{\textcolor{blue}{\bf The gauging of BV algebras} }  
\end{large}
\vskip.2cm
\centerline{by}
\vskip.2cm
\centerline{\textcolor{blue}{\bf\bf Roberto Zucchini}}
\centerline{\it Dipartimento di Fisica, Universit\`a degli Studi di Bologna}
\centerline{\it V. Irnerio 46, I-40126 Bologna, Italy}
\centerline{\it I.N.F.N., sezione di Bologna, Italy}
\centerline{\it E--mail: zucchinir@bo.infn.it}
\vskip.7cm
\hrule
\vskip.7cm
\centerline{\textcolor{blue}{\bf Abstract}}
\par\noindent
\vskip.2cm
A BV algebra is a formal framework within which the BV quantization algorithm 
is implemented.
In addition to the gauge symmetry, encoded in the BV master equation, 
the master action often exhibits further global symmetries, which may be in turn 
gauged. We show how to carry this out in a BV algebraic set up.
Depending on the nature of the global symmetry, the gauging involves coupling to 
a pure ghost system with a varying amount of ghostly supersymmetry.
Coupling to an $N=0$ ghost system yields an ordinary gauge theory
whose observables are appropriately classified by the invariant BV cohomology. 
Coupling to an $N=1$ ghost system leads to a topological gauge field theory
whose observables are classified by the equivariant BV cohomology. 
Coupling to higher $N$ ghost systems yields topological gauge field theories
with higher topological symmetry. In the latter case, however, problems of a 
completely new kind emerge, which call for a revision of the standard
BV algebraic framework.

\vfill\eject

\tableofcontents

\vfill\eject


\section{\normalsize \textcolor{blue}{Introduction}}\label{sec:intro}

~~~~The Batalin--Vilkovisky (BV) approach \cite{BV1,BV2}
is the most general and powerful quantization 
algorithm presently available. It is suitable for the quantization of ordinary 
gauge theories, such as Yang--Mills theory, as well as more complicated gauge theories
with open and/or reducible gauge symmetries. 
Its main feature consists in the introduction of ghost fields from the outset  
automatically incorporating in this way BRST symmetry.

The general structure of the BV formalism is as follows \cite{Sam,Fiorenza1}. 
Given a classical field theory with gauge symmetries, one introduces 
an antifield with opposite statistics for each field, including ghost fields, therefore doubling 
the total number of fields. The resulting field/antifield space $\matheul{F}$ 
is equipped with an odd Poisson bracket $\{\cdot,\cdot\}$, called antibracket, 
and acquires an odd phase space structure, in which fields and antifields are canonically conjugate.
At tree level in the quantum theory, the original classical action 
is extended to a new action 
$S_0$ defined on the whole content of $\matheul{F}$ and exhibiting 
an off shell odd symmetry corresponding to the gauge symmetry of the original field theory.
The gauge fixing is carried out by restricting the action $S_0$ to a suitable Lagrangian 
submanifold $\matheul{L}$ in $\matheul{F}$. Gauge independence, that is independence from the choice 
of $\matheul{L}$, is ensured if $S_0$ satisfies the classical BV master equation 
\begin{equation}
\{S_0,S_0\}=0.
 \label{intro1}
\end{equation}
At loop level, quantum corrections modify the action $S_0$ and turn it into a quantum action 
$S_\hbar$. Gauge independence is then ensured  
provided $S_\hbar$ satisfies the quantum BV master equation 
\begin{equation}
\hbar \Delta S_\hbar+\frac{1}{2}\{S_\hbar,S_\hbar\}=0,
\label{intro2}
\end{equation}
where $\Delta$ is a suitably regularized odd functional Laplacian in 
$\matheul{F}$. Violations of this correspond to gauge anomalies.

The observables of the field theory constructed in this way are characterized by having 
gauge independent correlators. The gauge independence of a correlator $\langle\psi_\hbar\rangle$ 
is ensured if $\psi_\hbar$ satisfies the equation
\begin{equation}
\delta_\hbar\psi_\hbar:=\hbar \Delta\psi_\hbar+\{S_\hbar,\psi_\hbar\}=0
\label{intro5}
\end{equation}
The solutions $\psi_\hbar$ of \eqref{intro5} are called quantum BV observables. 
The quantum BV operator $\delta_\hbar$ is nilpotent. Therefore, there is a cohomology associated with it,
the quantum BV cohomology. Since correlators of BV exact observables vanish, effectively distinct
BV observables are in one--to--one correspondence with the BV cohomology classes.

\vspace{-.02cm}
After this very brief review of BV theory, let us come to the topic of the paper. 
The algebraic structure consisting of the graded algebra of functionals on
the field/antifield space $\matheul{F}$, 
the antibracket $\{\cdot,\cdot\}$ and the odd Laplacian $\Delta$ is called 
a BV algebra. It provides the formal framework within which the BV quantization algorithm 
is implemented.
This has motivated a number of mathematical studies of BV algebras 
\cite{Schwarz1,Schwarz2,Zwiebach1}.

\vspace{-.02cm}
The classical field theory originally considered, even if it is a gauge theory, may still have 
global symmetries. In certain cases, one may wish to gauge these latter.  
In a BV framework, the gauging of a global symmetry consists in the coupling of the 
ungauged ``matter'' field theory and a suitable pure ``ghost'' field theory corresponding 
to the symmetry. (Ordinary ghost and gauge fields normally combine in ghost superfields.)
Two procedures of concretely working this out are possible in principle.

\vspace{-.02cm}
$i)$ One couples the matter and the ghost field theories at the classical level, by 
adding suitable interaction terms, obtaining a gauged classical field theory. 
Then, one quantizes this latter using the BV algorithm, by constructing the 
appropriate BV algebra and quantum BV master action. 

\vspace{-.02cm}
$ii)$ One separately quantizes the matter and the ghost field theories, 
by constructing the appropriate BV algebra and quantum BV master action 
of each of them. Then, one embeds the matter and ghost BV algebra structures
so obtained in a minimal gauged BV algebra structure and constructs a 
gauged quantum BV master action by adding the matter and ghost actions 
and suitable interaction terms in a way consistent with the quantum BV master equation.

\vspace{-.02cm}
We call these two approaches {\it classical gauging}
and {\it BV algebra gauging}, respectively.
Superficially, it may look like that classical gauging is more natural:
after all, BV theory was devised precisely to quantize classical gauge theories. 
In fact, in certain cases, BV gauging is more advantageous.

\vspace{-.02cm}
In a prototypical example, one efficient way of generating a sigma model on 
a non trivial manifold $X$ is the gauging of a sigma model on a simpler manifold 
$Y$ carrying the action 
of a Lie group $G$ such that $X\simeq Y/G$ \cite{Witten3,Spence1}. The target space 
of the gauged model turns out to be precisely $X$. 
In a BV formulation of the ungauged sigma model,
$G$ acts as a group of global symmetries. The gauging of these is performed
by coupling the ungauged model to a suitable ghost sigma model, yielding in a natural 
way a BV formulation of the gauged model \cite{Zucchini6,Zucchini7,Zucchini8}.

\vspace{-.02cm}
The Alexandrov--Kontsevich--Schwartz--Zaboronsky (AKSZ) formalism of ref. \cite{AKSZ} 
is a method of constructing solutions of the classical BV master equation directly, 
without starting from a classical action with a set of symmetries, as is originally done 
in the BV framework. When building models with gauged global symmetries in a AKSZ framework, 
BV algebra gauging is definitely more natural and transparent than classical gauging.

\vspace{-.02cm}
In this paper, we study in great detail the 
BV algebra gauging of a matter field theory with global symmetries.
For a certain global symmetry, the ghost field theory to be coupled to
the matter theory may have a varying amount of ``ghostly supersymmetry''.
Coupling, if feasible, to an $N=0$ ghost system yields an ordinary gauge field theory.
Coupling to an $N=1$ ghost system leads to a topological gauge field theory.
Coupling to higher $N$ ghost systems yields topological gauge field theories
with higher topological supersymmetry. In the latter case, however, problems of a 
completely new kind show up, which may require a major revision of the standard
BV algebraic framework.

\vspace{-.022cm}
Though BV algebra gauging is ultimately carried out within the framework of BV theory,
ordinary BV cohomology is not adequate for 
the classification of observables of the field theories constructed in this way.
If $\mathfrak{g}$ is the global symmetry Lie algebra, 
$\mathfrak{g}$--invariant BV cohomology in the $N=0$ case,  
$\mathfrak{g}$--equivariant BV cohomology in the $N=1$ case
and presumably some higher $\mathfrak{g}$--equivariant BV cohomologies for larger $N$
are required.

\vspace{-.022cm}
We shall carry out our analysis of BV gauging in a finite dimensional setting 
as in \cite{Schwarz1,Schwarz2,Zwiebach1}. This has its advantages and disadvantages. 
It allows one to focus on the essential features of gauging, especially those of an algebraic and 
geometric nature, on one hand, but it is of course no substitute for full--fledged field theory, 
which is essentially infinite dimensional, on the other. 
Nevertheless, with the due caution, one can presumably extend our considerations 
to realistic BV field theories. Further, it is known that certain BV field theories  
have finite dimensional 
reductions which capture some of their relevant structural features 
\cite{Mnev1,Mnev2,Bonechi1}. 

\vspace{-.022cm}
The plan of this paper is as follows.
In sect. \ref{sec:BValg}, we review the basics of BV algebra theory 
and set the notation used in the subsequent sections.
In sect. \ref{sec:BVacts}, we recall the definition and the main properties 
of the BV master action and observables. In sects. \ref{sec:BVgauge0}, \ref{sec:BVgauge1},
we illustrate how to carry out the $N=0$ and $N=1$ gauging of BV algebras and 
identify the relevant versions of BV cohomology. 
In sect. \ref{sec:BVgaugeN}, we tackle the problem of higher $N$ BV algebra gauging
highlighting the conceptual problems arising in this case.
In sect. \ref{sec:appls}, we illustrate a number of examples and applications of the 
theory developed in the preceding sections, showing in particular its relevance for the
finite dimensional reduction of the gauged Poisson sigma model of refs. \cite{Zucchini6,Zucchini7}.
In sect. \ref{sec:conclu}, we provide some concluding remarks.
Finally, in the appendices, we conveniently collect details on a few technical issues 
involved in our analysis.

\vspace{.1cm}

{\it Acknowledgments} 

We thank F. Bastianelli and P. Sundell for useful discussions.

\vspace{.1cm}

{\it Remarks on conventions and notation}

\vspace{.1cm}
 
In this paper, we use the following notations.
All spaces and algebras are over the field $\mathbb{R}$.

$a)$ Let $\mathcal{E}$ be $\mathbb{Z}$--graded vector space. 
$\mathcal{E}_k$ is the subspace of $\mathcal{E}$ of 
degree $k\in \mathbb{Z}$. If $x\in \mathcal{E}_k$ for some $k$, 
$x$ is said homogeneous and $\partial x=k$ is the degree of $x$.
If $p\in\mathbb{Z}$, $\mathcal{E}[p]$ is the $\mathbb{Z}$--graded vector 
space such that $\mathcal{E}[p]_k=\mathcal{E}_{k-p}$.
Similar conventions hold for a $\mathbb{Z}$--graded algebra $\mathcal{A}$.
In this case, we denote by $\mathcal{A}_{\mathrm{v}}$ the 
$\mathbb{Z}$--graded vector space underlying $\mathcal{A}$.

$b)$ Let $\mathcal{E}$, $\mathcal{F}$ be $\mathbb{Z}$--graded vector spaces.
$\mathcal{E}\otimes\mathcal{F}$ is the $\mathbb{Z}$--graded tensor product 
of $\mathcal{E}$, $\mathcal{F}$. Its grading is given by 
$(\mathcal{E}\otimes\mathcal{F})_k=\oplus_{l+m=k}\mathcal{E}_l\otimes\mathcal{F}_m$
for $k\in\mathbb{Z}$.
The tensor products $\mathcal{A}\otimes\mathcal{B}$ of two $\mathbb{Z}$--graded 
algebras $\mathcal{A}$, $\mathcal{B}$ is defined in the same fashion with the graded 
multiplication 
\begin{equation}
x\otimes u\,y\otimes v=(-1)^{\partial y\partial u}xy\otimes uv,
\label{intro4} 
\end{equation}
for homogeneous $x,y\in \mathcal{A}$, $u,v\in \mathcal{B}$. 

$c)$ Let $\mathcal{E}$, $\mathcal{F}$ be $\mathbb{Z}$--graded vector spaces.
$\Hom(\mathcal{E},\mathcal{F})$ is the $\mathbb{Z}$--graded vector space
of vector space homomorphisms of $\mathcal{E}$ into $\mathcal{F}$. Its grading 
is defined so that $X\in \Hom_k(\mathcal{E})$ if, for all $l\in \mathbb{Z}$, 
$X\mathcal{E}_l\subset \mathcal{F}_{k+l}$. 
The space $\End(\mathcal{E})=\Hom(\mathcal{E},\mathcal{E})$,
the set $\Iso(\mathcal{E},\mathcal{F})$ and the group 
$\Aut(\mathcal{E})=\Iso(\mathcal{E},\mathcal{E})$ are defined accordingly.
Similar notions hold for two $\mathbb{Z}$--graded algebras
$\mathcal{A}$, $\mathcal{B}$ with the proviso that 
algebra homomorphisms are concerned.

$d)$  Let $\mathcal{E}$ be a $\mathbb{Z}$--graded vector space.
For homogeneous $X,Y\in\End(\mathcal{E})$, the graded commutator of $X,Y$ is given by 
\begin{equation}
[X,Y]=XY-(-1)^{\partial X\partial Y}YX.
\label{intro6} 
\end{equation}
All commutators will always be assumed to be graded, unless otherwise stated. 

$e)$ Let $\mathcal{A}$ be a $\mathbb{Z}$--graded algebra. 
$\Der(\mathcal{A})$ is the graded vector subspace of 
$\End(\mathcal{A}_{\mathrm{v}})$ of graded derivations of $\mathcal{E}$.
If $k\in \mathbb{Z}$ and $D\in\Der_k(\mathcal{A})$, then  
\begin{equation}
D(xy)=Dxy+(-1)^{k\partial x}x Dy,
\label{intro3} 
\end{equation}
for homogeneous $x,y\in \mathcal{A}$.

$f)$ If $x$ is a formal graded variable, then $\partial_{Lx}=(\partial/\partial x)_L$
and $\partial_{Rx}=(\partial/\partial x)_R$, the subfixes $L,R$
indicating left, right graded differentiation. If $\phi$ is a function of $x$, then
$\partial_{Lx}\phi=(-1)^{(\partial\phi+1)\partial x}\partial_{Rx}\phi$.

$g)$ A differential space is a pair $(\mathcal{E},\delta)$, where 
$\mathcal{E}$ is a $\mathbb{Z}$--graded vector space,  
$\delta\in \End_1(\mathcal{E})$ and $\delta^2=0$. 
The associated cohomology $H^*(\mathcal{E},\delta)$ is a space.
A differential algebra is a pair $(\mathcal{A},\delta)$, where 
$\mathcal{A}$ is a $\mathbb{Z}$--graded algebra,  
$\delta\in \Der_1(\mathcal{E})$ and $\delta^2=0$. 
The associated cohomology $H^*(\mathcal{A},\delta)$ is then an algebra.

\vfill\eject

\section{\normalsize \textcolor{blue}{BV algebras}}\label{sec:BValg}

~~~~Batalin--Vilkovisky (BV) algebras are the formal structure underlying the BV 
quantization algorithm in quantum field theory \cite{BV1,BV2}. The BV algebra of a field theory
consists of a graded algebra of functions of fields and antifields,
an odd Poisson bracket defining the canonical structure of the field theory
at the classical level and an odd Laplacian required for implementing 
the field theory's quantization. BV algebras however can be treated in a completely formal
setting without invoking any concrete field theoretic realization \cite{Sam,Fiorenza1}.

A {\it BV algebra} is a triple $(\mathcal{A},\Delta,\{\cdot,\cdot\})$ consisting of the following
elements. 

\par\noindent 
~~~1) A $\mathbb{Z}$--graded commutative associative unital algebra $\mathcal{A}$. 

\par\noindent 
~~~2) A {\it BV Laplacian}, i. e. an element $\Delta\in\End_1(\mathcal{A})$ that is nilpotent,
\begin{equation}
\Delta^2=0.
\label{BValg1}
\end{equation}

\par\noindent 
~~~3) A {\it BV antibracket}, i.e. an $\mathbb{R}$--bilinear map 
$\{\cdot,\cdot\}:\mathcal{A}\times\mathcal{A}\rightarrow \mathcal{A}$ such that 
\begin{align}
&\Delta(\phi\psi)=\Delta\phi\psi+(-1)^{\partial\phi}\phi\Delta\psi+(-1)^{\partial\phi}\{\phi,\psi\},
\vphantom{\bigg]}
\label{BValg2}
\\
&\{\phi,\psi\upsilon\}=\{\phi,\psi\}\upsilon+(-1)^{(\partial\phi+1)\partial\psi}\psi\{\phi,\upsilon\},
\vphantom{\bigg]}
\label{BValg3}
\end{align}
for all homogeneous $\phi,\psi,\upsilon\in\mathcal{A}$. 

We notice that, by \eqref{BValg2}, $\{\cdot,\cdot\}$ is determined by 
$\Delta$. So, the notion of BV algebra could be defined in terms
of $\mathcal{A}$, $\Delta$ only. 

Several properties can be derived from the BV algebra axioms.

\par\noindent 
~~~$a$) One has 
$\partial\{\phi,\psi\}=\partial\phi+\partial\psi+1$ and 
\begin{align}
&\{\phi,\psi\}+(-1)^{(\partial\phi+1)(\partial\psi+1)}\{\psi,\phi\}=0,\hspace{5.1cm}
\vphantom{\bigg]}
\label{BValg4}
\\
&(-1)^{(\partial\phi+1)(\partial\upsilon+1)}\{\phi,\{\psi,\upsilon\}\}
+(-1)^{(\partial\psi+1)(\partial\phi+1)}\{\psi,\{\upsilon,\phi\}\}
\vphantom{\bigg]}
\label{BValg5}
\\
&\hspace{6cm}+(-1)^{(\partial\upsilon+1)(\partial\psi+1)}\{\upsilon,\{\phi,\psi\}\}=0,
\vphantom{\bigg]}
\nonumber
\end{align}
for all homogeneous $\phi,\psi,\upsilon\in\mathcal{A}$. These relations 
follow from \eqref{BValg2}. $\mathcal{A}$ with the multiplicative structure given by the 
bracket $\{\cdot,\cdot\}$ is a $\mathbb{Z}$--graded commutative algebra 
$\mathcal{A}_G$, called {\it Gerstenhaber (odd Poisson) algebra}.
The gradings of $\mathcal{A}$, $\mathcal{A}_G$ are such that 
$\mathcal{A}_{G\mathrm{v}}=\mathcal{A}_{\mathrm{v}}[1]$.

\par\noindent 
~~~$b$) On account of \eqref{BValg2}, $\Delta\not\in\Der_1(\mathcal{A})$. 
The Gerstenhaber bracket $\{\cdot,\cdot\}$ measures the failure of $\Delta$ being so.
However, one has 
\begin{equation}
\Delta\{\phi,\psi\}=\{\Delta\phi,\psi\}+(-1)^{\partial\phi+1}\{\phi,\Delta\psi\},
\label{BValg6} 
\end{equation}
for homogeneous $\phi,\psi\in\mathcal{A}$. This relation follows from combining \eqref{BValg1},
\eqref{BValg2}. Thus, $\Delta\in\Der_1(\mathcal{A}_G)$.  

\par\noindent 
~~~$c$) A derivation $D\in\Der_k(\mathcal{A})$ such that 
\begin{equation}
[D,\Delta]=0
\label{BValg6/2} 
\end{equation}
is called a {\it BV derivation}. If $D\in\Der_k(\mathcal{A})$, 
one has $D\not\in\Der_k(\mathcal{A}_G)$ in general.
However, if $D$ is a BV derivation, then $D\in\Der_k(\mathcal{A}_G)$ as well.
This follows straightforwardly from combining \eqref{BValg2}, \eqref{BValg6/2}.

\par\noindent 
~~~$d$) For $\alpha\in\mathcal{A}_{k-1}$, let us set 
\begin{equation}
\ad\alpha \,\phi=\{\alpha,\phi\},\qquad \phi\in\mathcal{A}.
\label{BValg6/1} 
\end{equation}
Then, simultaneously $\ad\alpha\in \Der_k(\mathcal{A})$,
$\ad\alpha\in \Der_k(\mathcal{A}_G)$. These properties follow 
directly from \eqref{BValg3}, \eqref{BValg5}, respectively.
A derivation $D\in\Der_k(\mathcal{A})$ is called {\it BV inner}, if it is of the form 
$D=\ad\alpha$ for some $\alpha\in\mathcal{A}_{k-1}$ such that 
\begin{equation}
\Delta\alpha=0.
\label{BValg6/5}
\end{equation}
Else, it is {\it BV outer}. By \eqref{BValg6}, \eqref{BValg6/5}, a BV inner $D$
fulfils \eqref{BValg6/2} and, so, is BV.

A BV algebra $(\mathcal{A},\Delta,\{\cdot,\cdot\})$ is a {\it BV subalgebra} 
of a BV algebra $(\mathcal{A}',\Delta',\{\cdot,\cdot\}')$ if 
$\mathcal{A}$ is a subalgebra of $\mathcal{A}'$ such that 
$\Delta'\mathcal{A}\subset\mathcal{A}$, $\{\mathcal{A},\mathcal{A}\}'\subset
\mathcal{A}$ and $\Delta=\Delta'|_{\mathcal{A}}$,
$\{\cdot,\cdot\}=\{\cdot|_{\mathcal{A}},\cdot|_{\mathcal{A}}\}'$.

\vfill\eject

There is a natural notion of homomorphism of BV algebras.
Let $(\mathcal{A},\Delta,\{\cdot,\cdot\})$, $(\mathcal{A}',\Delta',\{\cdot,\cdot\}')$
be BV algebras.
A map $T:\mathcal{A}\rightarrow\mathcal{A}'$ is a {\it BV algebra homomorphism}, if 

\par\noindent 
~~~1) $T\in\Hom_0(\mathcal{A},\mathcal{A}')$.

\par\noindent 
~~~2) $T$ intertwines the BV Laplacians $\Delta$, $\Delta'$, 
\begin{equation}
T\Delta=\Delta'T.
\label{BValg6/3}
\end{equation}

\par\noindent 
~~~3) $T$ intertwines the brackets $\{\cdot,\cdot\}$, $\{\cdot,\cdot\}'$, 
\begin{equation}
T\{\phi,\psi\}=\{T\phi,T\psi\}',
\label{BValg6/4}
\end{equation}
for $\phi,\psi\in\mathcal{A}$.

As a matter of fact, \eqref{BValg6/4} is not an independent condition, as it follows 
from \eqref{BValg2}, \eqref{BValg6/3}.
One can also define a {\it BV algebra monomorphism, epimorphism, isomorphism, endomorphism} and 
{\it automorphism} in obvious fashion.

A few properties can be deduced from the BV algebra homomorphism axioms.

\par\noindent 
~~~$a$) By \eqref{BValg6/4}, $T\in\Hom_0(\mathcal{A}_G,\mathcal{A}_G{}')$
as well. Indeed, $T$ is a homomorphism of the odd Poisson structures of 
$\mathcal{A}_G,\mathcal{A}_G{}'$.

\par\noindent 
~~~$b$) $\ker T$ is a subalgebra of $\mathcal{A}$ such that 
$\Delta\ker T\subset\ker T$, $\{\ker T,\ker T\}\subset\ker T$
and, so, with the BV algebra structure induced by $\mathcal{A}$, 
a BV subalgebra of $\mathcal{A}$.
Likewise, $\im T $ is a subalgebra of $\mathcal{A}'$ such that 
$\Delta'\im T\subset\im T$, $\{\im T,\im T\}'\subset\im T$
and, so, with the BV algebra structure induced by $\mathcal{A}'$, 
a BV subalgebra of $\mathcal{A}'$. This follows immediately from \eqref{BValg6/3},
\eqref{BValg6/4}.

Homomorphisms describe the natural relationships of BV algebras.

\par\noindent 
~~~$a$) Let $(\mathcal{A},\Delta,\{\cdot,\cdot\})$ be a BV subalgebra 
of the BV algebra $(\mathcal{A}',\Delta',\{\cdot,\cdot\}')$. Then, 
the natural injection
$I:\mathcal{A}\rightarrow\mathcal{A}'$ is a BV algebra monomorphism. 
 
\par\noindent 
~~~$b$) The automorphisms of a BV algebra $(\mathcal{A},\Delta,\{\cdot,\cdot\})$ 
represent the symmetries of this latter.
Let $\alpha\in \mathcal{A}_{-1}$ satisfy \eqref{BValg6/5}.
Define a map  $T_\alpha:\mathcal{A}\rightarrow\mathcal{A}$ by
\begin{equation}
T_\alpha=\exp(\ad\alpha),
\label{BValg6/6}
\end{equation}
the right hand side being defined by the usual exponential series.
It is assumed that that either the series terminates after a finite number of terms
by algebraic reasons or it converges in some natural topology of $\End(\mathcal{A}_{\mathrm{v}})$. 
Then, $T_\alpha$ is a BV algebra automorphism. The automorphisms of this type are called {\it BV inner}, since 
$\ad\alpha$ is a BV inner derivation of $\mathcal{A}$. Correspondingly, 
all other BV algebra automorphisms are called {\it BV outer}. 

The set of BV algebras can be organized as a category having BV algebras
homomorphisms as morphisms. 
One can define natural operations in this category. In particular,
there is a notion of {\it tensor product of BV algebras} that will be extensively used in the following.
Let $(\mathcal{A}',\Delta',\{\cdot,\cdot\}')$, $(\mathcal{A}'',\Delta'',\{\cdot,\cdot\}'')$
be BV algebras.
Construct a triple $(\mathcal{A},\Delta,\{\cdot,\cdot\})$ as follows.

\par\noindent 
~~~1) $\mathcal{A}=\mathcal{A}'\otimes\mathcal{A}''$, a 
tensor product of graded algebras.

\par\noindent 
~~~2) $\Delta\in\End_1(\mathcal{A})$ is defined by the relation
\begin{equation}
\Delta(\phi'\otimes\phi'')=\Delta'\phi'\otimes\phi''+(-1)^{\partial\phi'}\phi'\otimes\Delta''\phi'',
\label{BValg7} 
\end{equation}
for homogeneous $\phi'\in\mathcal{A}'$, $\phi''\in\mathcal{A}''$.

\par\noindent 
~~~3) $\{\cdot,\cdot\}:\mathcal{A}\times\mathcal{A}\rightarrow \mathcal{A}$ is defined by the relation
\begin{align}
&\{\phi'\otimes\phi'',\psi'\otimes\psi''\}
\vphantom{\bigg]}
\label{BValg8} 
\\
&\hspace{1cm}=(-1)^{(\partial\psi'+1)\partial\phi''}\{\phi',\psi'\}'\otimes\phi''\psi''
+(-1)^{(\partial\phi''+1)\partial\psi'}\phi'\psi'\otimes\{\phi'',\psi''\}'',
\vphantom{\bigg]}
\nonumber
\end{align}
for homogeneous $\phi',\psi'\in\mathcal{A}'$, $\phi'',\psi''\in\mathcal{A}''$.
Then, $(\mathcal{A},\Delta,\{\cdot,\cdot\})$ is a BV algebra.  
The verification of the basic 
relations \eqref{BValg1}--\eqref{BValg3} is straightforward. 
$(\mathcal{A},\Delta,\{\cdot,\cdot\})$ 
is called the 
{\it tensor product} of the BV algebras 
$(\mathcal{A}',\Delta',\{\cdot,\cdot\}')$, $(\mathcal{A}'',\Delta'',\{\cdot,\cdot\}'')$.

The maps $I':\mathcal{A}'\rightarrow \mathcal{A}$, $I'':\mathcal{A}''\rightarrow \mathcal{A}$
defined by $I'\phi'=\phi'\otimes 1''$, $\phi'\in \mathcal{A}'$,
$I''\phi''=1'\otimes\phi''$, $\phi''\in \mathcal{A}''$ are BV algebra monomorphisms.
Indeed, $I'$, $I''$ satisfy \eqref{BValg6/3}, \eqref{BValg6/4} on account of \eqref{BValg7},
\eqref{BValg8}. 
In this way, $\mathcal{A}'$, $\mathcal{A}''$ can be considered as BV subalgebras of 
$\mathcal{A}$.

Examples of BV algebras will be illustrated in the following sections.

\vfill\eject

\section{\normalsize \textcolor{blue}{Quantum BV master actions and observables}}\label{sec:BVacts}

~~~~ Let $(\mathcal{F},\Delta,\{\cdot,\cdot\})$ be the BV algebra relevant for a 
BV quantization problem. In general, quantization can be viewed as the addition 
to a classical quantity of a quantum correction expressed perturbatively as a formal power series 
in the Planck constant $\hbar$. For this reason, BV quantization requires working with the graded algebra 
$\mathcal{F}((\hbar))$ of formal power series $\phi_\hbar=\sss_{k\geq 0}\hbar^k\phi_{(k)}$
with $\phi_{(k)}\in\mathcal{F}$, where $\hbar$ is treated as a degree $0$ formal parameter.
The BV Laplacian $\Delta$ and antibracket $\{\cdot,\cdot\}$ extend by formal linearity to 
$\mathcal{F}((\hbar))$. $(\mathcal{F}((\hbar)),\Delta,\{\cdot,\cdot\})$ is then also a BV algebra. 
The natural injection of $\mathcal{F}$ into $\mathcal{F}((\hbar))$, 
defined by $\phi\rightarrow \sss_{k\geq 0}\hbar^k\delta_{k,0}\phi$,
is a BV algebra monomorphism and, so,
$\mathcal{F}$ can be viewed as a BV subalgebra of $\mathcal{F}((\hbar))$.
The quantum BV master action $S_\hbar$ and observables $\psi_\hbar$ are the solutions
of eqs. \eqref{intro2} and  \eqref{intro5} in $\mathcal{F}((\hbar))$. 
The corresponding classical approximations $S$ and $\psi$ are obtained by truncating
$S_\hbar$ and $\psi_\hbar$ to their components in $\mathcal{F}$. 

Keeping explicit the $\hbar$ dependence of the relevant quantities in the following analysis 
would lead to unnecessary notational complication. For this reason, we shall treat
the problems of quantization and classical approximation thereof more formally in the 
framework of a given BV algebra $(\mathcal{A},\Delta,\{\cdot,\cdot\})$.
It is tacitly understood that, in any physical realization, 
$\mathcal{A}$ must be correspondingly 
interpreted as either $\mathcal{F}((\hbar))$ or $\mathcal{F}$ for the relevant BV algebra 
$\mathcal{F}$. 

In the constructions of the following sections, other endomorphisms $f\in\End(\mathcal{A})$ 
will be considered beside $\Delta$. It is tacitly understood that, 
in any physical realization, the $f$ are independent from $\hbar$. 

Let $(\mathcal{A},\Delta,\{\cdot,\cdot\})$ be a BV algebra. 
An element $S\in\mathcal{A}_0$ is called a {\it quantum BV master action} of the BV algebra if
$S$ satisfies the {\it quantum BV master equation}
\begin{equation}
\Delta S+\frac{1}{2}\{S,S\}=0,
\label{BVacts2} 
\end{equation}
\eject\noindent
An element $\psi\in\mathcal{A}$
is a {\it quantum BV observable}, if it satisfies the equation 
\begin{equation}
\delta\psi=0,
\label{BVacts7} 
\end{equation}
where $\delta$ is the {\it quantum BV operator}
\begin{equation}
\delta=\Delta+\ad S. 
\label{BVacts8} 
\end{equation}
From the definition, using the master equation \eqref{BVacts2}, it can be easily verified  
that $\delta\in\End_1(\mathcal{A}_{\mathrm{v}})$ and that $\delta$ is nilpotent,
\begin{equation}
[\delta,\delta]=2\delta{}^2=0.
\label{BVacts9} 
\end{equation}
Hence, $(\mathcal{A},\delta)$ is a differential space. The associated 
cohomology  is the {\it quantum BV cohomology} space $H_{BV}{}^*(\mathcal{A})$. 
We note that $\delta$ is not a derivation, as $\Delta$ is not. 
So, even though $\mathcal{A}$ is an algebra, 
$(\mathcal{A},\delta)$ is only a differential space. 
Correspondingly, $H_{BV}{}^*(\mathcal{A})$ is only a cohomology space. 

The classical counterpart of the above is as follows. 
Let $(\mathcal{A},\Delta,\{\cdot,\cdot\})$ be a BV algebra.
An element $S\in\mathcal{A}_0$ is called a {\it classical BV master action} of the BV algebra if
$S$ satisfies the {\it classical BV master equation}
\begin{equation}
\{S,S\}=0.
\label{BVacts4} 
\end{equation}
An element $\psi\in\mathcal{A}$
is a {\it classical BV observable}, if it satisfies the equation 
\begin{equation}
\delta_c\psi=0,
\label{BVacts10} 
\end{equation}
where $\delta_c$ is the {\it classical BV operator}
\begin{equation}
\delta_c=\ad S.
\label{BVacts11} 
\end{equation}
From the definition, using the master equation \eqref{BVacts2}, it can be easily verified  
that $\delta_c\in\End_1(\mathcal{A}_{\mathrm{v}})$ and that $\delta_c$ is nilpotent,
\begin{equation}
[\delta_c,\delta_c]=2\delta_c{}^2=0.
\label{BVacts11/1} 
\end{equation}
So, $(\mathcal{A},\delta_c)$ is a differential algebra. The associated 
cohomology is the {\it classical BV cohomology} algebra $H_{cBV}{}^*(\mathcal{A})$.  
Recall that, in the quantum case,  
$(\mathcal{A},\delta)$ and $H_{BV}{}^*(\mathcal{A})$ 
are merely spaces.

Let $(\mathcal{A},\Delta,\{\cdot,\cdot\})$, $(\mathcal{A}',\Delta',\{\cdot,\cdot\}')$
be BV algebras  and let $T:\mathcal{A}\rightarrow\mathcal{A}'$ be a BV algebra homomorphism
(cf. sect. \ref{sec:BValg}). If $S$ be a quantum BV master action of the BV algebra
$(\mathcal{A},\Delta,\{\cdot,\cdot\})$, then
\begin{equation}
S'{}=TS
\label{BValg11/1}
\end{equation}
is a quantum BV master action of the BV algebra $(\mathcal{A}',\Delta',\{\cdot,\cdot\}')$.
This follows easily from \eqref{BVacts2}, \eqref{BValg6/3}, \eqref{BValg6/4}.
Similarly, if $\psi$ is a
quantum BV observable of the BV algebra $(\mathcal{A},\Delta,\{\cdot,\cdot\})$
and master action $S$, then 
\begin{equation}
\psi'{}=T\psi
\label{BValg11/2}
\end{equation}
is a quantum BV observable of the BV algebra $(\mathcal{A}',\Delta',\{\cdot,\cdot\}')$
and master action $S'{}$. This follows from \eqref{BVacts7}, \eqref{BVacts8}, 
\eqref{BValg6/3}, \eqref{BValg6/4}. In fact, one has
\begin{equation}
T\delta=\delta'{} T.
\label{BValg11/3}
\end{equation}
$T$ is therefore a chain map of the differential spaces $(\mathcal{A},\delta)$,
$(\mathcal{A}',\delta')$ and, so, it induces a homomorphism of the corresponding  
cohomology spaces 
$H_{BV}{}^*(\mathcal{A})$, $H_{BV}{}^*(\mathcal{A}')$.
Analogous  statements hold also in the classical case. 

Let $(\mathcal{A}',\Delta',\{\cdot,\cdot\}')$, $(\mathcal{A}'',\Delta'',\{\cdot,\cdot\}'')$
be BV algebras and let 
$(\mathcal{A},\Delta,\{\cdot,\cdot\})$ be their tensor product (cf. sect. \ref{sec:BValg}). 
If $S'$, $S''$ are quantum BV master actions of 
$(\mathcal{A}',\Delta',\{\cdot,\cdot\}')$, $(\mathcal{A}'',\Delta'',\{\cdot,\cdot\}'')$,
respectively, then 
\begin{equation}
S=S'\otimes 1''+1'\otimes S''
\label{BVacts5} 
\end{equation}
is a quantum BV master action of 
$(\mathcal{A},\Delta,\{\cdot,\cdot\})$. This property follows 
straightforwardly from applying \eqref{BValg7}, \eqref{BValg8}.        
Analogously, if $\psi'$, $\psi''$ are quantum BV observables 
of the BV algebras $(\mathcal{A}',\Delta',\{\cdot,\cdot\}')$, $(\mathcal{A}'',\Delta'',\{\cdot,\cdot\}'')$
and actions $S'$, $S''$, respectively, then \hphantom{xxxxxxxxxxxxxxxxxxx}
\begin{equation}
\psi=\psi'\otimes 1''+1'\otimes \psi''
\label{BVacts12} 
\end{equation}
is a quantum BV observable of $(\mathcal{A},\Delta,\{\cdot,\cdot\})$ and $S$.
The verification of this property is also straightforward. Again,
similar statements hold in the classical case. 

In quantum field theory, the above construction is simply the adjoining of two field
theories with no mutual interaction. From a physical point of view is therefore 
rather trivial. In interesting models, one requires adding to the non interacting action 
$S$ of eq. \eqref{BVacts5} interaction terms in a consistent way, that is without 
spoiling the quantum BV master equation \eqref{BVacts2}. BV gauging of a given field theory,
discussed in the next sections, is an important example of this procedure. 

Examples of BV master actions will be given in the following sections. 

\vfill\eject

\section{\normalsize \textcolor{blue}{N=0  BV gauging and N=0 ghost system}}\label{sec:BVgauge0}

~~~~Now, we are ready for starting the study of $N=0$ BV gauging and 
the $N=0$ ghost system. This will set the paradigm for $N=1$ and higher $N$ gaugings.

\vspace{.1cm}

{\it $N=0$ $\mathfrak{g}$--actions}

\vspace{.1cm}

Let $(\mathcal{A},\Delta,\{\cdot,\cdot\})$ be a BV algebra.
Let $\mathfrak{g}$ be a Lie algebra. 
An $N=0$ $\mathfrak{g}$--{\it action} on the BV algebra is 
a linear map  $l:\mathfrak{g}\rightarrow \Der_0(\mathcal{A})$
such that 
\begin{subequations}
\begin{align}
&[l_x,l_y]=l_{[x,y]},
\vphantom{\bigg]}
\label{BVgau01a} 
\\
&[l_x,\Delta]=0,
\vphantom{\bigg]}
\label{BVgau01b} 
\end{align}
\label{BVgau01} 
\end{subequations}
\hspace{-.29cm}
for $x,y\in\mathfrak{g}$. 
By \eqref{BVgau01b},  
for $x\in\mathfrak{g}$, $l_x\in\Der_0(\mathcal{A})$ is a BV derivation (cf. sect. \ref{sec:BValg}).

Let $S$ be a quantum BV master action of the BV algebra $(\mathcal{A},\Delta,\{\cdot,\cdot\})$. 
$S$ is said {\it invariant} under the 
$\mathfrak{g}$--action if 
\begin{equation}
l_xS=0,
\label{BVgau02} 
\end{equation} 
for all $x\in\mathfrak{g}$. This condition is compatible with \eqref{BVgau01a}, 
\eqref{BVgau01b} and the quantum BV master equation \eqref{BVacts2}. 
When $S$ is invariant, one has
\begin{equation}
[\delta,l_x]=0,
\label{BVgau02/1} 
\end{equation} 
where $\delta$ is the quantum BV operator (cf. sect. \eqref{sec:BVacts}, eq. \eqref{BVacts8}). 
By \eqref{BVacts9}, \eqref{BVgau01a}, \eqref{BVgau02/1},
$(\mathcal{A},\mathfrak{g},l,\delta)$ is an algebraic structure known as a 
differential $\mathfrak{g}$--module \cite{Grueb1}
(see appendix \ref{app:invcoh} for a 
review of differential Lie modules). 
By \eqref{BVgau02/1}, it is possible to define a $\mathfrak{g}$--invariant quantum BV cohomology, 
that is the cohomology of the differential space 
$(\mathcal{A}_{\mathrm{inv}},\delta)$, where 
$\mathcal{A}_{\mathrm{inv}}=\cap_{x\in\mathfrak{g}}\ker l_x\subset\mathcal{A}$. The same statements hold also 
for the classical BV operator and its cohomology. 

The $\mathfrak{g}$--action is called {\it BV Hamiltonian}, 
if there is a linear map $\lambda:\mathfrak{g}\rightarrow \mathcal{A}_{-1}$, 
called {\it BV moment map}, such that 
\begin{equation}
l_x=\ad\lambda_x,
\label{BVgau03} 
\end{equation} 
with $x\in\mathfrak{g}$, and that \hphantom{xxxxxxxxxxxxxxxx}
\vspace{-.1cm}
\begin{subequations}
\begin{align}
&\{\lambda_x,\lambda_y\}=\lambda_{[x,y]}, 
\vphantom{\bigg]}
\label{BVgau04a} 
\\
&\Delta\lambda_x=0,
\vphantom{\bigg]}
\label{BVgau04b} 
\end{align}
\label{BVgau04} 
\end{subequations}
\!\!with $x,y\in\mathfrak{g}$. 
\eqref{BVgau03} together with \eqref{BVgau04a}, \eqref{BVgau04b} 
are indeed sufficient for \eqref{BVgau01a}, \eqref{BVgau01b} to hold.
By \eqref{BVgau03}, \eqref{BVgau04b},  $l_x\in\Der_0(\mathcal{A})$ 
is a BV inner derivation (cf. sect. \ref{sec:BValg}).
Below, we consider only BV Hamiltonian  $\mathfrak{g}$--actions. 

If $S$ is a quantum BV master action of the BV algebra $(\mathcal{A},\Delta,\{\cdot,\cdot\})$
 invariant under the $\mathfrak{g}$--action, then \hphantom{xxxxxxxxxxxxxxxx}
\begin{equation}
\{\lambda_x,S\}=0,
\label{BVgau05} 
\end{equation} 
for all $x\in\mathfrak{g}$. 
By \eqref{BVgau04b}, \eqref{BVgau05}, 
$\lambda_x$ is a cocycle of both the classical and the quantum BV cohomology 
(cf. sect. \ref{sec:BVacts}). 

\vspace{.1cm}

{\it Gauging of a global $N=0$ $\mathfrak{g}$--symmetry}

\vspace{.1cm}

Consider a {\it matter} BV algebra $(\mathcal{A}_M,\Delta_M,\{\cdot,\cdot\}_M)$ 
carrying a BV Hamiltonian $N=0$ $\mathfrak{g}$--action $l_M$ with BV moment map $\lambda_M$
and a matter quantum BV master action $S_M$ invariant under the $\mathfrak{g}$--action.
By \eqref{BVgau05}, we may say that $S_M$ enjoys a {\it global $N=0$ $\mathfrak{g}$--symmetry}
and that $\lambda_M$ is the corresponding {\it symmetry charge}. 
We want to find a meaningful way of gauging these symmetries. 

The gauging proceeds in three steps. 

{\it 1.} We construct an $N=0$ {\it ghost} BV algebra
$(\mathcal{A}_{\mathfrak{g}|0},\Delta_{\mathfrak{g}|0},\{\cdot,\cdot\}_{\mathfrak{g}|0})$
with a BV Hamiltonian $N=0$ $\mathfrak{g}$--action $l_{\mathfrak{g}|0}$
with BV moment map $\lambda_{\mathfrak{g}|0}$
and an $N=0$ ghost quantum BV master action $S_{\mathfrak{g}|0}$ 
invariant under the $\mathfrak{g}$--action. The construction is canonical, 
in that it depends solely on $\mathfrak{g}$.

{\it 2.} 
We construct an $N=0$ {\it gauged matter} BV algebra
$(\mathcal{A}_{\mathfrak{g}|0M},\Delta_{\mathfrak{g}|0M}$, $\{\cdot,\cdot\}_{\mathfrak{g}|0M})$
and equip it with an appropriate BV Hamiltonian $N=0$ $\mathfrak{g}$--action $l_{\mathfrak{g}|0M}$
with BV moment map $\lambda_{\mathfrak{g}|0M}$.

{\it 3.} 
We construct an $N=0$ gauged matter quantum BV master action $S_{\mathfrak{g}|0M}$ 
\eject\noindent
of the gauged 
matter BV algebra invariant under the $\mathfrak{g}$--action.

{\it Step 1.} 
Given the Lie algebra $\mathfrak{g}$, define
\begin{equation}
\mathcal{A}_{\mathfrak{g}|0}=\Fun(\mathfrak{g}^\vee[-2]\oplus\mathfrak{g}[1]).
\label{BVgau06} 
\end{equation}
Denote by $b_i$, $c^i$, $i=1,\dots,\dim\mathfrak{g}$, the coordinates of 
$\mathfrak{g}^\vee[-2],\mathfrak{g}[1]$, 
respectively, 
corresponding to a chosen basis $\{t_i\}$ of $\mathfrak{g}$.
Then, $\mathcal{A}_{\mathfrak{g}|0}$ can be 
viewed as the $\mathbb{Z}$ 
graded commutative associative unital algebra of polynomials in the $b_i$, $c^i$. 
Define further the 2nd order differential operator
\begin{equation}
\Delta_{\mathfrak{g}|0}=\partial_{Lb}{}^i \partial_{Lci}
\label{BVgau07} 
\end{equation}
and the bilinear brackets
\begin{equation}
\{\phi,\psi\}_{\mathfrak{g}|0}=\partial_{Rb}{}^i\phi\partial_{Lci}\psi
-\partial_{Rci}\phi\partial_{Lb}{}^i\psi,
\qquad \phi,\psi\in \mathcal{A}_{\mathfrak{g}|0}.
\label{BVgau08} 
\end{equation}
Then, it is simple to check that
relations \eqref{BValg1}--\eqref{BValg3} are verified. It follows that 
$(\mathcal{A}_{\mathfrak{g}|0},\Delta_{\mathfrak{g}|0},\{\cdot,\cdot\}_{\mathfrak{g}|0})$
is a BV algebra, the  $N=0$  ghost BV algebra.

Let $f^i{}_{jk}$ be the structure constants of $\mathfrak{g}$ with respect to the basis 
$\{t_i\}$. Set
\begin{equation}
\lambda_{\mathfrak{g}|0i}=f^j{}_{ki}b_jc^k.
\label{BVgau09} 
\end{equation}
Since $\lambda_{\mathfrak{g}|0i}\in\mathcal{A}_{\mathfrak{g}|0-1}$,
it defines via \eqref{BVgau03} a linear map  $l_{\mathfrak{g}|0}:\mathfrak{g}
\rightarrow \Der_0(\mathcal{A}_{\mathfrak{g}|0})$.
If the Lie algebra $\mathfrak{g}$ is {\it unimodular}, that is 
\begin{equation}
f^j{}_{ji}=0,
\label{BVgau010} 
\end{equation} 
$l_{\mathfrak{g}|0}$ is a BV Hamiltonian $N=0$ $\mathfrak{g}$--action 
on the ghost BV algebra having $\lambda_{\mathfrak{g}|0}$
as BV moment map.
Indeed, using \eqref{BVgau07}, \eqref{BVgau08}, \eqref{BVgau09}, 
one finds that \eqref{BVgau04a} is verified and that
$\Delta_{\mathfrak{g}|0}\lambda_{\mathfrak{g}|0i}=f^j{}_{ji}$. 
So, \eqref{BVgau04b} is also verified, if \eqref{BVgau010} holds. 

The action of $l_{\mathfrak{g}|0}$ on $b_i$, $c^i$ is given by
\begin{subequations}
\begin{align}
l_{\mathfrak{g}|0i}b_j&=f^k{}_{ij}b_k,
\vphantom{\bigg]}
\label{BVgau011a} 
\\
l_{\mathfrak{g}|0i}c^j&=-f^j{}_{ik}c^k,
\vphantom{\bigg]}
\label{BVgau011b} 
\end{align}
\label{BVgau011} 
\end{subequations}
as follows readily from \eqref{BVgau03}, \eqref{BVgau08}, \eqref{BVgau09}.

The $N=0$ ghost algebra $\mathcal{A}_{\mathfrak{g}|0}$ contains an element
$S_{\mathfrak{g}|0}\in\mathcal{A}_{\mathfrak{g}|00}$ given by 
\begin{equation}
S_{\mathfrak{g}|0}=-\frac{1}{2}f^i{}_{jk}b_ic^jc^k.
\label{BVgau012} 
\end{equation}
$S_{\mathfrak{g}|0}$ satisfies the classical BV master equation 
\eqref{BVacts4} and, if $\mathfrak{g}$ is unimodular, 
also the quantum BV master equation \eqref{BVacts2}. 
Indeed, using \eqref{BVgau07}, \eqref{BVgau08}, \eqref{BVgau012}, one finds that 
\eqref{BVacts4} is verified and that $\Delta_{\mathfrak{g}|0}S_{\mathfrak{g}|0}=-f^i{}_{ij}c^j$.
So, \eqref{BVacts2} is also verified, if \eqref{BVgau010} holds.
$S_{\mathfrak{g}|0}$ is the $N=0$ ghost quantum BV master action.
$S_{\mathfrak{g}|0}$ is invariant under the $\mathfrak{g}$--action $l_{\mathfrak{g}|0}$. 
\eqref{BVgau08}, \eqref{BVgau09}, \eqref{BVgau012} indeed imply \eqref{BVgau05}.

The action of the quantum BV operator $\delta_{\mathfrak{g}|0}$ on $b_i$, $c^i$ is given by
\begin{subequations}
\begin{align}
&\delta_{\mathfrak{g}|0}b_i=f^k{}_{ji}b_kc^j,
\vphantom{\bigg]}
\label{BVgau013a} 
\\
&\delta_{\mathfrak{g}|0}c^i=-\frac{1}{2}f^i{}_{jk}c^jc^k,
\vphantom{\bigg]}
\label{BVgau013b} 
\end{align}
\label{BVgau013} 
\end{subequations}
\!\!as follows from the definition \eqref{BVacts8} and from \eqref{BVgau07}, \eqref{BVgau08}, 
\eqref{BVgau012}. Relations \eqref{BVgau013} are also the expressions 
of the action of the classical BV operator $\delta_{\mathfrak{g}|0 c}$
defined according to \eqref{BVacts11}. Recall however that the actions
of $\delta_{\mathfrak{g}|0}$ and $\delta_{\mathfrak{g}|0 c}$ on higher degree 
polynomials in $b_i$, $c^i$ are different because $\Delta_{\mathfrak{g}|0}$ acts 
non trivially on them in general. 


The fulfillment of the unimodularity condition \eqref{BVgau010} 
is required by $l_{\mathfrak{g}|0}$ being a Hamiltonian $\mathfrak{g}$--action 
and $S_{\mathfrak{g}|0}$ a quantum master action; it is 
thus crucial in the above BV construction. In full--fledged quantum field theory, 
\eqref{BVgau010} would be a quantum {\it anomaly cancellation} condition.

{\it Step 2.} 
The $N=0$ gauged matter BV algebra
$(\mathcal{A}_{\mathfrak{g}|0M},\Delta_{\mathfrak{g}|0M}$, $\{\cdot,\cdot\}_{\mathfrak{g}|0M})$
is the tensor product of the $N=0$ ghost BV algebra
$(\mathcal{A}_{\mathfrak{g}|0},\Delta_{\mathfrak{g}|0}$, $\{\cdot,\cdot\}_{\mathfrak{g}|0})$
and the matter BV algebra $(\mathcal{A}_M,\Delta_M,\{\cdot,\cdot\}_M)$
(cf. sect. \ref{sec:BValg}). 
Via \eqref{BVgau03}, the element $\lambda_{\mathfrak{g}|0Mi}\in\mathcal{A}_{\mathfrak{g}|0M-1}$ 
given by the expression
\begin{equation}
\lambda_{\mathfrak{g}|0Mi}=\lambda_{\mathfrak{g}|0i}\otimes 1_M+1_{\mathfrak{g}|0}\otimes\lambda_{Mi}
\label{BVgau014} 
\end{equation}
defines a linear map  $l_{\mathfrak{g}|0M}:\mathfrak{g}\rightarrow \Der_0(\mathcal{A}_{\mathfrak{g}|0M})$.
If, again, \eqref{BVgau010} is satisfied, $l_{\mathfrak{g}|0M}$ is a BV Hamiltonian $N=0$ 
$\mathfrak{g}$--action on the gauged matter BV algebra having $\lambda_{\mathfrak{g}|0M}$
as BV moment map. One just notices that $\lambda_{\mathfrak{g}|0M}$
satisfies \eqref{BVgau04a}, \eqref{BVgau04b} if simultaneously 
$\lambda_{\mathfrak{g}|0}$, $\lambda_M$ do, by \eqref{BValg7}, \eqref{BValg8}.
The $\mathfrak{g}$--action $l_{\mathfrak{g}|0M}$ extends trivially the $\mathfrak{g}$--actions 
$l_{\mathfrak{g}|0}$, $l_M$, in the sense that  
\begin{equation}
l_{\mathfrak{g}|0Mx}=l_{\mathfrak{g}|0x}\otimes 1_M+1_{\mathfrak{g}|0}\otimes l_{Mx},
\label{BVgau015} 
\end{equation}
for $x\in \mathfrak{g}$. 

{\it Step 3.}  The $N=0$ gauged matter algebra $\mathcal{A}_{\mathfrak{g}|0M}$ 
contains a distinguished element $S_{\mathfrak{g}|0M}\in\mathcal{A}_{\mathfrak{g}|0M0}$ given by 
\begin{equation}
S_{\mathfrak{g}|0M}=S_{\mathfrak{g}|0}\otimes 1_M+1_{\mathfrak{g}|0}\otimes S_M+c^i\otimes\lambda_{Mi}.
\label{BVgau016} 
\end{equation} 
The first two terms correspond to the trivial
non interacting ghost--matter action \eqref{BVacts5}. The third term 
is a genuine ghost--matter interaction term.
By explicit calculation, one can verify that, assuming again that \eqref{BVgau010} 
holds, $S_{\mathfrak{g}|0M}$ satisfies the quantum BV master equation 
\eqref{BVacts2}. One notice, using systematically \eqref{BValg7}, \eqref{BValg8}, that 
$S_{\mathfrak{g}|0M}$ satisfies \eqref{BVacts2}, if simultaneously $S_{\mathfrak{g}|0}$, $S_M$ 
satisfy \eqref{BVacts2}, $S_M$ satisfies \eqref{BVgau05}
and $\lambda_M$ satisfies \eqref{BVgau04a}, \eqref{BVgau04b}.
$S_{\mathfrak{g}|0M}$ is the $N=0$ gauged matter quantum BV master action.
Proceeding in a similar fashion, we find that $S_{\mathfrak{g}|0M}$ satisfies also \eqref{BVgau05},
so that $S_{\mathfrak{g}|0M}$ is invariant under the $\mathfrak{g}$--action $l_{\mathfrak{g}|0M}$. 

The coupling of ghosts and matter in the quantum master action $S_{\mathfrak{g}|0M}$ modifies 
the action of their respective quantum BV operators: $\delta_{\mathfrak{g}|0M}$ extends non trivially
$\delta_{\mathfrak{g}|0}$, $\delta_M$,  that is 
$\delta_{\mathfrak{g}|0M}\not=\delta_{\mathfrak{g}|0}\otimes 1_M+1_{\mathfrak{g}|0}\otimes \delta_M$. 
One has instead 
\begin{subequations}
\begin{align}
&\delta_{\mathfrak{g}|0M}(b_i\otimes 1_M)
=\delta_{\mathfrak{g}|0}b_i\otimes 1_M+ 1_{\mathfrak{g}|0}\otimes\lambda_{Mi},
\vphantom{\bigg]}
\label{BVgau017a}
\\
&\delta_{\mathfrak{g}|0M}(c^i\otimes 1_M)
=\delta_{\mathfrak{g}|0}c^i\otimes 1_M,
\vphantom{\bigg]}
\label{BVgau017b}
\\
&\delta_{\mathfrak{g}|0M}(1_{\mathfrak{g}|0}\otimes\phi)
=1_{\mathfrak{g}|0}\otimes\delta_M\phi+c^i\otimes l_{Mi}\phi, 
\vphantom{\bigg]}
\label{BVgau017c}
\end{align}
\label{BVgau017} 
\end{subequations}
\!\!
where $\delta_{\mathfrak{g}|0}b_i$, $\delta_{\mathfrak{g}|0}c^i$ are given by \eqref{BVgau013a},
\eqref{BVgau013b}, respectively.


\vspace{.1cm}

{\it Analysis of BV cohomology}.

\vspace{.1cm}

On physical grounds, not all the observables of the original matter
system remain such upon gauging the global symmetry. Only those which
are invariant under the symmetry do. They represent classes of 
the matter invariant quantum BV cohomology $H_{BV\mathrm{inv}}{}^*(\mathcal{A}_{M})$.  
So, {\it it is the invariant BV cohomology that is relevant rather than the 
ordinary one}. 

The map $\Upsilon_0:\mathcal{A}_M\rightarrow\mathcal{A}_{\mathfrak{g}|0M}$ defined by
\begin{equation}
\Upsilon_0\phi=1_{\mathfrak{g}|0}\otimes\phi, \qquad\phi\in\mathcal{A}_M,
\label{BVgaux03} 
\end{equation}
yields a natural embedding of 
$\mathcal{A}_M$ into $\mathcal{A}_{\mathfrak{g}|0M}$.
It is immediate to check that $\Upsilon_0$ 
is a monomorphism of BV algebras (cf. sect. \ref{sec:BValg}). 
Further, we have
\begin{equation}
l_{\mathfrak{g}|0Mx}\Upsilon_0=\Upsilon_0l_{Mx},
\label{BVgaux04} 
\end{equation}
for $x\in\mathfrak{g}$, and \hphantom{xxxxxxxxxxxxxxxxxxxxx} 
\begin{equation}
\delta_{\mathfrak{g}|0M}\Upsilon_0=\Upsilon_0\delta_M+c^i\otimes 1_M\cdot \Upsilon_0l_{Mi}.
\label{BVgaux05} 
\end{equation}
By \eqref{BVgaux04}, $\Upsilon_0$ maps the matter invariant subalgebra
$\mathcal{A}_{M\mathrm{inv}}$ into the gauged matter invariant 
subalgebra $\mathcal{A}_{\mathfrak{g}|0M\mathrm{inv}}$.
By \eqref{BVgaux05}, $\Upsilon_0|_{\mathcal{A}_{M\mathrm{inv}}}$ is
a chain map of the matter and gauged matter invariant differential spaces 
$(\mathcal{A}_{M\mathrm{inv}},\delta_M)$, 
$(\mathcal{A}_{\mathfrak{g}|0M\mathrm{inv}},\delta_{\mathfrak{g}|0M})$.
So, $\Upsilon_0|_{\mathcal{A}_{M\mathrm{inv}}}$ induces 
a homomorphism of the matter and gauged matter invariant quantum BV cohomology spaces 
$H_{BV\mathrm{inv}}{}^*(\mathcal{A}_{M})$,
$H_{BV\mathrm{inv}}{}^*(\mathcal{A}_{\mathfrak{g}|0M})$.
The homomorphism is not a monomorphism in general and, so, 
$H_{BV\mathrm{inv}}{}^*(\mathcal{A}_{M})$ is not naturally 
embedded in $H_{BV\mathrm{inv}}{}^*(\mathcal{A}_{\mathfrak{g}|0M})$.
This renders the study of the observables in the gauged matter theory 
a bit problematic. The way out is the following.

From \eqref{BVgau06}, we notice that the $N=0$ ghost algebra $\mathcal{A}_{\mathfrak{g}|0}$ 
contains as a subalgebra the Chevalley--Eilenberg algebra
\begin{equation}
CE(\mathfrak{g})=\Fun(\mathfrak{g}[1])
\label{BVgaux1} 
\end{equation}
\!\!\cite{Chevalley1,Koszul1}.
$CE(\mathfrak{g})$ is generated by the $c^i$. 
By \eqref{BVgau011b}, $CE(\mathfrak{g})$ is stable
under the $\mathfrak{g}$--action $l_{\mathfrak{g}|0}$.
By \eqref{BVgau013b}, $CE(\mathfrak{g})$ is also stable
under the BV operator $\delta_{\mathfrak{g}|0}$. 
Thus, $(CE(\mathfrak{g}),\mathfrak{g},l_{\mathfrak{g}|0},\delta_{\mathfrak{g}|0})$ is a differential 
$\mathfrak{g}$--module (see appendix \ref{app:invcoh}). 
Inspecting \eqref{BVgau013b}, we realize that the BV cohomology $H_{BV}{}^*(CE(\mathfrak{g}))$ 
is the Chevalley--Eilenberg cohomology $H_{CE}{}^*(\mathfrak{g})$ of $\mathfrak{g}$. 
Similarly, from \eqref{BVgau011b}, \eqref{BVgau013b}, we see that the invariant 
BV cohomology $H_{BV\mathrm{inv}}{}^*(CE(\mathfrak{g}))$ 
is the invariant Chevalley--Eilenberg cohomology $H_{CE\mathrm{inv}}{}^*(\mathfrak{g})$ of $\mathfrak{g}$. 
$H_{CE}{}^*(\mathfrak{g})$ is not known in general, but it is under
the weak assumption that $\mathfrak{g}$ is reductive, i. e. the direct
sum of a semisimple and an Abelian Lie algebra, in which case 
$H_{CE}{}^*(\mathfrak{g})\simeq CE(\mathfrak{g})_{\mathrm{inv}}$,
the invariant subalgebra of $CE(\mathfrak{g})$. 
We recall that reductive Lie algebras are unimodular. So, this result fits 
usefully in the theory developed above.
$H_{CE\mathrm{inv}}{}^*(\mathfrak{g})\simeq CE(\mathfrak{g})_{\mathrm{inv}}$
always. 
We note that the classical and quantum BV operators are equal on $CE(\mathfrak{g})$, since, 
by \eqref{BVgau07}, $\Delta_{\mathfrak{g}|0}$ vanishes on $CE(\mathfrak{g})$ and, so, 
the classical and the quantum BV cohomologies coincide. 

The $N=0$ gauged matter algebra $\mathcal{A}_{\mathfrak{g}|0M}$ contains as a subalgebra 
\begin{equation}
\mathcal{A}^+{}_{\mathfrak{g}|0M}=CE(\mathfrak{g})\otimes \mathcal{A}_M.
\label{BVgaux2} 
\end{equation}
By \eqref{BVgau015}, \eqref{BVgau011b}, $\mathcal{A}^+{}_{\mathfrak{g}|0M}$ is stable
under the $\mathfrak{g}$--action $l_{\mathfrak{g}|0M}$.
Similarly, by \eqref{BVgau017b}, \eqref{BVgau017c}, \eqref{BVgau013b}, 
$\mathcal{A}^+{}_{\mathfrak{g}|0M}$ is stable under the BV operator $\delta_{\mathfrak{g}|0M}$. 
Thus, $(\mathcal{A}^+{}_{\mathfrak{g}|0M},\mathfrak{g},l_{\mathfrak{g}|0M},\delta_{\mathfrak{g}|0M})$ 
is a differential 
$\mathfrak{g}$--module. By \eqref{BVgau017b}, \eqref{BVgau017c}, \eqref{BVgau013b}, 
the quantum BV cohomology $H_{BV}{}^*(\mathcal{A}^+{}_{\mathfrak{g}|0M})$ 
is the Chevalley--Eilenberg  cohomology $H_{CE}{}^*(\mathfrak{g},\mathcal{A}_M)$ of 
$\mathfrak{g}$ with coefficients in the differential space $(\mathcal{A}_M,\delta_M)$. 
Similarly, by \eqref{BVgau015}, \eqref{BVgau011b}, 
\eqref{BVgau017b}, \eqref{BVgau017c}, \eqref{BVgau013b}, 
the invariant quantum BV cohomology
$H_{BV\mathrm{inv}}{}^*(\mathcal{A}^+{}_{\mathfrak{g}|0M})$ 
is the invariant Chevalley--Eilenberg cohomology 
$H_{CE\mathrm{inv}}{}^*(\mathfrak{g},\mathcal{A}_M)$ of $\mathfrak{g}$
with coefficients in the differential $\mathfrak{g}$--module $(\mathcal{A}_M,\mathfrak{g},l_M,\delta_M)$.
Unlike for the pure ghost system, the quantum and classical BV operators are generally 
different in the matter sector and, so, it is necessary 
to distinguish the classical and the quantum BV cohomologies.  
Anyway, analogous statements hold in the classical case, with the proviso that 
$(\mathcal{A}_M,\delta_{M{ c}})$ is a differential algebra in this case. 

Let us now come back to the problem of the cohomological analysis of observables
in the gauged matter theory. 
We notice that the range of the BV algebra homomorphism
$\Upsilon_0:\mathcal{A}_M\rightarrow\mathcal{A}_{\mathfrak{g}|0M}$ 
is contained in $\mathcal{A}^+{}_{\mathfrak{g}|0M}$. 
By \eqref{BVgaux04}, \eqref{BVgaux05}, $\Upsilon_0|_{\mathcal{A}_{M\mathrm{inv}}}$ is
a chain map of the invariant differential spaces 
$(\mathcal{A}_{M\mathrm{inv}},\delta_M)$, 
$(\mathcal{A}^+{}_{\mathfrak{g}|0M\mathrm{inv}},\delta_{\mathfrak{g}|0M})$.
Thus, $\Upsilon_0|_{\mathcal{A}_{M\mathrm{inv}}}$ induces 
a homomorphism of the invariant BV cohomology spaces 
$H_{BV\mathrm{inv}}{}^*(\mathcal{A}_{M})$,
$H_{BV\mathrm{inv}}{}^*(\mathcal{A}^+{}_{\mathfrak{g}|0M})$,
which can be shown to be a monomorphism. So, 
$H_{BV\mathrm{inv}}{}^*(\mathcal{A}_{M})$ is naturally 
embedded in $H_{BV\mathrm{inv}}{}^*(\mathcal{A}^+{}_{\mathfrak{g}|0M})$.
In this way, {\it the study of the observables in the gauged matter theory is
naturally framed in that of the invariant 
BV cohomology $H_{BV\mathrm{inv}}{}^*(\mathcal{A}^+{}_{\mathfrak{g}|0M})$
of $\mathcal{A}^+{}_{\mathfrak{g}|0M}$}.
In fact, more can be shown \cite{Grueb1}. {\it If the Lie algebra $\mathfrak{g}$ is reductive, then} 
\begin{equation}
H_{BV\mathrm{inv}}{}^*(\mathcal{A}^+{}_{\mathfrak{g}|0M})
\simeq CE(\mathfrak{g})_{\mathrm{inv}}\otimes H_{BV\mathrm{inv}}{}^*(\mathcal{A}_{M}).
\label{BVgau0x6} 
\end{equation}
A self--contained proof of \eqref{BVgau0x6} is given in appendix \ref{app:invcoh}. 
\eqref{BVgau0x6} indicates that {\it the gauged matter algebra $\mathcal{A}_{\mathfrak{g}|0M}$
contains objects which may reasonably considered to be observables, but which do not arise 
from the original matter algebra $\mathcal{A}_M$}. They are the {\it pure gauge theoretic observables}.

\vfill\eject

\section{\normalsize \textcolor{blue}{N=1 BV gauging and N=1 ghost system}}\label{sec:BVgauge1}

~~~~$N=1$ gauging follows in outline the same steps as $N=0$ gauging, 
though the $N=1$ ghost system 
has a larger amount of ghost supersymmetry than its $N=0$ counterpart.  
However, there are some significant differences, the most 
conspicuous of which is that the unimodularity of the symmetry 
Lie algebra $\mathfrak{g}$ is no longer required for the consistency 
of the construction.

\vspace{.1cm}

{\it $N=1$ $\mathfrak{g}$--actions}

\vspace{.1cm}

Let $(\mathcal{A},\Delta,\{\cdot,\cdot\})$ be a BV algebra.
Let $\mathfrak{g}$ be a Lie algebra. 
An $N=1$ $\mathfrak{g}$--{\it action} on the BV algebra is 
a pair of linear maps $i:\mathfrak{g}\rightarrow \Der_{-1}(\mathcal{A})$,
$l:\mathfrak{g}\rightarrow \Der_0(\mathcal{A})$
satisfying the 
commutation relations
\begin{subequations}
\begin{align}
&[i_x,i_y]=0,
\vphantom{\bigg]}
\label{BVgau11a} 
\\
&[l_x,i_y]=i_{[x,y]},
\vphantom{\bigg]}
\label{BVgau11b} 
\\
&[l_x,l_y]=l_{[x,y]},
\vphantom{\bigg]}
\label{BVgau11c} 
\\
&[i_x,\Delta]=0,
\vphantom{\bigg]}
\label{BVgau11d} 
\\
&[l_x,\Delta]=0,
\vphantom{\bigg]}
\label{BVgau11e} 
\end{align}
\label{BVgau11}
\end{subequations}
\!\!with $x,y\in\mathfrak{g}$.
Note that an $N=1$ action is automatically also an $N=0$ action
(cf. sect. \ref{sec:BVgauge0}). 
By \eqref{BVgau11d}, \eqref{BVgau11e}, for $x\in\mathfrak{g}$, 
$i_x\in\Der_{-1}(\mathcal{A})$, $l_x\in\Der_0(\mathcal{A})$
are both BV derivations (cf. sect. \ref{sec:BValg}).

Let $S$ be a quantum BV master action of the BV algebra $(\mathcal{A},\Delta,\{\cdot,\cdot\})$. 
$S$ is said {\it invariant} under the $\mathfrak{g}$--action if 
\begin{subequations}
\begin{align}
&l_x=\ad i_xS, 
\vphantom{\bigg]}
\label{BVgau12a} 
\\
&l_xS=0,
\vphantom{\bigg]}
\label{BVgau12b} 
\end{align} 
\label{BVgau12}
\end{subequations}
\!\!for all $x\in\mathfrak{g}$.
These conditions are compatible with \eqref{BVgau11a}--\eqref{BVgau11e} and 
the quantum 
\vfill\eject\noindent
BV master equation \eqref{BVacts2}. Note that this notion of invariance is more restrictive
than the corresponding one of the $N=0$ case: it is not simply a condition on $S$, but also 
on $l$. When $S$ is invariant, one has
\begin{subequations}
\begin{align}
&[\delta,i_x]=l_x,
\vphantom{\bigg]}
\label{BVgau12xa} 
\\
&[\delta,l_x]=0,
\vphantom{\bigg]}
\label{BVgau12xb} 
\end{align} 
\label{BVgau12x}
\end{subequations}
\!\!where $\delta$ is the quantum BV operator (cf. sect. \eqref{sec:BVacts}, eq. \eqref{BVacts8}). 
By \eqref{BVacts9}, \eqref{BVgau11a}--\eqref{BVgau11c}, \eqref{BVgau12xa}, \eqref{BVgau12xb},
$(\mathcal{A},\mathfrak{g},i,l,\delta)$ is an algebraic structure known as a 
$\mathfrak{g}$--operation \cite{Grueb1} (see appendix \ref{app:equivcoh} for a 
review of Lie operations).
By \eqref{BVgau12xa}, \eqref{BVgau12xb}, it is possible to
define a $\mathfrak{g}$--basic quantum BV cohomology, that is the 
cohomology of the differential space 
$(\mathcal{A}_{\mathrm{bas}},\delta)$, where $\mathcal{A}_{\mathrm{bas}}=
\cap_{x\in\mathfrak{g}}(\ker i_x\cap \ker l_x)\subset\mathcal{A}$. 
The same statements hold also for the classical BV 
operator and its cohomology. 

The action is called {\it BV Hamiltonian}, 
if there exists a pair of linear maps $\iota:\mathfrak{g}\rightarrow \mathcal{A}_{-2}$, 
$\lambda:\mathfrak{g}\rightarrow \mathcal{A}_{-1}$, called below {\it BV premoment} 
and {\it moment map}, respectively, such that \hphantom{xxxxxxxxxxxxxxxxxxxxxxxxxx}
\begin{subequations}
\begin{align}
&i_x=\ad\iota_x,  
\vphantom{\bigg]}
\label{BVgau13a} 
\\
&l_x=\ad\lambda_x, 
\vphantom{\bigg]}
\label{BVgau13b} 
\end{align}
\label{BVgau13}
\end{subequations}
\!\!with $x\in\mathfrak{g}$ and that \hphantom{xxxxxxxxxxxxxxxxxxxxxxxxxxx}
\begin{subequations}
\begin{align}
&\{\iota_x,\iota_y\}=0,
\vphantom{\bigg]}
\label{BVgau14a} 
\\
&\{\lambda_x,\iota_y\}=\iota_{[x,y]},
\vphantom{\bigg]}
\label{BVgau14b} 
\\
&\{\lambda_x,\lambda_y\}=\lambda_{[x,y]},
\vphantom{\bigg]}
\label{BVgau14c} 
\\
&\Delta\iota_x=0,
\vphantom{\bigg]}
\label{BVgau14d}
\\
&\Delta\lambda_x=0,
\vphantom{\bigg]}
\label{BVgau14e} 
\end{align}
\label{BVgau14}
\end{subequations}
\!\!with $x,y\in\mathfrak{g}$. 
\eqref{BVgau13a}, \eqref{BVgau13b} together with \eqref{BVgau14a}--\eqref{BVgau14e} 
are indeed sufficient for 
\vfill\eject\noindent
\eqref{BVgau11a}--\eqref{BVgau11e} to hold.
By \eqref{BVgau13a}, \eqref{BVgau13b}, \eqref{BVgau14d},
\eqref{BVgau14e},  $i_x\in\Der_{-1}(\mathcal{A})$, $l_x\in\Der_0(\mathcal{A})$ 
are both BV inner derivations (cf. sect. \ref{sec:BValg}). 
Note that the underlying $N=0$ $\mathfrak{g}$--action is also Hamiltonian
(cf. sect. \ref{sec:BVgauge0}).
Below, we consider only BV Hamiltonian $N=1$ actions.

A quantum BV master action $S$ of the BV algebra $(\mathcal{A},\Delta,\{\cdot,\cdot\})$
is {\it Hamiltonian invariant} under the $\mathfrak{g}$--action, if $S$ satisfies  
\begin{subequations}
\begin{align}
&\{\iota_x,S\}=\lambda_x,
\vphantom{\bigg]}
\label{BVgau15a} 
\\
&\{\lambda_x,S\}=0,
\vphantom{\bigg]}
\label{BVgau15b} 
\end{align} 
\label{BVgau15}
\end{subequations}
\!\!for all $x\in\mathfrak{g}$. Hamiltonian invariance is stricter than simple invariance.
If $S$ were simply invariant, \eqref{BVgau15a} would hold only up to a central element
of the Gerstenhaber algebra 
$\mathcal{A}_G$, 
as  \eqref{BVgau12a}, \eqref{BVgau13a}, \eqref{BVgau13b} imply only that 
$\lambda_x-\{\iota_x,S\}$ is central. 
In the $N=0$ case,  
there is no similar distinction between simple and Hamiltonian invariance.
\eqref{BVgau14d}, \eqref{BVgau15a} combined imply that 
$\lambda_x$ is a coboundary of both the classical and the quantum BV cohomology 
(cf. sect. \ref{sec:BVacts}). 
Recall that in the $N=0$ case, $\lambda_x$ is a cocycle in general. 
\eqref{BVgau15b} is not an independent relation; it 
follows from \eqref{BVgau14d}, \eqref{BVgau14e}, \eqref{BVgau15a} and the master equation
\eqref{BVacts2}. 

\vspace{.1cm}

{\it Gauging of a global $N=1$ $\mathfrak{g}$--symmetry}

\vspace{.1cm}

Consider a {\it matter} BV algebra $(\mathcal{A}_M,\Delta_M,\{\cdot,\cdot\}_M)$ 
carrying a BV Hamiltonian $N=1$ $\mathfrak{g}$--action $i_M$, $l_M$ with BV (pre)moment 
maps $\iota_M$, $\lambda_M$ and a matter quantum BV master action $S_M$ invariant under 
the $\mathfrak{g}$--action. By \eqref{BVgau15a}, \eqref{BVgau15b}, 
we may say that $S_M$ enjoys a {\it global $N=1$ $\mathfrak{g}$--symmetry}
and that $\lambda_M$ is the corresponding {\it symmetry charge}, extending,
perhaps with some abuse, the terminology of the $N=0$ case. 
However, here, the symmetry is derived in the sense that 
\eqref{BVgau15b} is actually a consequence of the more basic relation
\eqref{BVgau15a}, unlike for $N=0$.
We want to gauge this symmetry in a way that reflects this richer structure. 

As the in $N=0$ case, the gauging proceeds in three steps. 

{\it 1.} We construct an $N=1$ {\it ghost} BV algebra
$(\mathcal{A}_{\mathfrak{g}|1},\Delta_{\mathfrak{g}|1},\{\cdot,\cdot\}_{\mathfrak{g}|1})$
with a BV Hamiltonian $N=1$ $\mathfrak{g}$--action $i_{\mathfrak{g}|1}$, $l_{\mathfrak{g}|1}$
with BV (pre)moment map $\iota_{\mathfrak{g}|1}$, $\lambda_{\mathfrak{g}|1}$
and an $N=1$ ghost quantum BV master action $S_{\mathfrak{g}|1}$ 
invariant under the $\mathfrak{g}$--action. The construction is canonical, 
depending on $\mathfrak{g}$ only.

{\it 2.} 
We construct an $N=1$ {\it gauged matter} BV algebra
$(\mathcal{A}_{\mathfrak{g}|1M},\Delta_{\mathfrak{g}|1M}$, $\{\cdot,\cdot\}_{\mathfrak{g}|1M})$
and equip it with an appropriate BV Hamiltonian $N=1$ $\mathfrak{g}$--action $i_{\mathfrak{g}|1M}$,
$l_{\mathfrak{g}|1M}$ with BV (pre)moment maps $\iota_{\mathfrak{g}|1M}$, $\lambda_{\mathfrak{g}|1M}$.

{\it 3.} 
We construct an $N=1$ gauged matter quantum BV master action $S_{\mathfrak{g}|1M}$ of the gauged 
matter BV algebra invariant under the $\mathfrak{g}$--action.

{\it Step 1}. For a Lie algebra $\mathfrak{g}$, define 
\begin{equation}
\mathcal{A}_{\mathfrak{g}|1}=\Fun(\mathfrak{g}^\vee[-2]
\oplus\mathfrak{g}[1]\oplus\mathfrak{g}^\vee[-3]\oplus\mathfrak{g}[2])
\label{BVgau16} 
\end{equation}
Denote by $b_i$, $c^i$, $B_i$, $C^i$, $i=1,\dots,\dim\mathfrak{g}$, the coordinates of 
$\mathfrak{g}^\vee[-2],\mathfrak{g}[1],\mathfrak{g}^\vee[-3]$, $\mathfrak{g}[2]$,
respectively, 
corresponding to a chosen basis $\{t_i\}$ of $\mathfrak{g}$.
Then, $\mathcal{A}_{\mathfrak{g}|1}$ can be 
viewed as the $\mathbb{Z}$ graded commutative associative 
unital algebra of polynomials in the $b_i$, $c^i$, $B_i$, $C^i$. 
Define next the 2nd order differential operator
\begin{equation}
\Delta_{\mathfrak{g}|1}=\partial_{Lb}{}^i \partial_{Lci}-\partial_{LB}{}^i \partial_{LCi}
\label{BVgau17} 
\end{equation}
and the bilinear bracket
\begin{align}
\{\phi,\psi\}_{\mathfrak{g}|1}&=\partial_{Rb}{}^i\phi\partial_{Lci}\psi-\partial_{Rci}\phi\partial_{Lb}{}^i\psi
\hspace{3.1cm}
\vphantom{\bigg]}
\label{BVgau18}
\\
&\hspace{3.1cm}+\partial_{RB}{}^i\phi\partial_{LCi}\psi-\partial_{RCi}\phi\partial_{LB}{}^i\psi,
\qquad \phi,\psi\in \mathcal{A}_{\mathfrak{g}|1}.
\vphantom{\bigg]}
\nonumber
\end{align}
Then, it is simple to check that
relations \eqref{BValg1}--\eqref{BValg3} are verified. It follows that 
$(\mathcal{A}_{\mathfrak{g}|1},\Delta_{\mathfrak{g}|1},\{\cdot,\cdot\}_{\mathfrak{g}|1})$
is a BV algebra, the $N=1$  ghost BV algebra.

Let $f^i{}_{jk}$ be the structure constants of $\mathfrak{g}$ with respect to the basis 
$\{t_i\}$. Set
\begin{subequations}
\begin{align}
&\iota_{\mathfrak{g}|1i}=b_i
\vphantom{\bigg]}
\label{BVgau19a} 
\\
&\lambda_{\mathfrak{g}|1i}=f^j{}_{ki}b_jc^k+f^j{}_{ki}B_jC^k.
\vphantom{\bigg]}
\label{BVgau19b} 
\end{align}
\label{BVgau19}
\end{subequations}
\!\!Since $\iota_{\mathfrak{g}|1i}\in\mathcal{A}_{\mathfrak{g}|1-2}$, 
$\lambda_{\mathfrak{g}|1i}\in\mathcal{A}_{\mathfrak{g}|1-1}$, 
they define via \eqref{BVgau13a}, \eqref{BVgau13b} linear maps
$i_{\mathfrak{g}|1}:\mathfrak{g}\rightarrow \Der_{-1}(\mathcal{A}_{\mathfrak{g}|1})$,
$l_{\mathfrak{g}|1}:\mathfrak{g}\rightarrow \Der_0(\mathcal{A}_{\mathfrak{g}|1})$.
The pair $i_{\mathfrak{g}|1}$, $l_{\mathfrak{g}|1}$ is a BV Hamiltonian $N=1$ $\mathfrak{g}$--action 
on the ghost BV algebra having $\iota_{\mathfrak{g}|1}$, $\lambda_{\mathfrak{g}|1}$
as BV (pre)moment maps. Indeed, by 
\eqref{BVgau17}, \eqref{BVgau18}, \eqref{BVgau19a}, \eqref{BVgau19b}, 
relations \eqref{BVgau14a}--\eqref{BVgau14e} are verified. 
The Lie algebra $\mathfrak{g}$ no longer needs to be unimodular
(cf. eq. \eqref{BVgau010}), as in the $N=0$ case, 
due to the cancellation of the offending terms $f^j{}_{ji}$ of the $bc$ and 
$BC$ sectors. 

The action of $i_{\mathfrak{g}|1}$, $l_{\mathfrak{g}|1}$ on $b_i$, $c^i$, $B_i$, $C^i$  
is given by
\begin{subequations}
\begin{align}
&i_{\mathfrak{g}|1i}b_j=0,
\vphantom{\bigg]}
\label{BVgau110a} 
\\
&i_{\mathfrak{g}|1i}c^j=\delta_i{}^j,
\vphantom{\bigg]}
\label{BVgau110b} 
\\
&i_{\mathfrak{g}|1i}B_j=0,
\vphantom{\bigg]}
\label{BVgau110c} 
\\
&i_{\mathfrak{g}|1i}C^j=0,
\vphantom{\bigg]}
\label{BVgau110d} 
\\
&l_{\mathfrak{g}|1i}b_j=f^k{}_{ij}b_k,
\vphantom{\bigg]}
\label{BVgau110e} 
\\
&l_{\mathfrak{g}|1i}c^j=-f^j{}_{ik}c^k,
\vphantom{\bigg]}
\label{BVgau110f} 
\\
&l_{\mathfrak{g}|1i}B_j=f^k{}_{ij}B_k,
\vphantom{\bigg]}
\label{BVgau110g} 
\\
&l_{\mathfrak{g}|1i}C^j=-f^j{}_{ik}C^k,
\vphantom{\bigg]}
\label{BVgau110h} 
\end{align}
\label{BVgau110} 
\end{subequations} 
as follows readily from \eqref{BVgau13a}, \eqref{BVgau13b}, \eqref{BVgau18}, \eqref{BVgau19a},
\eqref{BVgau19b}. 

The $N=1$ ghost algebra $\mathcal{A}_{\mathfrak{g}|1}$ contains an element
$S_{\mathfrak{g}|1}\in\mathcal{A}_{\mathfrak{g}|10}$ given by 
\begin{equation}
S_{\mathfrak{g}|1}=-\frac{1}{2}f^i{}_{jk}b_ic^jc^k+b_iC^i+f^i{}_{jk}B_ic^jC^k.
\label{BVgau111} 
\end{equation}
$S_{\mathfrak{g}|1}$ satisfies both the classical and the quantum BV master equation 
\eqref{BVacts4}. This follows from the definition \eqref{BVgau111} using  \eqref{BVgau17}, \eqref{BVgau18}.
Again, $\mathfrak{g}$ needs not to be unimodular, as in the $N=0$ case, 
due to the cancellation of the anomalous terms $f^i{}_{ij}c^j$ originating in the  
$bc$ and $BC$ sectors. $S_{\mathfrak{g}|1}$ is the $N=1$ ghost quantum BV master action.
$S_{\mathfrak{g}|1}$ is invariant under the $\mathfrak{g}$--action $i_{\mathfrak{g}|1}$, $l_{\mathfrak{g}|1}$. 
\eqref{BVgau18}, \eqref{BVgau19a}, \eqref{BVgau19b}, \eqref{BVgau111} indeed imply 
\eqref{BVgau15a}, \eqref{BVgau15b}.

The action of the quantum BV operator $\delta_{\mathfrak{g}|1}$ on $b_i$, $c^i$, $B_i$, $C^i$ reads
\begin{subequations}
\begin{align}
&\delta_{\mathfrak{g}|1}b_i=f^k{}_{ji}b_kc^j+f^k{}_{ji}B_kC^j,
\vphantom{\bigg]}
\label{BVgau112a} 
\\
&\delta_{\mathfrak{g}|1}c^i=C^i-\frac{1}{2}f^i{}_{jk}c^jc^k.
\vphantom{\bigg]}
\label{BVgau112b} 
\\
&\delta_{\mathfrak{g}|1}B_i=-b_i-f^k{}_{ji}B_kc^j,
\vphantom{\bigg]}
\label{BVgau112ax} 
\\
&\delta_{\mathfrak{g}|1}C^i=-f^i{}_{jk}c^jC^k,
\vphantom{\bigg]}
\label{BVgau112bx} 
\end{align}
\label{BVgau112} 
\end{subequations}
as follows by direct application of the definition 
\eqref{BVacts8}. Relations \eqref{BVgau112} are also the expressions 
of the action of the classical BV operator $\delta_{\mathfrak{g}|1 c}$
defined according to \eqref{BVacts11}. Analogously to the $N=0$ case, the action
of $\delta_{\mathfrak{g}|1}$ and $\delta_{\mathfrak{g}|1 c}$ on higher
polynomials in $b_i$, $B_i$, $c^i$, $C^i$ is different 
because $\Delta_{\mathfrak{g}|1}$ acts on them non trivially in general. 

The $N=1$ ghost system has an elegant superfield formulation that is illustrated in appendix
\ref{sec:N=1ghost}.

{\it Step 2.} 
The $N=1$ gauged matter BV algebra
$(\mathcal{A}_{\mathfrak{g}|1M},\Delta_{\mathfrak{g}|1M}$, $\{\cdot,\cdot\}_{\mathfrak{g}|1M})$
is the tensor product of the $N=1$ ghost BV algebra
$(\mathcal{A}_{\mathfrak{g}|1},\Delta_{\mathfrak{g}|1}$, $\{\cdot,\cdot\}_{\mathfrak{g}|1})$
and the matter BV algebra $(\mathcal{A}_M,\Delta_M,\{\cdot,\cdot\}_M)$
(cf. sect. \ref{sec:BValg}), analogously to the $N=0$ case. 
The elements $\iota_{\mathfrak{g}|1Mi}\in\mathcal{A}_{\mathfrak{g}|1M-2}$,
$\lambda_{\mathfrak{g}|1Mi}\in\mathcal{A}_{\mathfrak{g}|1M-1}$ given by 
\begin{subequations}
\begin{align}
&\iota_{\mathfrak{g}|1Mi}=\iota_{\mathfrak{g}|1i}\otimes 1_M,
\vphantom{\bigg]}
\label{BVgau117a} 
\\
&\lambda_{\mathfrak{g}|1Mi}=\lambda_{\mathfrak{g}|1i}\otimes 1_M+1_{\mathfrak{g}|1}\otimes\lambda_{Mi}
\vphantom{\bigg]}
\label{BVgau117b} 
\end{align}
\label{BVgau117} 
\end{subequations}
\!\!define via \eqref{BVgau13a}, \eqref{BVgau13b} 
linear maps $i_{\mathfrak{g}|1M}:\mathfrak{g}\rightarrow \Der_{-1}(\mathcal{A}_{\mathfrak{g}|1M})$,
$l_{\mathfrak{g}|1M}:\mathfrak{g}\rightarrow \Der_0(\mathcal{A}_{\mathfrak{g}|1M})$.
The pair $i_{\mathfrak{g}|1M}$, $l_{\mathfrak{g}|1M}$ is a BV Hamiltonian $N=1$ 
$\mathfrak{g}$--action on the gauged matter BV algebra having $\iota_{\mathfrak{g}|1M}$,
$\lambda_{\mathfrak{g}|1M}$ as BV (pre)moment maps. One just notices that $\iota_{\mathfrak{g}|1M}$,
$\lambda_{\mathfrak{g}|1M}$ satisfy \eqref{BVgau14a}--\eqref{BVgau14e} if 
$\iota_{\mathfrak{g}|1}$, 
$\lambda_{\mathfrak{g}|1}$, $\lambda_M$ do, by \eqref{BValg7}, \eqref{BValg8}. 
The $\mathfrak{g}$--action $i_{\mathfrak{g}|1M}$ $l_{\mathfrak{g}|1M}$ extends the $\mathfrak{g}$--actions 
$i_{\mathfrak{g}|1}$, $l_{\mathfrak{g}|1}$, $i_M$, $l_M$ as 
\begin{subequations}
\begin{align}
&i_{\mathfrak{g}|1Mi}=i_{\mathfrak{g}|1i}\otimes 1_M,
\vphantom{\bigg]}
\label{BVgau118a} 
\\
&l_{\mathfrak{g}|1Mi}=l_{\mathfrak{g}|1i}\otimes 1_M+1_{\mathfrak{g}|1}\otimes l_{Mi}
\vphantom{\bigg]}
\label{BVgau118b} 
\end{align}
\label{BVgau118} 
\end{subequations}
\!\!for $x\in \mathfrak{g}$. 
Unlike the $N=0$ case, this extension is non trivial:
in the right hand side of \eqref{BVgau118a}, a term $1_{\mathfrak{g}|1}\otimes i_{Mi}$
is absent. 

{\it Step 3.}  The $N=1$ gauged matter algebra $\mathcal{A}_{\mathfrak{g}|1M}$ 
contains a distinguished element $S_{\mathfrak{g}|1M}\in\mathcal{A}_{\mathfrak{g}|1M0}$ given by 
\begin{equation}
S_{\mathfrak{g}|1M}=S_{\mathfrak{g}|1}\otimes 1_M+1_{\mathfrak{g}|1}\otimes S_M+c^i\otimes\lambda_{Mi}
-C^i\otimes\iota_{Mi}.
\label{BVgau119} 
\end{equation} 
As in the $N=0$ case, the first two terms correspond to the trivial
non interacting ghost--matter action \eqref{BVacts5} while the third and fourth terms 
are genuine ghost--matter interaction terms.
By explicit calculation, one can verify that $S_{\mathfrak{g}|1M}$ satisfies the quantum 
BV master equation \eqref{BVacts2}, noticing that, by systematic use of \eqref{BValg7}, \eqref{BValg8}, 
that $S_{\mathfrak{g}|1M}$ satisfies \eqref{BVacts2}, if simultaneously $S_{\mathfrak{g}|1}$, $S_M$ 
satisfy \eqref{BVacts2}, $S_M$ satisfies \eqref{BVgau15a}, \eqref{BVgau15b},  
and $\iota_M$, $\lambda_M$ satisfy \eqref{BVgau14a}--\eqref{BVgau14e}.
$S_{\mathfrak{g}|1M}$ is the $N=1$ gauged matter quantum BV master action.
As $S_{\mathfrak{g}|1M}$ satisfies also \eqref{BVgau15a}, \eqref{BVgau15b},
$S_{\mathfrak{g}|1M}$ is invariant under the $\mathfrak{g}$--action $l_{\mathfrak{g}|1M}$. 

As in the $N=0$ case, the coupling of ghosts and matter in the quantum master action $S_{\mathfrak{g}|1M}$ 
modifies 
the action of their respective quantum BV operators: 
$\delta_{\mathfrak{g}|1M}\not=\delta_{\mathfrak{g}|1}\otimes 1_M+1_{\mathfrak{g}|1}\otimes \delta_M$ 
and, so, $\delta_{\mathfrak{g}|1M}$ extends non trivially
$\delta_{\mathfrak{g}|1}$, $\delta_M$. One has instead 
\begin{subequations}
\begin{align}
&\delta_{\mathfrak{g}|1M}(b_i\otimes 1_M)
=\delta_{\mathfrak{g}|1}b_i\otimes 1_M+1_{\mathfrak{g}|1}\otimes\lambda_{Mi}, \hspace{1.91cm}
\vphantom{\bigg]}
\label{BVgau122}
\\
&\delta_{\mathfrak{g}|1M}(c^i\otimes 1_M)
=\delta_{\mathfrak{g}|1}c^i\otimes 1_M,
\vphantom{\bigg]}
\label{BVgau124}
\\
&\delta_{\mathfrak{g}|1M}(B_i\otimes 1_M)
=\delta_{\mathfrak{g}|1}B_i\otimes 1_M+1_{\mathfrak{g}|1}\otimes\iota_{Mi},
\vphantom{\bigg]}
\label{BVgau123}
\end{align}
\begin{align}
&\delta_{\mathfrak{g}|1M}(C^i\otimes 1_M)
=\delta_{\mathfrak{g}|1}C^i\otimes 1_M,
\vphantom{\bigg]}
\label{BVgau125}
\\
&\delta_{\mathfrak{g}|1M}(1_{\mathfrak{g}|1}\otimes\phi)
=1_{\mathfrak{g}|1}\otimes\delta_M\phi+c^i\otimes l_{Mi}\phi
-C^i\otimes i_{Mi}\phi.
\vphantom{\bigg]}
\label{BVgau126}
\end{align}
\label{BVgau122-26} 
\end{subequations}
\!\!where $\delta_{\mathfrak{g}|0}b_i$, $\delta_{\mathfrak{g}|0}c^i$, $\delta_{\mathfrak{g}|0}B_i$, 
$\delta_{\mathfrak{g}|0}C^i$  are given by the expressions \eqref{BVgau112a}--\eqref{BVgau112bx}, 
respectively. 

\vspace{.1cm}

{\it Analysis of BV cohomology}

\vspace{.1cm}

Before beginning the study of BV cohomology, the following remarks are in order.
The construction illustrated above is modeled on topological gauge field theory.
In the Mathai--Quillen formulation of topological field theory \cite{Mathai1}, 
the computation of a topological correlator is reduced to that
of an integral of the form $\int_{\matheul{Z}(s)}\omega$, where $\omega$ is a closed form of the field space 
$\matheul{M}$ and $\matheul{Z}(s)$ is the submanifold of $\matheul{M}$ of solutions of a certain 
field equation $s=0$ (a phenomenon called localization).
Now, it turns out that $\int_{\matheul{Z}(s)}\omega=\int_{\matheul{M}}\omega\wedge e(\matheul{E})$, 
where $e(\matheul{E})$ is a closed form of $\matheul{M}$ representing 
the Euler class of an oriented Riemannian vector bundle $\matheul{E}$ 
over $\matheul{M}$, of which $s$ is a section. 
It is known that $e(\matheul{E})=s^*t(\matheul{E})$, where
$t(\matheul{E})$ is a closed form of $\matheul{E}$ representing the Thom class of $\matheul{E}$
(a distinguished element of the vertical rapid decrease cohomology of $\matheul{E}$). 
$t(\matheul{E})$ in turn yields the closed form $\pi^*t(\matheul{E})$ 
of the natural principal $G$--bundle $\pi:\matheul{P}\times V\rightarrow \matheul{E}$, 
where $\matheul{P}$ and $V$ are the oriented orthogonal frame principal bundle and the 
typical fiber of $\matheul{E}$, respectively, and $G\simeq SO(V)$ is the structure group 
of $\matheul{P}$. $\matheul{P}$ is endowed with a canonical $\mathfrak{g}$--operation, 
where $\mathfrak{g}$ is the Lie algebra of $G$ \cite{Grueb1}.  
The operation allows one to define basic forms of $\matheul{P}\times V$.
$\pi^*t(\matheul{E})$ is closed and basic. In this way, 
the problem of the computation of the original topological correlator can be formulated in terms 
of the basic cohomology and the closely related equivariant cohomology of 
the principal bundle $\matheul{P}\times V$.
See \cite{Moore1,Guillemin1} for up to date reviews of this subject matter.  

As observed at the beginning of this section, 
$N=1$ $\mathfrak{g}$--actions on BV algebras are instances of $\mathfrak{g}$--operations. 
So, it is reasonable to suppose that, in a BV algebraic
formulation, a topological field theory should be realized as matter BV algebra
with an $N=1$ $\mathfrak{g}$--action and an invariant BV master action.
For the reasons explained above, among all the observables of the matter system,
those which are basic under the $\mathfrak{g}$--action have a central role.
In certain topological field theories, the relevant basic observables
turn out to be non local. The way to restore locality is precisely the gauging of the 
$N=1$ $\mathfrak{g}$--symmetry.

Let us thus assume that the observables of the original matter theory which are relevant upon gauging 
are the basic ones. They represent classes of the matter basic quantum BV cohomology 
$H_{BV\mathrm{bas}}{}^*(\mathcal{A}_{M})$.  
So, {\it it is the basic BV cohomology that is relevant rather than the 
ordinary one}. 

We now proceed similarly as we did in the $N=0$ case (cf. sect.  \ref{sec:BVgauge0}). 
The map $\Upsilon_1:\mathcal{A}_M\rightarrow\mathcal{A}_{\mathfrak{g}|1M}$ defined by
\begin{equation}
\Upsilon_1\phi=1_{\mathfrak{g}|1}\otimes\phi,\qquad\phi\in\mathcal{A}_M,
\label{BVgaux13} 
\end{equation}
yields a natural embedding of 
$\mathcal{A}_M$ into $\mathcal{A}_{\mathfrak{g}|1M}$,
which is a monomorphism of BV algebras (cf. sect. \ref{sec:BValg}). 
Further, we have
\begin{subequations}
\begin{align}
&i_{\mathfrak{g}|1Mx}\Upsilon_1=0,
\vphantom{\bigg]}
\label{BVgaux14a} 
\\
&l_{\mathfrak{g}|1Mx}\Upsilon_1=\Upsilon_1l_{Mx},
\vphantom{\bigg]}
\label{BVgaux14b} 
\end{align}
\label{BVgaux14} 
\end{subequations}
\!\!for $x\in\mathfrak{g}$, and 
\hphantom{xxxxxxxxxxxxxxxxxxxxx}
\begin{equation}
\delta_{\mathfrak{g}|1M}\Upsilon_1=\Upsilon_1\delta_M
+c^i\otimes 1_M\cdot \Upsilon_1l_{Mi}-C^i\otimes 1_M \cdot\Upsilon_1i_{Mi}.
\label{BVgaux15} 
\end{equation}
By \eqref{BVgaux14a}, \eqref{BVgaux14b}, $\Upsilon_1$ maps the matter basic subalgebra
$\mathcal{A}_{M\mathrm{bas}}$ into the gauged matter basic 
subalgebra $\mathcal{A}_{\mathfrak{g}|1M\mathrm{bas}}$.
By \eqref{BVgaux15}, $\Upsilon_1|_{\mathcal{A}_{M\mathrm{bas}}}$ is
a chain map of the matter and gauged matter basic differential spaces 
$(\mathcal{A}_{M\mathrm{bas}},\delta_M)$, 
$(\mathcal{A}_{\mathfrak{g}|1M\mathrm{bas}},\delta_{\mathfrak{g}|1M})$.
So, $\Upsilon_1|_{\mathcal{A}_{M\mathrm{bas}}}$ induces 
a homomorphism of the matter and gauged matter basic quantum BV cohomologies
$H_{BV\mathrm{bas}}{}^*(\mathcal{A}_{M})$,
$H_{BV\mathrm{bas}}{}^*(\mathcal{A}_{\mathfrak{g}|1M})$.
As in the $N=0$ case, the homomorphism is not a monomorphism in general and, so, 
$H_{BV\mathrm{bas}}{}^*(\mathcal{A}_{M})$ is not naturally 
embedded in $H_{BV\mathrm{bas}}{}^*(\mathcal{A}_{\mathfrak{g}|1M})$.
As in the $N=0$ case again, this renders the study of the observables in the gauged matter theory 
problematic. The way out is similar in spirit.

From \eqref{BVgau16}, we observe that the $N=1$ ghost algebra $\mathcal{A}_{\mathfrak{g}|1}$ 
contains as a subalgebra the Weil algebra 
\begin{equation}
W(\mathfrak{g})=\Fun(\mathfrak{g}[1]\oplus \mathfrak{g}[2])
\label{BVgaux11} 
\end{equation}
\!\!\cite{Weil1,HCartan2,HCartan1}.
$W(\mathfrak{g})$ is generated by the $c^i$, $C^i$. 
By \eqref{BVgau110b}, \eqref{BVgau110d}, \eqref{BVgau110f}, \eqref{BVgau110h}, 
$W(\mathfrak{g})$ is stable
under the $\mathfrak{g}$--action $i_{\mathfrak{g}|1}$, $l_{\mathfrak{g}|1}$.
By \eqref{BVgau112b}, \eqref{BVgau112bx}, $W(\mathfrak{g})$ is also stable
under the BV operator $\delta_{\mathfrak{g}|1}$. 
Thus, $(W(\mathfrak{g}),\mathfrak{g},i_{\mathfrak{g}|1}, l_{\mathfrak{g}|1},\delta_{\mathfrak{g}|1})$ 
is a $\mathfrak{g}$--operation (see appendix \ref{app:equivcoh}). 
Upon inspecting \eqref{BVgau112b}, \eqref{BVgau112bx}, we recognize that the BV cohomology 
$H_{BV}{}^*(W(\mathfrak{g}))$ is the Weil algebra cohomology $H_W{}^*(\mathfrak{g})$ of $\mathfrak{g}$. 
Similarly, from \eqref{BVgau110b}, \eqref{BVgau110d}, \eqref{BVgau110f}, \eqref{BVgau110h}, 
\eqref{BVgau112b}, \eqref{BVgau112bx}, we see that the basic
BV cohomology $H_{BV\mathrm{bas}}{}^*(CE(\mathfrak{g}))$ coincides with the basic
Weil algebra cohomology $H_{W\mathrm{bas}}{}^*(\mathfrak{g})$ of $\mathfrak{g}$. 
It is known that $H_W{}^*(\mathfrak{g})\simeq \mathbb{R}\delta_{*,0}$, i.e. the Weil cohomology
is trivial. $H^{*}{}_{W \mathrm{bas}}(\mathfrak{g})$ is instead non trivial and
concentrated in even degree, namely
$H_{W \mathrm{bas}}{}^*(\mathfrak{g})\simeq W(\mathfrak{g})_{\mathrm{bas}}=\Fun(\mathfrak{g}[2])_{\mathrm{inv}}$,
the basic subalgebra of $W(\mathfrak{g})$.
As in the $N=0$ case, there is no distinction
between classical and quantum BV operators, since, by \eqref{BVgau17}, 
$\Delta_{\mathfrak{g}|1}$ vanishes on $W(\mathfrak{g})$ and, so, 
there is also no distinction between classical and quantum BV cohomologies. 

The $N=1$ gauged matter algebra $\mathcal{A}_{\mathfrak{g}|1M}$ contains as a subalgebra 
\begin{equation}
\mathcal{A}^+{}_{\mathfrak{g}|1M}=W(\mathfrak{g})\otimes \mathcal{A}_M.
\label{BVgaux12} 
\end{equation}
By \eqref{BVgau118a}, \eqref{BVgau118b}, 
\eqref{BVgau110b}, \eqref{BVgau110d}, \eqref{BVgau110f}, \eqref{BVgau110h},
$\mathcal{A}^+{}_{\mathfrak{g}|1M}$ is stable
under the $\mathfrak{g}$--action $i_{\mathfrak{g}|1M}$, $l_{\mathfrak{g}|1M}$.
Similarly, by \eqref{BVgau124}, \eqref{BVgau125}, \eqref{BVgau126}, 
$\mathcal{A}^+{}_{\mathfrak{g}|1M}$ is stable under the BV operator $\delta_{\mathfrak{g}|1M}$. 
Thus, $(\mathcal{A}^+{}_{\mathfrak{g}|1M},\mathfrak{g},i_{\mathfrak{g}|1M}, l_{\mathfrak{g}|1M},\delta_{\mathfrak{g}|1M})$ 
is a $\mathfrak{g}$--operation. 
From \eqref{BVgau124}, \eqref{BVgau125}, \eqref{BVgau126}, we realize that 
the quantum BV cohomology $H_{BV}{}^*(\mathcal{A}^+{}_{\mathfrak{g}|1M})$ is 
the Weil cohomology $H_W{}^*(\mathfrak{g},\mathcal{A}_M)$ of 
$\mathfrak{g}$ with coefficients in the differential space $(\mathcal{A}_M,\delta_M)$. 
Similarly, from \eqref{BVgau118a}, \eqref{BVgau118b}, 
\eqref{BVgau110b}, \eqref{BVgau110d}, \eqref{BVgau110f}, \eqref{BVgau110h},
\eqref{BVgau124}, \eqref{BVgau125}, \eqref{BVgau126}, we find that 
the quantum basic BV cohomology 
$H_{BV\mathrm{bas}}{}^*(\mathcal{A}^+{}_{\mathfrak{g}|1M})$ 
is the basic Weil cohomology 
$H_{W\mathrm{bas}}{}^*(\mathfrak{g},\mathcal{A}_M)$ of $\mathfrak{g}$
with coefficients in the $\mathfrak{g}$--operation $(\mathcal{A}_M,\mathfrak{g},i_M,l_M,\delta_M)$.
Unlike for the pure ghost system, the quantum and classical BV operators are generally 
different in the matter sector and, so, it is necessary 
to distinguish the classical and the quantum BV cohomologies, as in the $N=0$ case. 
Analogous statements hold in the classical case,  
$(\mathcal{A}_M,\delta_{M c})$ being a differential algebra in this case. 


Now, we can solve the problem of the cohomological analysis of observables
in the gauged matter theory. 
We notice that the range of the BV algebra homomorphism
$\Upsilon_1:\mathcal{A}_M\rightarrow\mathcal{A}_{\mathfrak{g}|1M}$ 
is contained in $\mathcal{A}^+{}_{\mathfrak{g}|1M}$. 
From \eqref{BVgaux14a}, \eqref{BVgaux14b}, \eqref{BVgaux15}, $\Upsilon_1|_{\mathcal{A}_{M\mathrm{bas}}}$ is 
a chain map of the basic differential spaces 
$(\mathcal{A}_{M\mathrm{bas}},\delta_M)$, 
$(\mathcal{A}^+{}_{\mathfrak{g}|1M\mathrm{bas}},\delta_{\mathfrak{g}|1M})$.
Thus, $\Upsilon_1|_{\mathcal{A}_{M\mathrm{bas}}}$ induces 
a homomorphism of the basic BV cohomology spaces 
$H_{BV\mathrm{bas}}{}^*(\mathcal{A}_{M})$,
$H_{BV\mathrm{bas}}{}^*(\mathcal{A}^+{}_{\mathfrak{g}|1M})$. 
It can be shown that  this is in fact as an isomorphism \cite{Grueb1,Guillemin1},
\begin{equation}
H_{BV\mathrm{bas}}{}^*(\mathcal{A}^+{}_{\mathfrak{g}|1M})\simeq H_{BV\mathrm{bas}}{}^*(\mathcal{A}_{M})
\label{BVgaux23} 
\end{equation}
under the mild assumption that the $\mathfrak{g}$--operation
$\mathcal{A}_M$ admits a {\it connection}. 
A self--contained proof of \eqref{BVgaux23} is given in appendix \ref{app:equivcoh}.
\eqref{BVgaux23} is to be compared with its $N=0$ counterpart,
eq. \eqref{BVgau0x6}, from which it differs qualitatively in two ways.
First, \eqref{BVgaux23} holds with no restriction on the Lie algebra
$\mathfrak{g}$, whilst \eqref{BVgau0x6} holds provided $\mathfrak{g}$
is reductive. Second, by \eqref{BVgaux23}, $H_{BV\mathrm{bas}}{}^*(\mathcal{A}_{M})$ 
is actually naturally isomorphic to $H_{BV\mathrm{bas}}{}^*(\mathcal{A}^+{}_{\mathfrak{g}|1M})$,
whilst, by  \eqref{BVgau0x6}, $H_{BV\mathrm{inv}}{}^*(\mathcal{A}_{M})$ is simply naturally 
embedded in $H_{BV\mathrm{inv}}{}^*(\mathcal{A}^+{}_{\mathfrak{g}|0M})$.
The reason for this can be ultimately traced back to the triviality
of the Weil cohomology. 
In this way, {\it the study of the observables in the gauged matter theory is
fully reduced to that of the basic
BV cohomology $H_{BV\mathrm{bas}}{}^*(\mathcal{A}^+{}_{\mathfrak{g}|1M})$
of $\mathcal{A}^+{}_{\mathfrak{g}|1M}$}. 


$H_{BV\mathrm{bas}}{}^*(\mathcal{A}^+{}_{\mathfrak{g}|1M})
\simeq H_{W\mathrm{bas}}{}^*(\mathfrak{g},\mathcal{A}_M)$ is known as the 
$\mathfrak{g}$--{\it equivariant cohomology}
$H_{\mathrm{equiv}}{}^*(\mathcal{A}_M)$ of $\mathcal{A}_M$ \cite{Grueb1}. 
Equivariant cohomology is defined usually for differential algebras
$\mathcal{A}_M$. In our case, $\mathcal{A}_M$ is a differential algebra
in the classical but not in the quantum case (cf. sect. \ref{sec:BVacts}). 
However, $H_{\mathrm{equiv}}{}^*(\mathcal{A}_M)$ can still be defined.


As is well--known, there are three models of equivariant cohomology: 
the original models of Weil and Cartan of refs. \cite{Weil1,HCartan2,HCartan1} 
and the so-called BRST model of ref. \cite{Ouvry1}. 
The three models are in fact equivalent. 
The most direct and efficient way to show this was found in ref.
\cite{Kalkman1}, where the author proves that the Cartan and Weil model 
can be obtained from the BRST model
via reduction of and application of a
suitable inner automorphism to the algebra $\mathcal{A}^+{}_{\mathfrak{g}|1M}$, respectively.
The formal structure of the underlying algebra $\mathcal{A}^+{}_{\mathfrak{g}|1M}$,
$\mathfrak{g}$--action $i_{\mathfrak{g}|1M}$, $l_{\mathfrak{g}|1M}$ and differential 
$\delta_{\mathfrak{g}|1M}$ reproduces very closely that of the corresponding objects 
of the BRST model of equivariant cohomology. 
Thus,  mimicking the treatment of \cite{Kalkman1}, 
one may try to generate the counterparts of the Cartan and Weil model in 
the present BV algebraic framework by {\it reduction} and action of a suitable 
{\it BV inner automorphism} (cf. sect. \ref{sec:BValg}), respectively. 


The Cartan model relies on the algebra $\mathcal{C}^+{}_{\mathfrak{g}|1M}
=\Fun(\mathfrak{g}[2])\otimes\mathcal{A}_M$ instead of $\mathcal{A}^+{}_{\mathfrak{g}|1M}$.
$\mathcal{C}^+{}_{\mathfrak{g}|1M}$ is the subalgebra of the elements of $\mathcal{A}^+{}_{\mathfrak{g}|1M}$
containing no occurrences of the $c^i$. The Cartan model can be obtained from BRST model by 
observing that 
$\mathcal{A}^+{}_{\mathfrak{g}|1M\mathrm{bas}}=\mathcal{C}^+{}_{\mathfrak{g}|1M\mathrm{inv}}$.
So, the Cartan model is in a sense an ``effective'' reduction of the BRST model
in which the $c^i$ have been eliminated 
from the outset. 

The Weil model can be derived from the BRST model as follows. Define 
\begin{equation}
\alpha_{\mathfrak{g}|1M}=c^i\otimes\iota_{Mi}.
\label{BVgaux3}
\end{equation}
Clearly, $\alpha_{\mathfrak{g}|1M}\in\mathcal{A}_{\mathfrak{g}|1M-1}$.
Further, by \eqref{BVgau14d}, \eqref{BVgau17}, we have 
\begin{equation}
\Delta_{\mathfrak{g}|1M}\alpha_{\mathfrak{g}|1M}=0.
\label{BVgaux6}
\end{equation}
Therefore, as shown in sect. \ref{sec:BValg}, \hphantom{xxxxxxxxxxxxxxxxxxxxx}
\begin{equation}
T_{\mathfrak{g}|1M}=\exp(-\ad \alpha_{\mathfrak{g}|1M})
\label{BVgaux7}
\end{equation}
is a BV algebra inner automorphism. 
It is indeed the BV algebra analog of the automorphism defined and exploited in ref. \cite{Kalkman1}
to show the equivalence of the BRST and Weil models. 
By an elementary calculation, we find 
\begin{subequations}
\begin{align}
&\iota'{}_{\mathfrak{g}|1Mi}:=T_{\mathfrak{g}|1M}\iota_{\mathfrak{g}|1Mi}
=\iota_{\mathfrak{g}|1i}\otimes 1_M+1_{\mathfrak{g}|1}\otimes\iota_{Mi},
\vphantom{\bigg]}
\label{BVgaux4a}
\\
&\lambda'{}_{\mathfrak{g}|1Mi}:=T_{\mathfrak{g}|1M}\lambda_{\mathfrak{g}|1Mi}
=\lambda_{\mathfrak{g}|1i}\otimes 1_M+1_{\mathfrak{g}|1}\otimes\lambda_{Mi},
\vphantom{\bigg]}
\label{BVgaux4b}
\end{align}
\label{BVgaux4}
\end{subequations}
\!\!where $\iota_{\mathfrak{g}|1Mi}$, $\lambda_{\mathfrak{g}|1Mi}$ are given by \eqref{BVgau19a}, \eqref{BVgau19b}. 
In this way, the $\mathfrak{g}$--action $i'{}_{\mathfrak{g}|1M}$, $l'{}_{\mathfrak{g}|1M}$ 
resulting from the application of $T_{\mathfrak{g}|1M}$ is a trivial extension of the actions
$i_{\mathfrak{g}|1}$, $l_{\mathfrak{g}|1}$  and $i_M$, $l_M$.  
Another simple calculation shows that
\begin{equation}
S'{}_{\mathfrak{g}|1M}=T_{\mathfrak{g}|1M}S_{\mathfrak{g}|1M}
=S_{\mathfrak{g}|1}\otimes 1_M+1_{\mathfrak{g}|1}\otimes S_M,
\label{BVgaux5}
\end{equation}
where $S_{\mathfrak{g}|1M}$ is given by \eqref{BVgau119}. 
So, the BV master action $S'{}_{\mathfrak{g}|1M}$ resulting from the application of $T_{\mathfrak{g}|1M}$ 
is the trivial non interacting one. Correspondingly, 
the quantum BV operator $\delta'{}_{\mathfrak{g}|1M}$ yielded by $T_{\mathfrak{g}|1M}$
is a trivial extension of  the BV operators $\delta_{\mathfrak{g}|1}$, $\delta_M$.
The formal structure of the underlying algebra $\mathcal{A}^+{}_{\mathfrak{g}|1M}$,
$\mathfrak{g}$--action $i'{}_{\mathfrak{g}|1M}$, $l'{}_{\mathfrak{g}|1M}$ and differential 
$\delta'{}_{\mathfrak{g}|1M}$ obtained in this way reproduces closely that of the corresponding objects 
of the Weil model of equivariant cohomology. 

In this way, the interaction of the matter and gauge sector can be absorbed
by means of an inner BV automorphism. This would seem to trivialize
the gauged matter model. However, recall that in the quantum field theoretic 
realizations of the construction, the automorphism may introduce non locality.

\vfill\eject

\section{\normalsize \textcolor{blue}{Higher N BV gaugings and ghost systems}}\label{sec:BVgaugeN}

~~~~In sect. \ref{sec:BVgauge1}, we found out that $N=1$ BV gauging is at the basis 
of topological gauge field theory. The topological models concerned here
have $N=1$ topological supersymmetry. There are also topological models 
having $N=2$ topological supersymmetry, which were first systematically studied 
by Dijkgraaf and Moore in
ref. \cite{Dijkgraaf1}, where they were called balanced. The problem
then arises of describing their gauging in a BV framework as done in the $N=1$ 
case. However, when attempting this, problems of a new kind show up, 
as we now explain.

The basic elements of $N=1$ BV gauging treated in sect. \ref{sec:BVgauge1}
are a BV algebra $(\mathcal{A},\Delta,\{\cdot,\cdot\})$ equipped a quantum BV operator $\delta$ 
and a $N=1$ $\mathfrak{g}$--action $i$, $l$ organized in an algebraic structure, called
a $\mathfrak{g}$--operation in the terminology of \cite{Grueb1}. This structure underlies 
the Mathai--Quillen formulation of $N=1$ topological field theory \cite{Mathai1}.
From now on, we shall refer to it as an $N=1$ $\mathfrak{g}$--operation.
In ref. \cite{Dijkgraaf1}, the authors showed that the Mathai--Quillen 
formulation can be generalized to $N=2$ topological field theory. 
Their construction hinges on an algebraic framework generalizing
that of $N=1$ $\mathfrak{g}$--operation and thus called 
$N=2$ $\mathfrak{g}$--operation henceforth. 

If we tried to implement $N=2$ BV gauging following \cite{Dijkgraaf1} and 
mimicking the $N=1$ case, the basic elements
would be a BV algebra $(\mathcal{A},\Delta,\{\cdot,\cdot\})$ equipped with a doublet
of quantum BV operator $\delta_A$, $A=1,2$ satisfying 
\begin{equation}
[\delta_A,\delta_B]=0.
\label{BVgaugeN1}
\end{equation}
In addition, we would have an $N=2$ $\mathfrak{g}$--action, which is a set of linear  
maps $j:\mathfrak{g}\rightarrow \Der_{-2}(\mathcal{A})$,
$i_A:\mathfrak{g}\rightarrow \Der_{-1}(\mathcal{A})$, $A=1,2$, 
$l:\mathfrak{g}\rightarrow \Der_0(\mathcal{A})$
satisfying the following commutation relations \hphantom{xxxxxxxxxxxxxxxxx}
\begin{subequations}
\begin{align}
&[j_x,j_y]=0, \hspace{.9cm}
\vphantom{\bigg]}
\label{BVgaugeN2a}
\end{align}
\begin{align}
&[j_x,i_{Ay}]=0,
\vphantom{\bigg]}
\label{BVgaugeN2b}
\\
&[i_{Ax},i_{By}]=\epsilon_{AB}j_{[x,y]},
\vphantom{\bigg]}
\label{BVgaugeN2c}
\\
&[l_x,j_y]=j_{[x,y]},
\vphantom{\bigg]}
\label{BVgaugeN2d}
\\
&[l_x,i_{Ay}]=i_{A[x,y]},
\vphantom{\bigg]}
\label{BVgaugeN2e}
\\
&[l_x,l_y]=l_{[x,y]}, 
\vphantom{\bigg]}
\label{BVgaugeN2f}
\end{align}
\label{BVgaugeN2}
\end{subequations}
\!\!with $x,y\in\mathfrak{g}$, where $\epsilon_{AB}$ is the two dimensional antisymmetric symbol. 
Finally, the derivations $j_x$, $i_{Ax}$, $l_x$ would be related as 
\begin{subequations}
\begin{align}
&[\delta_A,j_x]=i_{Ax},
\vphantom{\bigg]}
\label{BVgaugeN3a}
\\
&[\delta_A,i_{Bx}]=-\epsilon_{AB}l_x,
\vphantom{\bigg]}
\label{BVgaugeN3b}
\\
&[\delta_A,l_x]=0, 
\vphantom{\bigg]}
\label{BVgaugeN3c}
\end{align}
\label{BVgaugeN3}
\end{subequations}
\!\!with $x\in\mathfrak{g}$.
Relations \eqref{BVgaugeN2a}--\eqref{BVgaugeN2f} and \eqref{BVgaugeN3a}--\eqref{BVgaugeN3c}
define an $N=2$ $\mathfrak{g}$--operation.  
(Compare with relations \eqref{BVgau11a}--\eqref{BVgau11c} and \eqref{BVgau12xa}, \eqref{BVgau12xb}
defining an $N=1$ $\mathfrak{g}$--operation). 
$N=2$ $\mathfrak{g}$--operations  
were systematically studied in ref. \cite{Zucchini0}. One of their main properties
is the existence of an internal $\mathfrak{sl}(2,\mathbb{R})\oplus \mathbb{R}$ 
algebra of automorphisms, an "$R$--symmetry" in physical parlance.

In a BV framework, the existence of a doublet of quantum BV operators $\delta_A$
is intriguing. It apparently implies the 
corresponding existence of a doublet of quantum BV master actions $S_A$. 
However, a relation of the form
\begin{equation}
\delta_A=\Delta+\ad S_A
\label{BVgaugeN4}
\end{equation}
is incompatible with the internal $\mathfrak{sl}(2,\mathbb{R})$--symmetry.
This indicates that the ordinary approach based on BV algebras is inadequate 
for the construction we are attempting. 
If we wish to remedy this changing as little as possible our BV framework,
a doublet of degree $1$ BV Laplacians $\Delta_A$ rather a single one $\Delta$ 
is required in addition to the bracket $\{\cdot,\cdot\}$. So, 
instead of a customary BV algebra $(\mathcal{A},\Delta,\{\cdot,\cdot\})$, we should have
some structure of the form $(\mathcal{A},\Delta_A,\{\cdot,\cdot\})$.

$\Delta_A$ and $\{\cdot,\cdot\}$ should fulfill certain conditions
generalizing those defining a BV algebra in natural fashion.
Presumably, they are the following. 
First, the bracket $\{\cdot,\cdot\}$ satisfy the graded Leibniz
relation \eqref{BValg3} and 
the Gerstenhaber relations \eqref{BValg4}, \eqref{BValg5}.
Second, the BV Laplacians $\Delta_A$ are nilpotent 
and anticommute  
\begin{equation}
[\Delta_A,\Delta_B]=0
\label{BVgaugeN6}
\end{equation}
(compare with \eqref{BValg1}).
Third, the $\Delta_A$ are degree $1$ derivations of $\mathcal{A}_G$,
\begin{equation}
\Delta_A\{\phi,\psi\}=\{\Delta_A\phi,\psi\}+(-1)^{\partial\phi+1}\{\phi,\Delta_A\psi\},
\label{BVgaugeN7}
\end{equation}
with $\phi,\psi\in\mathcal{A}$ (compare with \eqref{BValg6}).
There is no 
extension of relation \eqref{BValg2}.

In the resulting extended BV algebraic framework, \eqref{BVgaugeN4} is improved as 
\begin{equation}
\delta_A=\Delta_A+\ad S_A.
\label{BVgaugeN5}
\end{equation}
In order \eqref{BVgaugeN1} to be satisfied, it is sufficient that 
\begin{equation}
\Delta_AS_B+\Delta_BS_A+\{S_A,S_B\}=0.
\label{BVgaugeN8}
\end{equation}
This is the resulting generalization of the master equation \eqref{BVacts2}.
Its field theoretic origin, if any, is not clear at all.

Let us assume that the $N=2$ $\mathfrak{g}$--action is BV Hamiltonian
in the following sense. There exist linear BV moment maps $\eta:\mathfrak{g}
\rightarrow \mathcal{A}_{-3}$, 
$\iota_A:\mathfrak{g}\rightarrow \mathcal{A}_{-2}$, $A=1,2$,  
$\lambda:\mathfrak{g}\rightarrow \mathcal{A}_{-1}$ such that 
\begin{subequations}
\begin{align}
&j_x=\ad\eta_x,  
\vphantom{\bigg]}
\label{BVgaugeN9a} 
\\
&i_{Ax}=\ad\iota_{Ax},  
\vphantom{\bigg]}
\label{BVgauNge9b} 
\\
&l_x=\ad\lambda_x, 
\vphantom{\bigg]}
\label{BVgaugeN9c} 
\end{align}
\label{BVgaugeN9}
\end{subequations}
and satisfying the relations 
\begin{subequations}
\begin{align}
&\{\eta_x,\eta_y\}=0, 
\vphantom{\bigg]}
\label{BVgaugeN10a}
\\
&\{\eta_x,\iota_{Ay}\}=0,
\vphantom{\bigg]}
\label{BVgaugeN10b}
\\
&\{\iota_{Ax},\iota_{By}\}=\epsilon_{AB}\eta_{[x,y]},
\vphantom{\bigg]}
\label{BVgaugeN10c}
\\
&\{\lambda_x,\eta_y\}=\eta_{[x,y]},
\vphantom{\bigg]}
\label{BVgaugeN10d}
\\
&\{\lambda_x,\iota_{Ay}\}=\iota_{A[x,y]},
\vphantom{\bigg]}
\label{BVgaugeN10e}
\\
&\{\lambda_x,\lambda_y\}=\lambda_{[x,y]}, 
\vphantom{\bigg]}
\label{BVgaugeN10f}
\\
&\Delta_A\eta_x=0,
\vphantom{\bigg]}
\label{BVgaugeN10g} 
\\
&\Delta_A\iota_{Bx}=0,
\vphantom{\bigg]}
\label{BVgaugeN10h}
\\
&\Delta_A\lambda_x=0,
\vphantom{\bigg]}
\label{BVgaugeN10i} 
\end{align}
\label{BVgaugeN10}
\end{subequations} 
\!\!with $x,y\in\mathfrak{g}$. 
Assuming that \eqref{BVgaugeN9a}--\eqref{BVgaugeN9c} hold, 
\eqref{BVgaugeN10a}--\eqref{BVgaugeN10f} ensure that  
\eqref{BVgaugeN2a}--\eqref{BVgaugeN2f} are fulfilled.
On account of \eqref{BVgaugeN5}, \eqref{BVgaugeN10g}--\eqref{BVgaugeN10i}
ensure that \eqref{BVgaugeN3a}--\eqref{BVgaugeN3c} are also fulfilled if 
the master action doublet $S_A$ satisfies 
\begin{subequations}
\begin{align}
&\{S_A,\eta_x\}=\iota_{Ax},
\vphantom{\bigg]}
\label{BVgaugeN11a} 
\\
&\{S_A,\iota_{Bx}\}=-\epsilon_{AB}\lambda_x,
\vphantom{\bigg]}
\label{BVgauN11db}
\\
&\{S_A,\lambda_x\}=0,
\vphantom{\bigg]}
\label{BVgaugeN11c} 
\end{align}
\label{BVgaugeN11}
\end{subequations}
\!\!with $x\in\mathfrak{g}$. We may take this as the definition of invariance
of the action doublet $S_A$ under the $N=2$ BV Hamiltonian $\mathfrak{g}$--action.

We shall not attempt to fully generalize the constructions of sects.
\ref{sec:BVgauge0}, \ref{sec:BVgauge1} to obtain $N=2$ BV 
gauging. The construction is algebraically complicated, on one hand,
and its eventual relevance in field theoretic applications still
doubtful, on the other. Moreover, the definitions of the relevant structures do 
not seem to be unique.
We shall limit ourselves to a broad outline of the $N=2$ gauging procedure and the
structure of the $N=2$ ghost system and its coupling to a matter system.

Consider a {\it matter} extended BV algebra $(\mathcal{A}_M,\Delta_{MA},\{\cdot,\cdot\}_M)$ 
carrying a BV Hamiltonian $N=2$ $\mathfrak{g}$--action $j_M$, $i_{MA}$, $l_M$ with BV moment 
maps $\eta_M$, $\iota_{MA}$, $\lambda_M$ and a matter quantum BV master action doublet $S_{MA}$ 
invariant under the $\mathfrak{g}$--action. We want to gauge the $\mathfrak{g}$--symmetry. 

The gauging proceeds in three steps, as usual. 

{\it 1.} We construct an $N=2$ {\it ghost} extended BV algebra
$(\mathcal{A}_{\mathfrak{g}|2},\Delta_{\mathfrak{g}|2A},\{\cdot,\cdot\}_{\mathfrak{g}|2})$
with a BV Hamiltonian $N=2$ $\mathfrak{g}$--action $j_{\mathfrak{g}|2}$, $i_{\mathfrak{g}|2A}$, 
$l_{\mathfrak{g}|2}$
with BV moment maps $\eta_{\mathfrak{g}|2}$, $\iota_{\mathfrak{g}|2A}$, $\lambda_{\mathfrak{g}|2}$
and an $N=2$ ghost quantum BV master action doublet $S_{\mathfrak{g}|2A}$ 
invariant under the $\mathfrak{g}$--action. The construction is canonical, 
depending on $\mathfrak{g}$ only.

{\it 2.} 
We construct an $N=2$ {\it gauged matter} extended BV algebra
$(\mathcal{A}_{\mathfrak{g}|2M},\Delta_{\mathfrak{g}|2MA}$, $\{\cdot,\cdot\}_{\mathfrak{g}|2M})$
and equip it with an appropriate BV Hamiltonian $N=2$ $\mathfrak{g}$--action 
$j_{\mathfrak{g}|2M}$, $i_{\mathfrak{g}|2MA}$,
$l_{\mathfrak{g}|2M}$ with BV moment maps $\eta_{\mathfrak{g}|2M}$, 
$\iota_{\mathfrak{g}|2MA}$, $\lambda_{\mathfrak{g}|2M}$.

{\it 3.} 
We construct an $N=2$ gauged matter action doublet $S_{\mathfrak{g}|2MA}$ of the gauged 
matter BV algebra invariant under the $\mathfrak{g}$--action.

The $N=2$ ghost system was introduced originally in ref. 
\cite{Dijkgraaf1} and studied in detail in ref. \cite{Zucchini0}.
It consists of a degree $1$ $\mathfrak{g}$--valued doublet $c_A{}^i$, 
a degree $2$ $\mathfrak{g}$--valued singlet $c^i$, 
a degree $2$ $\mathfrak{g}$--valued triplet $C_{AB}{}^i$ symmetric 
in $A,B$ and a degree $3$ $\mathfrak{g}$--valued doublet $C_A{}^i$.
In the extended BV framework, these are conjugated to 
a degree $-2$ $\mathfrak{g}^\vee$--valued doublet $b^A{}_i$, 
a degree $-3$ $\mathfrak{g}^\vee$--valued singlet $b_i$, 
a degree $-3$ $\mathfrak{g}^\vee$--valued triplet $B^{AB}{}_i$ symmetric in $A,B$ 
and a degree $-4$ $\mathfrak{g}^\vee$--valued doublet $B^A{}_i$, respectively.
They span a graded algebra $\mathcal{A}_{\mathfrak{g}|2}$.
Apparently, the only consistent choice of the BV Laplacians $\Delta_{\mathfrak{g}|2A}$ in  
$\mathcal{A}_{\mathfrak{g}|2}$ is the trivial one 
\begin{equation}
\Delta_{\mathfrak{g}|2A}=0.
\label{BVgaugeN12}
\end{equation}
$\mathcal{A}_{\mathfrak{g}|2}$ has instead a natural non trivial bracket
\begin{align}
\{\phi,\psi\}_{\mathfrak{g}|2}&=\partial_{RbA}{}^i\phi\partial_{Lc}{}^A{}_i\psi
-\partial_{Rc}{}^A{}_i\phi\partial_{LbA}{}^i\psi
\hspace{3cm}
\vphantom{\bigg]}
\label{BVgaugeN13}
\\
&~~+\partial_{RBAB}{}^i\phi\partial_{LC}{}^{AB}{}_i\psi-\partial_{RC}{}^{AB}{}_i\phi\partial_{LBAB}{}^i\psi
\vphantom{\bigg]}
\nonumber
\\
&~~+\partial_{Rb}{}^i\phi\partial_{Lci}\psi-\partial_{Rci}\phi\partial_{Lb}{}^i\psi
\vphantom{\bigg]}
\nonumber
\\
&~~+\partial_{RBA}{}^i\phi\partial_{LC}{}^A{}_i\psi-\partial_{RC}{}^A{}_i\phi\partial_{LBA}{}^i\psi,
\qquad \phi,\psi\in \mathcal{A}_{\mathfrak{g}|2}.
\vphantom{\bigg]}
\nonumber
\end{align}
The construction of the moment maps $\eta_{\mathfrak{g}|2}$, $\iota_{\mathfrak{g}|2A}$, $\lambda_{\mathfrak{g}|2}$
of the appropriate Hamiltonian $N=2$ $\mathfrak{g}$--action on $\mathcal{A}_{\mathfrak{g}|2}$ and of the correct
$N=2$ ghost master action doublet $S_{\mathfrak{g}|2A}$ satisfying the invariance conditions \eqref{BVgaugeN11}
and the master equation \eqref{BVgaugeN8} is an open problem. 
A superfield formulation of the $N=2$ ghost system is possible in principle, as in the $N=1$ case.

The $N=2$ gauged matter extended BV algebra
$(\mathcal{A}_{\mathfrak{g}|2M},\Delta_{\mathfrak{g}|2MA}$, $\{\cdot,\cdot\}_{\mathfrak{g}|2M})$
is the tensor product of the $N=2$ ghost BV algebra
$(\mathcal{A}_{\mathfrak{g}|2},\Delta_{\mathfrak{g}|2A}$, $\{\cdot,\cdot\}_{\mathfrak{g}|2})$
and the matter BV algebra $(\mathcal{A}_M,\Delta_{MA},\{\cdot,\cdot\}_M)$. The tensor product 
of extended BV algebras is defined by a straightforward generalization of the definition of 
tensor product of ordinary BV algebras given in sect. \ref{sec:BValg}.
We expect that, in a BRST model, the appropriate Hamiltonian $N=2$ $\mathfrak{g}$--action of the gauged matter 
BV algebra to be some non trivial extension of those of its ghost and matter factors,
as in the $N=1$ case. The precise definition of the corresponding moment maps 
$\eta_{\mathfrak{g}|2M}$, $\iota_{\mathfrak{g}|2MA}$, $\lambda_{\mathfrak{g}|2M}$ is a further open 
problem.

If we tried to generalize \eqref{BVgau119} in the extended BV framework 
illustrated above, the gauged matter action doublet would be something like
\begin{align}
&S_{\mathfrak{g}|2MA}=S_{\mathfrak{g}|2A}\otimes 1_M+1_{\mathfrak{g}|2}\otimes S_{MA}
+c_A{}^i\otimes\lambda_{Mi}
\vphantom{\bigg]}
\label{BVgaugeNx}
\\ 
&\hspace{5cm}-\epsilon^{BC}(C_{AB}{}^i-\epsilon_{AB}c^i)\otimes\iota_{MCi}
+C_A{}^i\otimes\eta_{Mi},
\vphantom{\bigg]}
\nonumber
\end{align}
the last three terms being interaction terms. 
The fulfilment of the invariance con-
\eject\noindent
ditions \eqref{BVgaugeN11} and the master equation \eqref{BVgaugeN8} 
cannot be ascertained as long as the explicit form
of the ghost BV action $S_{\mathfrak{g}|2A}$ is not known.

It is reasonable to expect that the appropriate classification of the observables of a 
theory described by an extended BV algebra $(\mathcal{A},\Delta_A,\{\cdot,\cdot\})$ 
and a quantum BV master action doublet $S_A$ is encoded in the cohomology of the 
bidifferential space $(\mathcal{A},\delta_A)$. However, this cohomology cannot have 
the customary form of a $\mathbb{Z}$--bigraded cohomology. $\mathcal{A}$ has no 
$\mathbb{Z}$--bigrading such that there are two independent linear combinations of the 
$\delta_A$ each of which raises one of the underlying $\mathbb{Z}$--gradings by one unit and 
leaves invariant the other one. Rather, the observables are classified by the cohomology 
of any non vanishing linear combination of the $\delta_A$, the internal
$\mathfrak{sl}(2,\mathbb{R})\oplus\mathbb{R}$ guaranteeing the independence of the 
cohomology from the choice of the combination.

If the extended BV algebra $(\mathcal{A},\Delta_A,\{\cdot,\cdot\})$ is equipped with 
$N=2$ $\mathfrak{g}$--action $j$, $i_A$, $l$ under which the BV action doublet $S_A$
is invariant in the sense that \eqref{BVgaugeN3a}--\eqref{BVgaugeN3c} are satisfied, 
one may define an $N=2$ $\mathfrak{g}$--basic quantum BV cohomology. This is the 
cohomology, as defined in the previous paragraph, 
of the bidifferential space $(\mathcal{A}_{\mathrm{bas}},\delta_A)$, where 
$\mathcal{A}_{\mathrm{bas}}=\cap_{x\in\mathfrak{g}}(\ker j_x\cap\cap_A\ker i_{Ax}\cap \ker l_x)
\subset\mathcal{A}$. When carrying out the gauging of a matter extended BV algebra with 
an invariant matter action doublet as outlined above, a corresponding notion of $N=2$ 
$\mathfrak{g}$--equivariant cohomology should appear.

The above analysis presumably generalizes to higher values of $N$. 
To the best of our knowledge, virtually nothing is known about $N\geq 3$ 
$\mathfrak{g}$--operations and ghost systems. However, we expect 
the inadequacy of the customary BV algebraic framework to emerge
again.

\vfill\eject

\section{\normalsize \textcolor{blue}{Applications and examples}}\label{sec:appls}

~~~~In this section, we shall present a few applications of the formalism developed
in the preceding sections. Our examples are drown from Lie algebroid  
and Poisson geometry, which cover a broad spectrum of cases. 
We concentrate on the well understood $N=0$ and $N=1$ gauging. 

\vspace{.1cm}

{\it The BV algebra of a Lie algebroid and its gauging}

\vspace{.1cm}

A {\it Lie algebroid} is a vector bundle $E$ over a manifold $M$ 
equipped with a bundle map $\rho_E:E\rightarrow TM$, called the {\it anchor}, and 
an $\mathbb{R}$--linear bracket $[\cdot,\cdot]_E:\Gamma(E)\times \Gamma(E)
\rightarrow \Gamma(E)$ with the following properties.

\par\noindent 
~~~1) $[\cdot,\cdot]$ is a Lie bracket so that $\Gamma(E)$ is a Lie algebra:
\begin{align}
&[X,Y]_E+[Y,X]_E=0,
\vphantom{\bigg]}
\label{LAA1}
\\
&[X,[Y,Z]_E]_E+[Y,[Z,X]_E]_E+[Z,[X,Y]_E]_E=0,
\vphantom{\bigg]}
\label{LAA2}
\end{align}
for $X,Y,Z\in \Gamma(E)$. 

\par\noindent 
~~~2) $\rho$ defines a Lie algebra homomorphism of $\Gamma(E)$ into $\Gamma(TM)$: 
\begin{equation}
\rho([X,Y]_E)=[\rho(X),\rho(Y)]_{TM},
\label{LAA3}
\end{equation}
for $X,Y\in \Gamma(E)$, where $[\cdot,\cdot]_{TM}$ is the usual Lie bracket
of vector fields of $M$. 

\par\noindent 
~~~3) The generalized Leibniz rule holds:
\begin{equation}
[X,fY]_E=f[X,Y]_E+(\rho(X)f)Y,
\label{LAA4}
\end{equation}
for $f\in C^\infty(M)$ and $X,Y\in \Gamma(E)$.

The prototype Lie algebroid over $M$ is the tangent bundle $TM$: the anchor is the
identity $\id_{TM}$ and the bracket is the usual Lie bracket $[\cdot,\cdot]_{TM}$. 
Lie algebroids generalize Lie algebras: a Lie algebra can be viewed as 
a Lie algebroid over the singleton manifold $M=\mathrm{pt}$.

Let $\{e_r\}$ be a local frame of $E$. Then, one has
\begin{align}
&\rho_E(e_r)=\rho_r{}^a\partial_a,
\vphantom{\bigg]}
\label{LAA5}
\\
&[e_r,e_s]_E=c^t{}_{rs}e_t.
\vphantom{\bigg]}
\label{LAA6}
\end{align}
Here, $a,b, c,\dots$ are base coordinate indices 
while $r,s,t,\dots$ are fiber coordinate indices.
$\rho_r{}^a$, $c^t{}_{rs}$ are called the anchor and structure functions of $E$, respectively.
From \eqref{LAA1}--\eqref{LAA4}, they satisfy
\begin{subequations}
\begin{align}
&c^r{}_{st}+c^r{}_{ts}=0,
\vphantom{\bigg]}
\label{LAA7}
\\
&c^r{}_{sv}c^v{}_{tu}+c^r{}_{tv}c^v{}_{us}+c^r{}_{uv}c^v{}_{st}
+\rho_s{}^a\partial_ac^r{}_{tu}+\rho_t{}^a\partial_ac^r{}_{us}
+\rho_u{}^a\partial_ac^r{}_{st}=0,
\vphantom{\bigg]}
\label{LAA8}
\\
&\rho_r{}^b\partial_b\rho_s{}^a-\rho_s{}^b\partial_b\rho_r{}^a-c^t{}_{rs}\rho_t{}^a=0.
\vphantom{\bigg]}
\label{LAA9}
\end{align}
\label{LAA7,8,9}
\end{subequations}
\!\!\eqref{LAA7}--\eqref{LAA9} are the structure relations of $E$.

The Lie algebroid $E$ is characterized by a natural cohomology. We shall define 
this conveniently in the language of graded geometry \cite{Vaintrob1}.
Consider the parity shifted bundle $E[1]$ and
the $\mathbb{Z}$ graded algebra $\Fun(E[1])$ of functions on $E[1]$.
There exists a degree $1$ derivation $d_E$ of $\Fun(E[1])$ defined by 
\begin{equation}
d_E=\rho_r{}^a(x)\xi^r\partial_{La}-\frac{1}{2}c^r{}_{st}(x)\xi^s\xi^t\partial_{Lr},
\label{LAA10}
\end{equation}
where $x^a,\xi^r$ are the base and fiber coordinates of a generic 
trivialization of $E[1]$ with degree $0, 1$, respectively, and 
$\partial_a=\partial/\partial x^a$, $\partial_r=\partial/\partial \xi^r$.
Using the relations \eqref{LAA7,8,9}, one checks easily that
$d_E$ is nilpotent and is therefore 
a differential
\begin{equation}
d_E{}^2=0.
\label{LAA13}
\end{equation}
The cohomology of the differential space $(\Fun(E[1]),d_E)$
is the Lie algebroid cohomology of $E$, $H_{LA}{}^*(E)$.
When $E=TM$, $d_E$ reduces to the ordinary de Rham differential  
and $H_{LA}{}^*(E)$ reduces to the familiar de Rham cohomology.

With any section $X\in\Gamma(E)$, there are associated two derivations of $\Fun(E[1])$ of
degree $-1$, $0$ defined by 
\begin{subequations}
\begin{align}
i_{EX}&=X^r(x)\partial_{Lr}
\vphantom{\bigg]}
\label{LAA11}
\\
l_{EX}&=\rho_r{}^aX^r(x)\partial_{La}
+(\rho_r{}^a\partial_aX^s+c^s{}_{rt}X^t)(x)\xi^r\partial_{Ls}.
\vphantom{\bigg]}
\label{LAA12}
\end{align}
\label{LAA11,12}
\end{subequations}
\!\!They generalize the interior and Lie derivatives  
 of de Rham theory and reduce to those when $E=TM$.

It is simple to check that the above derivations satisfy 
\begin{subequations}
\begin{align}
&[d_E,d_E]=0,
\vphantom{\bigg]}
\label{LAA14}
\\
&[d_E,i_{EX}]=l_{EX},
\vphantom{\bigg]}
\label{LAA15}
\\
&[d_E,l_{EX}]=0,
\vphantom{\bigg]}
\label{LAA16}
\\
&[i_{EX},i_{EY}]=0,
\vphantom{\bigg]}
\label{LAA17}
\\
&[l_{EX},i_{EY}]=i_{E[X,Y]_E},
\vphantom{\bigg]}
\label{LAA18}
\\
&[l_{EX},l_{EY}]=l_{E[X,Y]_E},
\vphantom{\bigg]}
\label{LAA19}
\end{align}
\label{LAA14-19}
\end{subequations}
\!\!with $X,Y\in\Gamma(E)$, generalizing the well--known Cartan relations.

Let $\mathfrak{g}$ be a Lie algebra and let $\varphi:\mathfrak{g}\rightarrow\Gamma(E)$ be 
a fiducial Lie algebra homomorphism. Then, for $x\in \mathfrak{g}$, the degree $-1$, $0$ derivations 
of $\Fun(E[1])$ 
\begin{subequations}
\begin{align}
&i_{Ex}=i_{E\varphi(x)},
\vphantom{\bigg]}
\label{LAA43}
\\
&l_{Ex}=l_{E\varphi(x)}
\vphantom{\bigg]}
\label{LAA44}
\end{align}
\label{LAA43,44}
\end{subequations} 
\!\!are defined. By \eqref{LAA14}, \eqref{LAA16}, \eqref{LAA19}, $(\Fun(E[1]),\mathfrak{g},l_{E},d_E)$
is a differential $\mathfrak{g}$--module (cf. app. \ref{app:invcoh}). 
The associated invariant cohomology is the invariant Lie algebroid cohomology 
$H_{LA\mathrm{inv}}{}^*(E)$ of $E$ \cite{Kubarski}.
Analogously, by \eqref{LAA14}--\eqref{LAA19}, $(\Fun(E[1]),\mathfrak{g},i_{E},l_{E}, d_E)$
is a $\mathfrak{g}$--operation (cf. app. \ref{app:equivcoh}).
The basic cohomology as-
\vfill\eject\noindent
sociated with it is the basic Lie algebroid cohomology 
$H_{LA\mathrm{bas}}{}^*(E)$ of $E$ \cite{Ginzburg}.

The $\mathbb{Z}$--graded algebra $\Fun(T^*[-1]E[1])$ of functions
on the parity shifted co\-tangent bundle $T^*[-1]E[1]$ of $E[1]$ 
can be given a structure of BV algebra.
This BV algebra extends, in an appropriate sense to be specified, the algebra
$\Fun(E[1])$ considered above. 

$T^*[-1]E[1]$ has the canonical degree $-1$ symplectic structure
\begin{equation}
\omega_E=dy_adx_a+d\eta_rd\xi^r,
\label{LAA20}
\end{equation}
where $x^a,\xi^r,y_a,\eta_r$ are the base and fiber coordinates of a generic 
trivialization of $T^*[-1]E[1]$ with degree $0, 1,-1,-2$, respectively.
Let us assume now that the orientation line bundle $Q_E=\wedge^nT^*M\otimes\wedge^qE$,
where $n=\dim M$ and $q=\rank E$ is trivial. 
There then exists a nowhere vanishing section
$\gamma\in \Gamma(Q_E)$, which can be used to construct  
a volume form on $T^*[-1]E[1]$, 
\begin{equation}
\mu_{E\gamma}=\gamma^2(x)dx^1\cdots dx^nd\xi^1\cdots d\xi^rdy_1\cdots dy_nd\eta_1\cdots d\eta_r.
\label{LAA26}
\end{equation}
These geometrical objects allow us to endow $\Fun(T^*[-1]E[1])$ with the structure of 
BV algebra. The construction is standard and is illustrated 
in the literature (see e. g. ref. \cite{Zwiebach1}).
The BV Laplacian $\Delta_{E\gamma}$ is given by 
\begin{align}
\Delta_{E\gamma}&=\gamma^{-1}(x)\partial_{La}\gamma(x)\partial_L{}^a
-\partial_{Lr}\partial_L{}^r
\vphantom{\bigg]}
\label{LAA25}
\\
&=\big(\partial_{La}+\partial_a\ln\gamma(x)\big)\partial_L{}^a
-\partial_{Lr}\partial_L{}^r,
\vphantom{\bigg]}
\nonumber
\end{align}
where $\partial_a=\partial/\partial x^a$, $\partial_r=\partial/\partial \xi^r$,
$\partial^a=\partial/\partial y_a$, $\partial^r=\partial/\partial \eta_r$.
The BV antibracket has the standard form 
\begin{equation}
\{\phi,\psi\}_E=\partial_{Ra}\phi\partial_L{}^a\psi-\partial_R{}^a\phi\partial_{La}\psi
+\partial_{Rr}\phi\partial_L{}^r\psi-\partial_R{}^r\phi\partial_{Lr}\psi,
\label{LAA21}
\end{equation}
with $\phi,\psi\in \Fun(T^*[-1]E[1])$. It is easy to check that 
the triple $(\Fun(T^*[-1]E[1])$, $\Delta_{E\gamma},\{\cdot,\cdot\}_E)$ satisfies
\eqref{BValg1}--\eqref{BValg3} and is therefore a BV algebra
as announced. 

\vfill\eject

The bundle projection $\pi_E:T^*[-1]E[1]\rightarrow E[1]$ induces a degree $0$
graded algebra monomorphism $\pi_E{}^*:\Fun(E[1])\rightarrow\Fun(T^*[-1]E[1])$.
In this way, $\Fun(E[1])$ can be viewed as a subalgebra of $\Fun(T^*[-1]E[1])$.
From \eqref{LAA25}, \eqref{LAA21}, one has $\Delta_{E\gamma}|_{\Fun(E[1])}=0$
and $\{\cdot|_{\Fun(E[1])},\cdot|_{\Fun(E[1])}\}_E=0$.
It follows that $\Fun(E[1])$, equipped with the trivial BV algebra structure, is a BV subalgebra
of the BV algebra $\Fun(T^*[-1]E[1])$ (cf. sect. \ref{sec:BValg}). 
Indeed, the BV antibracket structure of $\Fun(T^*[-1]E[1])$ is closely related to the 
bracket structure of the ``big bracket'' formulation of Lie algebroid theory 
\cite{KosmannSchwarzbach1,Roytenberg1,Roytenberg2}. 

With applications of the theory of the preceding sections
in mind, we want to equip the BV algebra $\Fun(T^*[-1]E[1])$ 
with a quantum BV master action with global symmetries. This is achieved 
by the following construction.

$\Fun(T^*[-1]E[1])$ contains the degree $0$ element 
\begin{equation}
S_E=\rho_r{}^a(x)y_a\xi^r+\frac{1}{2}c^r{}_{st}(x)\xi^s\xi^t\eta_r
\label{LAA22}
\end{equation}
and, for $X\in\Gamma(E)$, the degree $-2$, $-1$ elements 
\begin{subequations}
\begin{align}
\iota_{EX}&=-X^r(x)\eta_r,
\vphantom{\bigg]}
\label{LAA23}
\\
\lambda_{EX}&=-\rho_r{}^aX^r(x)y_a-(\rho_r{}^a\partial_aX^s+c^s{}_{rt}X^t)(x)\xi^r\eta_s.
\vphantom{\bigg]}
\label{LAA24}
\end{align}
\label{LAA23,24}
\end{subequations} 
\!\! By a straightforward calculation, one finds the brackets
\begin{subequations}
\begin{align}
&\{S_E,S_E\}_E=0,     
\vphantom{\bigg]}
\label{LAA27}
\\
&\{\iota_{EX},S_E\}_E=\lambda_{EX},
\vphantom{\bigg]}
\label{LAA28}
\\
&\{\lambda_{EX},S_E\}_E=0,
\vphantom{\bigg]}
\label{LAA29}
\\
&\{\iota_{EX},\iota_{EY}\}_E=0,
\vphantom{\bigg]}
\label{LAA30}
\\
&\{\lambda_{EX},\iota_{EY}\}_E=\iota_{E[X,Y]_E},
\vphantom{\bigg]}
\label{LAA31}
\end{align}
\begin{align}
&\{\lambda_{EX},\lambda_{EY}\}_E=\lambda_{E[X,Y]_E}.
\vphantom{\bigg]}
\label{LAA32}
\end{align}
\label{LAA27,32}
\end{subequations} 
\!\!One also shows that the relations 
\vspace{-.25cm}
\begin{subequations}
\begin{align}
&\Delta_{E\gamma}S_E=0,
\vphantom{\bigg]}
\label{LAA39}
\\
&\Delta_{E\gamma}\iota_{EX}=0,
\vphantom{\bigg]}
\label{LAA33}
\\
&\Delta_{E\gamma}\lambda_{EX}=0.
\vphantom{\bigg]}
\label{LAA34}
\end{align}
\label{LAA33,34}
\end{subequations}
\!\!hold, provided $\gamma$ satisfies the condition 
\vspace{-.1cm}
\begin{equation}
\partial_a\rho_r{}^a+\rho_r{}^a\partial_a\ln\gamma-c^s{}_{sr}=0.
\label{LAA38}
\end{equation}

In general, the chosen $\gamma\in\Gamma(Q_E)$ does not fulfil \eqref{LAA38}.
Note, however, that $\gamma$ is determined only up to a rescaling by a 
factor of the form $\mathrm{e}^f$ with $f\in \Fun(M)$.
Hence, if, instead of \eqref{LAA38}, $\gamma$ satisfies the weaker condition
\begin{equation}
\partial_a\rho_r{}^a+\rho_r{}^a\partial_a\ln\gamma-c^s{}_{sr}+\rho_r{}^a\partial_af=0
\label{LAA40}
\end{equation}
for some function $f\in \Fun(M)$, then, after redefining $\gamma$ into $\mathrm{e}^f\gamma$, 
one can make $\gamma$ fulfil \eqref{LAA38}.
It can be shown that this is the case precisely when the Lie algebroid $E$ 
is {\it unimodular}, i. e. its modular class $\theta_E$, a distinguished 
element of the degree $1$ cohomology $H_{LA}{}^1(E)$, 
vanishes \cite{Weinstein}. Indeed, the first three terms in left hand side of 
\eqref{LAA40} constitute the local expression of a generic 
representative of $\theta_E$ and \eqref{LAA40} is the statement that this representative 
is exact. 
See appendix \ref{app:modular} for a review of the definition and the main 
properties of the modular class.  
The relevance of unimodularity in BV theory 
has been recently emphasised in ref. \cite{Bonechi1}.

When a $\gamma\in\Gamma(Q_E)$ satisfying \eqref{LAA38} exists, it may not be unique. 
We are still free to redefine $\gamma$ into $\mathrm{e}^f\gamma$ for
any function $f\in \Fun(M)$ such that
\begin{equation}
\rho_r{}^a\partial_af=0.
\label{LAA45}
\end{equation}
Note that eq. \eqref{LAA45} reads compactly as $d_Ef=0$. 
So, its solutions span the degree $0$ cohomology $H_{LA}{}^0(E)$.

Henceforth, we assume 
that a nowhere vanishing 
$\gamma\in\Gamma(Q_E)$ satisfying \eqref{LAA38} exists
and has been chosen. Naturalness requires that all the relevant 
BV structures do not depend on this choice, 
a property that must be carefully checked. 

The BV algebra $\Fun(T^*[-1]E[1])$ is now equipped with the degree $1$ derivation 
\begin{equation}
\overline{d}_E=\ad_ES_E, 
\label{LAA37}
\end{equation}
and, for $X\in\Gamma(E)$, the degree $-1$, $0$ derivations
\vspace{-.1cm}
\begin{subequations}
\begin{align}
&\overline{i}_{EX}=\ad_E\iota_{EX}, 
\vphantom{\bigg]}
\label{LAA35}
\\
&\overline{l}_{EX}=\ad_E\lambda_{EX}, 
\vphantom{\bigg]}
\label{LAA36}
\end{align}
\label{LAA35,36}
\end{subequations}
\!\!where $\ad_E$ is defined according to \eqref{BValg6/1}. 
By \eqref{LAA27}--\eqref{LAA32},
$\overline{d}_E$, $\overline{i}_{EX}$, $\overline{l}_{EX}$ satisfy the Cartan relations
\eqref{LAA14}--\eqref{LAA19}. Further, by \eqref{LAA39}--\eqref{LAA34}, 
$\overline{d}_E$, $\overline{i}_{EX}$, $\overline{l}_{EX}$ are BV inner derivations 
(cf. sect. \ref{sec:BValg}). 

Inspecting \eqref{LAA10}, \eqref{LAA11}, \eqref{LAA12}, we observe that $d_E=\overline{d}_E|_{\Fun(E[1])}$,
$i_{EX}=\overline{i}_{EX}|_{\Fun(E[1])}$, $l_{EX}=\overline{l}_{EX}|_{\Fun(E[1])}$,
with $X\in\Gamma(E)$. 
Therefore, the derivations $\overline{d}_E$, $\overline{i}_E$, $\overline{l}_E$ extend 
$d_E$, $i_E$, $l_E$ from $\Fun(E[1])$ to $\Fun(T^*[-1]E[1])$.

By \eqref{LAA27}, \eqref{LAA39}, 
$S_E$ satisfies the quantum BV master equation \eqref{BVacts2} 
and is therefore a quantum BV master action of the BV algebra $\Fun(T^*[-1]E[1])$. 
The quantum BV operator is $\delta_{E\gamma}=\Delta_{E\gamma}+\ad_ES_E$ 
(cf. eq. \eqref{BVacts8}). 
$\delta_{E\gamma}$ depends explicitly on $\gamma$. The quantum BV cohomology 
$H_{BV}{}^*(\Fun(T^*[-1]E[1]))$, conversely, does not
up to isomorphism, since, for $f\in\Fun(M)$ satisfying \eqref{LAA45}, one has 
$\delta_{E\mathrm{e}^f\gamma}=\mathrm{e}^{-f}\delta_{E\gamma}\mathrm{e}^f$.
The classical BV operator is $\delta_{E\gamma c}=\ad_ES_E=\overline{d}_E$
(cf. eq. \eqref{BVacts11}). It is manifestly independent from $\gamma$. 
Hence, the classical BV cohomology $H_{cBV}{}^*(\Fun(T^*[-1]E[1]))$ also is.

Since $\delta_{E\gamma}|_{\Fun(E[1])}=\delta_{E\gamma c}|_{\Fun(E[1])}=d_E$,
the algebra inclusion $\pi_E{}^*:\Fun(E[1])$ $\rightarrow\Fun(T^*[-1]E[1])$
induces a homomorphism of the Lie algebroid cohomology $H_{LA}{}^*(E)$ into the quantum 
BV cohomology $H_{BV}{}^*(\Fun(T^*[-1]$ $E[1]))$ as well as 
the classical BV cohomology $H_{cBV}{}^*(\Fun(T^*[-1]E[1]))$.
Thus, each Lie algebroid cohomology class gives rise to
a well--defined BV observable.

Let $\mathfrak{g}$ be a Lie algebra and let $\varphi:\mathfrak{g}\rightarrow \Gamma(E)$ be 
a fiducial Lie algebra homomorphism. For any $x\in \mathfrak{g}$, let us define
\vspace{-.1cm}
\begin{subequations}
\begin{align}
&\iota_{Ex}=\iota_{E\varphi(x)},
\vphantom{\bigg]}
\label{LAA41}
\\
&\lambda_{Ex}=\lambda_{E\varphi(x)},
\vphantom{\bigg]}
\label{LAA42}
\end{align}
\label{LAA41,42}
\end{subequations}
\!\!and then define derivations $\overline{i}_{Ex}$, $\overline{l}_{Ex}$ on $\Fun(T^*[-1]E[1])$ via 
\eqref{LAA35}, \eqref{LAA36}. 

Suppose we keep only the Lie derivations $\overline{l}_{Ex}$ and forget about the interior
derivations $\overline{i}_{Ex}$. From \eqref{LAA36}, \eqref{LAA32}, \eqref{LAA34}, 
it follows immediately that   
\eqref{BVgau03}, \eqref{BVgau04}  are satisfied.
Hence, the BV algebra $\Fun(T^*[-1]E[1])$ 
carries an $N=0$ BV Hamiltonian $\mathfrak{g}$--action $\overline{l}_E$
having $\lambda_E$ as BV moment map. 
Further, by \eqref{LAA29}, the master action $S_E$ satisfies \eqref{BVgau05}
and is thus invariant under 
the $\mathfrak{g}$--action. Therefore,
we can perform the $N=0$ gauging of the BV algebra
following the scheme described in sect. \ref{sec:BVgauge0}. 

The relevant invariant quantum BV cohomology $H_{BV\mathrm{inv}}{}^*(\Fun(T^*[-1]E[1]))$ 
(cf. sect. \ref{sec:BVgauge0}) is independent from
the choice of $\gamma$, like $H_{BV}{}^*(\Fun(T^*[-1]E[1]))$.
In fact, for $f\in\Fun(M)$ satisfying \eqref{LAA45}, one has 
$\delta_{E\mathrm{e}^f\gamma}=\mathrm{e}^{-f}\delta_{E\gamma}\mathrm{e}^f$
and, for $x\in \mathfrak{g}$, $\overline{l}_{Ex}=\mathrm{e}^{-f}\overline{l}_{Ex}\mathrm{e}^f$,
as $l_{Ex}f=0$. Obviously,
the invariant classical BV cohomology $H_{cBV\mathrm{inv}}{}^*(\Fun(T^*[-1]E[1]))$ is 
independent from the choice of $\gamma$.

Since $\delta_{E\gamma}|_{\Fun(E[1])}=\delta_{E\gamma c}|_{\Fun(E[1])}=d_E$
and, for $x\in \mathfrak{g}$, $l_{Ex}=\overline{l}_{Ex}|_{\Fun(E[1])}$, 
the algebra inclusion $\pi_E{}^*:\Fun(E[1])$ $\rightarrow\Fun(T^*[-1]E[1])$
induces a homomorphism of the invariant Lie algebroid cohomology $H_{LA\mathrm{inv}}{}^*(E)$ into the 
invariant quantum BV cohomology $H_{BV\mathrm{inv}}{}^*(\Fun(T^*[-1]E[1]))$ as well as 
the invariant classical BV cohomology $H_{cBV\mathrm{inv}}{}^*(\Fun(T^*[-1]E[1]))$.
Thus, each invariant Lie algebroid cohomology class gives rise to
an invariant BV observable.

\vfill\eject

Suppose we keep both the interior derivations 
$\overline{i}_{Ex}$ and Lie derivations $\overline{l}_{Ex}$.
From \eqref{LAA35}, \eqref{LAA36}, \eqref{LAA30}--\eqref{LAA32}, \eqref{LAA33}, \eqref{LAA34}, 
it follows immediately that
\eqref{BVgau13}, \eqref{BVgau14} are satisfied. 
Hence, the BV algebra $\Fun(T^*[-1]E[1])$ 
carries an $N=1$ BV Hamiltonian $\mathfrak{g}$--action $\overline{i}_E$, $\overline{l}_E$
having $\iota_E$, $\lambda_E$  as BV (pre)moment maps. 
Further by \eqref{LAA28}, \eqref{LAA29}, 
the master action $S_E$ satisfies \eqref{BVgau15}
and is thus (Hamiltonian) invariant under the $\mathfrak{g}$--action. Therefore,
the $N=1$ gauging of the BV algebra can be carried out
along the lines illustrated in sect. \ref{sec:BVgauge1},

The relevant basic quantum BV cohomology $H_{BV\mathrm{bas}}{}^*(\Fun(T^*[-1]E[1]))$ 
(cf. sect. \ref{sec:BVgauge1}) is independent from the choice of $\gamma$, 
like $H_{BV}{}^*(\Fun(T^*[-1]E[1]))$, analogously to the $N=0$ case.
In fact, for $f\in\Fun(M)$ satisfying \eqref{LAA45}, one has 
$\delta_{E\mathrm{e}^f\gamma}=\mathrm{e}^{-f}\delta_{E\gamma}\mathrm{e}^f$
and, for $x\in \mathfrak{g}$, $\overline{i}_{Ex}=\mathrm{e}^{-f}\overline{i}_{Ex}\mathrm{e}^f$,
$\overline{l}_{Ex}=\mathrm{e}^{-f}\overline{l}_{Ex}\mathrm{e}^f$, as $i_{Ex}f$ $=l_{Ex}f=0$. 
The basic classical BV cohomology $H_{cBV\mathrm{bas}}{}^*(\Fun(T^*[-1]E[1]))$ is of course 
independent from the choice of $\gamma$.

As $\delta_{E\gamma}|_{\Fun(E[1])}=\delta_{E\gamma c}|_{\Fun(E[1])}=d_E$
and, for $x\in \mathfrak{g}$, $i_{Ex}=\overline{i}_{Ex}|_{\Fun(E[1])}$, $l_{Ex}=\overline{l}_{Ex}|_{\Fun(E[1])}$, 
the algebra inclusion $\pi_E{}^*:\Fun(E[1])$ $\rightarrow\Fun(T^*[-1]E[1])$
induces a homomorphism of the basic Lie algebroid cohomology $H_{LA\mathrm{bas}}{}^*(E)$ into the 
basic quantum BV cohomology $H_{BV\mathrm{bas}}{}^*(\Fun(T^*[-1]$ $E[1]))$ as well as 
the basic classical BV cohomology $H_{cBV\mathrm{bas}}{}^*(\Fun(T^*[-1]E[1]))$,
analogously to the $N=0$ case. So, 
each basic Lie algebroid cohomology class gives rise to
a basic BV observable.

The general construction expounded above exhibits a rich geometry, but, 
from the point of view of BV gauging, is kind of trivial: one can gauge 
any Lie algebra $\mathfrak{g}$ under the mild assumption that a Lie algebra 
homomorphism $\varphi:\mathfrak{g}\rightarrow \Gamma(E)$ is available.  
In physical problems, there virtually always are restrictions on the symmetries 
that one can gauge. The Poisson Lie algebroid is a special case of the above general
construction, in which such restrictions emerge naturally.
 
\vspace{.2cm}

{\it The Poisson Lie algebroid BV algebra and its gauging}

\vspace{.2cm}

Suppose that $M$ is a Poisson manifold and that $P\in\Gamma(\wedge^2TM)$ 
is its Poisson 
\vfill\eject\noindent
bivector \cite{Vaisman}. Then, $P$ satisfies the Poisson condition 
\begin{equation}
P^{ad}\partial_dP^{bc}+P^{bd}\partial_dP^{ca}+P^{cd}\partial_dP^{ab}=0.
\label{Poi0}
\end{equation}

As is well--known, the Poisson structure of $M$ endows the cotangent bundle $T^*M$ 
of $M$ with the structure of Lie algebroid. For simplicity, we shall mark all objects 
referring to this algebroid with a suffix $P$.
The anchor and structure functions of $T^*M$ 
are given by $\rho^{ab}=P^{ab}$ and $c_a{}^{ab}=\partial_aP^{bc}$. 
The Lie algebroid cohomology of $T^*M$ is the 
Poisson--Lichnerowicz cohomology of $P$.

Proceeding as explained in detail
above, we construct the associated BV algebra $(\Fun(T^*[-1]T^*[1]M), \Delta_{P\gamma},
\{\cdot,\cdot\}_P)$, the quantum BV master action $S_P$ and, for $\alpha\in\Gamma(T^*M)$, 
the (pre)moments $\iota_{P\alpha}$, $\lambda_{P\alpha}$
of the interior and Lie derivations $\overline{i}_{P\alpha}$, $\overline{l}_{P\alpha}$ of
$\Fun(T^*[-1]T^*[1]M)$.

The master action $S_P$ of, defined according to \eqref{LAA22}, reads as 
\begin{equation}
S_P=P^{ab}(x)y_b\xi_a+\frac{1}{2}\partial_aP^{bc}(x)\xi_b\xi_c\eta^a.
\label{Poi1}
\end{equation}
Similarly, for $\alpha\in\Gamma(T^*M)$, the (pre)moments $\iota_{P\alpha}$, $\lambda_{P\alpha}$
defined according to  \eqref{LAA23}, \eqref{LAA24}, are given by
\begin{subequations}
\begin{align}
\iota_{P\alpha}&=-\alpha_a(x)\eta^a,
\vphantom{\bigg]}
\label{Poi2}
\\
\lambda_{P\alpha}&=-P^{ab}\alpha_a(x)y_b-(P^{ac}\partial_c\alpha_b+\partial_bP^{ac}\alpha_c)(x)\xi_a\eta^b.
\vphantom{\bigg]}
\label{Poi3}
\end{align}
\label{Poi2,3}
\end{subequations}  
\hspace{-.25cm}
The orientation line bundle $Q_P=(\wedge^nT^*M)^{\otimes 2}$ is always trivial. 
Letting $\gamma\in\Gamma(Q_P)$ be a nowhere vanishing section, 
the unimodularity condition \eqref{LAA38} reads 
\begin{equation}
-2\gamma^{-1/2}\partial_b(\gamma^{1/2}P^{ba})=0., 
\label{Poi8}
\end{equation}
\eqref{Poi8} determines $\gamma$ only up to a rescaling by a factor $\mathrm{e}^f$,  
where $f\in \Fun(M)$ is a Casimir function of $P$ (that is  $P^{ab}\partial_bf=0$, 
cf. eq. \eqref{LAA45}). 
The action $S_P$ coincides with the reduced action used in the semiclassical
computation of the correlators of quantum observables for the Poisson sigma model
on the sphere in ref. \cite{Bonechi1}. 

Now, we shall assess whether it is possible to perform a non trivial gauging
of the Poisson BV algebra just constructed on the lines of ref. \cite{Zucchini7}. 
To this end, we make the following assumptions.

\begin{enumerate}

\item A compact connected Lie group $G$ with Lie algebra $\mathfrak{g}$ is given.

\item $M$ carries a smooth effective left $G$--action. 


\item The $G$--action is Hamiltonian.

\end{enumerate}

\noindent
As is well--known, the fundamental vector fields of the $G$--action organize as a section 
$u\in \Gamma(TM\otimes \mathfrak{g}^\vee)$. $u$ is $G$--equivariant, that is 
\begin{equation}
u_i{}^b\partial_bu_j{}^a-u_j{}^b\partial_bu_i{}^a=f^k{}_{ij}u_k{}^a,
\label{Poi4}
\end{equation}
where $f^k{}_{ij}$ are the structure constants of $\mathfrak{g}$.
As the $G$--action is Hamiltonian, there exists a moment map
$\mu\in \Gamma(\mathfrak{g}^\vee)$ of it. $\mu$ is $G$--equivariant, that is 
\begin{equation}
u_i{}^b\partial_b\mu_j=f^k{}_{ij}\mu_k,
\label{Poi5}
\end{equation}
and has the property that \hphantom{xxxxxxxxxxxxxxxxxxxxxxxx}
\begin{equation}
u_i{}^a=-P^{ab}\partial_b\mu_i.
\label{Poi6}
\end{equation}
Being the fundamental vector fields Hamiltonian, they leave the Poisson 
$2$--vector invariant, $l_{Mu_i}P^{ab}=0$. 

Now, define a section $\varphi\in\Gamma(T^*M\otimes \mathfrak{g}^\vee)$ by
\begin{equation}
\varphi_{ia}=\partial_a\mu_i
\label{Poi7}
\end{equation}
A simple calculation based on \eqref{Poi5}, \eqref{Poi6} shows that 
\begin{equation}
[\varphi_i,\varphi_j]_{Pa}=P^{bc}(\varphi_{ib}\partial_c\varphi_{ja}-\varphi_{jb}\partial_c\varphi_{ia})
+\partial_aP^{bc}\varphi_{ib}\varphi_{jc}=f^k{}_{ij}\varphi_{ka}.
\label{Poi9}
\end{equation}
Therefore, $\varphi:\mathfrak{g}\rightarrow \Gamma(T^*M)$ is a Lie algebra homomorphism.
The (pre)moments $\iota_{Pi}$, $\lambda_{Pi}$, defined according to \eqref{LAA41}, \eqref{LAA42},
are obtained by substituting $\varphi_i$ for $\alpha$ in \eqref{Poi2},\eqref{Poi3},
\begin{subequations}
\begin{align}
\iota_{Pi}&=-\partial_a\mu_i(x)\eta^a,
\vphantom{\bigg]}
\label{Poi10}
\\
\lambda_{Pi}&=P^{ab}\partial_b\mu_i(x)y_a-\partial_b(P^{ac}\partial_c\mu_i)(x)\xi_a\eta^b.
\vphantom{\bigg]}
\label{Poi11}
\end{align}
\label{Poi10,11}
\end{subequations}  
\!\!As explained in the first part of this section, we can construct in this way
an $N=0$ and an $N=1$ gauging of the BV algebra $\Fun(T^*[-1]T^*[1]M)$
with invariant master action $S_P$.
In the $N=0$ case, the invariant Poisson--Lichenrowicz cohomology of $P$ is contained, 
in the sense precisely defined above, in the invariant BV cohomology of $\Fun(T^*[-1]T^*[1]M)$.
Similarly, in the $N=1$ case, the basic Poisson--Lichenrowicz cohomology of $P$ is contained
in the basic BV cohomology of $\Fun(T^*[-1]T^*[1]M)$. 

\vspace{.1cm}

{\it Relation to the Poisson--Weil sigma model}

\vspace{.1cm}

The Poisson--Weil sigma model is a gauged version of the Poisson sigma model.
It has been studied in an AKSZ framework in refs. \cite{Zucchini6,Zucchini7} and further generalized 
in ref. \cite{Zucchini8}. The target space of the model is 
a Poisson manifold $M$ with a Hamiltonian effective left $G$---action as described above.
The fields of the model are de Rham superfields, 
that is sections of suitable bundles on the parity shifted tangent bundle
$T[1]\Sigma$ of the $2$--dimensional world sheet $\Sigma$. In the simplest version of the model, 
the field content is as follows

\begin{enumerate}

\item $\mathbfs{b}\in\Gamma(T[1]\Sigma,\mathfrak{g}^\vee[0])$.

\item $\mathbfs{c}\in\Gamma(T[1]\Sigma,\mathfrak{g}[1])$.

\item $\mathbfs{B}\in\Gamma(T[1]\Sigma,\mathfrak{g}^\vee[-1])$.

\item $\mathbfs{C}\in\Gamma(T[1]\Sigma,\mathfrak{g}[2])$

\item $\mathbfs{x}\in\Map(T[1]\Sigma,M)$.

\item $\mathbfs{y}\in\Gamma(T[1]\Sigma,\mathbfs{x}^*T^*[1]M)$. 

\end{enumerate}

The classical BV master action of the Poisson--Weil sigma model is 
\begin{align}
S_{PW}=\int_{T[1]\Sigma}\varrho\Big[&\mathbfs{b}_i\Big(\mathbfs{d}\mathbfs{c}^i
-\frac{1}{2}f^i{}_{jk}\mathbfs{c}^j\mathbfs{c}^k+\mathbfs{C}^i\Big)
-\mathbfs{B}_i\big(\mathbfs{d}\mathbfs{C}^i-f^i{}_{jk}\mathbfs{c}^j\mathbfs{C}^k\big)
\vphantom{\bigg]}
\label{PWBVact}
\\
&+\mathbfs{y}_a\big(\mathbfs{d}\mathbfs{x}^a+u_i{}^a(\mathbfs{x})\mathbfs{c}^i\big)
-\mu_i(\mathbfs{x})\mathbfs{C}^i
-\frac{1}{2}P^{ab}(\mathbfs{x})\mathbfs{y}_a\mathbfs{y}_b\Big],
\vphantom{\bigg]}
\nonumber
\end{align}
where $\varrho$ is the invariant supermeasure on $T[1]\Sigma$. It is not known whether
$S_{PW}$ satisfies also the appropriate quantum BV master equation, though it is known this to be the
case for the pure Poisson sigma model \cite{Cattaneo1}
\footnote{$\vphantom{\bigg[}$ In \cite{Zucchini6,Zucchini7}, $\mu_i$ and $P^{ab}$ 
have opposite sign.}.

The Poisson--Weil sigma model has a finite dimensional reduction defined as follows.
Denote by $\mathbfs{1}$ and $\mathbfs{\omega}$ the unit and a volume form of $\Sigma$, 
viewed respectively as a degree $0$ element and a nowhere vanishing degree $2$ element 
of $\Fun(T[1]\Sigma)$. We assume further that $\mathbfs{\omega}$ is normalized as 
\begin{equation}
\int_{T[1]\Sigma}\varrho\,\mathbfs{\omega}=1.
\label{Poi13}
\end{equation}
Take the superfields of the model to be of the form
\begin{subequations}
\begin{align}
\mathbfs{b}_i&=b_i\,\mathbfs{\omega},
\vphantom{\bigg]}
\label{Poi14a}
\\
\mathbfs{c}^i&=c^i\,\mathbfs{1},
\vphantom{\bigg]}
\label{Poi14b}
\\
\mathbfs{B}_i&=B_i\,\mathbfs{\omega},
\vphantom{\bigg]}
\label{Poi14c}
\\
\mathbfs{C}^i&=C^i\,\mathbfs{1},
\vphantom{\bigg]}
\label{Poi14d}
\\
\mathbfs{x}^a&=x^a\,\mathbfs{1}-\eta^a\,\mathbfs{\omega},
\vphantom{\bigg]}
\label{Poi14e}
\\
\mathbfs{y}_a&=\xi_a\,\mathbfs{1}+y_a\,\mathbfs{\omega},
\vphantom{\bigg]}
\label{Poi14f}
\end{align}
\label{Poi14}
\end{subequations} 
\!\!where $(b_i,c^i)$, $(B_i,C^i)$, $(x^a,y_a)$, $(\xi_a,\eta^a)$
are BV conjugate pairs of variables of degrees $(-2,1)$, $(-3,2)$, $(0,-1)$, $(1,-2)$, 
respectively. Substituting \eqref{Poi14a}--\eqref{Poi14f} into \eqref{PWBVact}, 
we get a finite dimensional reduction $S_{PW}{}^0$ of $S_{PW}$. 
This reads as
\begin{equation}
S_{PW}{}^0=S_{\mathfrak{g}|1}+S_P+c^i\lambda_{Pi}-C^i\iota_{Pi},
\label{Poi12} 
\end{equation} 
where $S_{\mathfrak{g}|1}$, $S_P$, $\iota_{Pi}$, $\lambda_{Pi}$ are given by 
\eqref{BVgau111}, \eqref{Poi1}, \eqref{Poi10}, \eqref{Poi11}, respectively.
Upon comparing with \eqref{BVgau119}, we immediately realize that $S_{PW}{}^0$ is nothing but the 
$N=1$ gauged matter BV master action 
of the finite dimensional Poisson Lie algebroid model described above 
\begin{equation}
S_{PW}{}^0=S_{\mathfrak{g}|1P}.
\label{Poi15} 
\end{equation} 
(In eq. \eqref{Poi12}, the tensor product symbol $\otimes$ is omitted.)
Presumably, the finite dimensional model can be used to compute correlators of the 
Poisson--Weil model along the lines described in ref. \cite{Bonechi1}.

\vfill\eject

\section{\normalsize \textcolor{blue}{Conclusions}}\label{sec:conclu}

~~~~In this paper, we have explored certain less known features of BV algebras, which
have not been the object of a systematic study so far. We have pointed out that
a BV master action may possess global symmetries not directly related to the 
gauge symmetries which underlie the BV symmetry and which may be interesting to 
gauge for a variety of reasons. 
We have seen that the gauging can be carried out in a purely BV framework.
The global symmetry of the master action organizes as a Lie algebra action
with a varying amount of supersymmetry, which determines directly the amount of 
ghost supersymmetry and the procedure of gauging.
We have found that $N=0$ and $N=1$ gauging correspond to ordinary gauging 
and to topological gauging, respectively. For higher $N$, the situation 
is not clear yet. The ordinary formal structure of BV algebras seems to be inadequate to 
treat these cases and, though sensible algebraic constructions can be carried out, 
their eventual field theoretic origin or underpinning is not clear.
This may be the object of future investigation.

We feel that the BV algebraic framework is more versatile than it has so far 
been realized. It would be certainly worth the effort to explore the full range of 
its applications.

\vfill\eject

\appendix

\section{\normalsize \textcolor{blue}{Differential Lie modules and their invariant cohomology}}
\label{app:invcoh}

In this appendix, we recall the basic properties of differential Lie modules and their cohomology. 
We further provide a self-contained proof of the important cohomology isomorphism
\eqref{inv5}. See ref. \cite{Grueb1} for background material. 

A {\it differential Lie module} is a quadruplet $(\mathcal{E}, \mathfrak{g},l,\delta)$,
where $\mathcal{E}$ is a $\mathbb{Z}$--graded vector space, $\mathfrak{g}$ is a Lie algebra and 
$l:\mathfrak{g}\rightarrow \End_0(\mathcal{E})$ is a linear map and $\delta\in \End_1(\mathcal{E})$ 
satisfying the graded commutation relations 
\begin{subequations}
\begin{align}
&[l_i,l_j]=f^k{}_{ij}l_k,
\vphantom{\bigg]}
\label{inv1a}
\\
&[\delta,l_i]=0,
\vphantom{\bigg]}
\label{inv1b}
\\
&[\delta,\delta]=0,
\vphantom{\bigg]}
\label{inv1c}
\end{align} 
\label{inv1}
\end{subequations}
\!\!with respect to a chosen basis $\{t_i\}$ of $\mathfrak{g}$.
We note that neither $\mathcal{E}$ is supposed to be an algebra
nor $l_i$, $\delta$ are supposed to be graded derivations.
If $\mathcal{F}\subset\mathcal{E}$ is a subspace, 
$\mathcal{F}_{\mathrm{inv}}=\mathcal{F}\cap(\cap_i\ker l_i)$ is called the 
{\it invariant} component of $\mathcal{F}$. 

The pairs $(\mathcal{E},\delta)$, $(\mathcal{E}_{\mathrm{inv}},\delta)$ 
are both differential spaces. Their associated cohomologies 
$H^*(\mathcal{E})=H^*(\mathcal{E},\delta)$, 
$H_{\mathrm{inv}}{}^*(\mathcal{E})=H^*(\mathcal{E}_{\mathrm{inv}},\delta)$
are the {\it ordinary} and the {\it invariant cohomology} of the differential Lie 
module $(\mathcal{E}, \mathfrak{g},l,\delta)$.

With the Lie algebra $\mathfrak{g}$, there is associated a canonical differential Lie
module $(CE(\mathfrak{g}),\mathfrak{g},l_{\mathfrak{g}},\delta_{\mathfrak{g}})$,
called {\it the Chevalley--Eilenberg Lie module}. $CE(\mathfrak{g})$ is 
\begin{equation}
CE(\mathfrak{g})=\Fun(\mathfrak{g}[1]),
\label{inv2}
\end{equation}
the algebra of polynomials of the coordinates 
$c^i$  of $\mathfrak{g}[1]$  with respect to the basis $\{t_i\}$.
$l_{\mathfrak{g}},\delta_{\mathfrak{g}}$ are defined by the relations 
\begin{subequations}
\begin{align}
&l_{\mathfrak{g}i}c^j=-f^j{}_{ik}c^k,
\vphantom{\bigg]}
\label{inv3a}
\\
&\delta_{\mathfrak{g}}c^i=-\frac{1}{2}f^i{}_{jk}c^jc^k.
\vphantom{\bigg]}
\label{inv3b}
\end{align}
\label{inv3}
\end{subequations}

For a given differential Lie module $(\mathcal{E}, \mathfrak{g},l,\delta)$, let us set
\begin{equation}
\mathcal{E}'=CE(\mathfrak{g})\otimes\mathcal{E}.
\label{inv9}
\end{equation}
We can define endomorphisms of $\mathcal{E}'$ by  
\begin{subequations}
\begin{align}
l'{}_i&=l_{\mathfrak{g}i}\otimes 1+1_{\mathfrak{g}}\otimes l_i,
\vphantom{\bigg]}
\label{inv4a} 
\\
\delta'&=\delta_{\mathfrak{g}}\otimes 1+1_{\mathfrak{g}}\otimes \delta+c^i\otimes l_i.
\vphantom{\bigg]}
\label{inv4b} 
\end{align}
\label{inv4}
\end{subequations}
\!\!Then, $(\mathcal{E}', \mathfrak{g},l',\delta')$ is a differential Lie module.

{\it If $\mathfrak{g}$ is a reductive Lie algebra, then}
\begin{equation}
H_{\mathrm{inv}}{}^*(\mathcal{E}')\simeq 
CE(\mathfrak{g})_{\mathrm{inv}}\otimes H_{\mathrm{inv}}{}^*(\mathcal{E}).
\label{inv5}
\end{equation}
Recall that $\mathfrak{g}$ is reductive if $\mathfrak{g}$  is the direct sum of an Abelian and
a semisimple Lie algebra.
The rest of this appendix is devoted to the sketch of the proof of the above result.

To begin with, we note that 
\begin{equation}
\mathcal{E}'{}_{\mathrm{inv}}=(CE(\mathfrak{g})\otimes\mathcal{E})_{\mathrm{inv}}.
\label{inv10}
\end{equation}
This suggests defining the following subspaces of $\mathcal{E}'{}_{\mathrm{inv}}$ 
\begin{align}
\mathcal{C}_n&=(CE(\mathfrak{g})_n\otimes\mathcal{E})_{\mathrm{inv}},
\vphantom{\bigg]}
\label{inv11} 
\\
\mathcal{D}_n&=\bigoplus_{0\leq m\leq n}\mathcal{C}_m,
\vphantom{\bigg]}
\label{inv12} 
\end{align}
where $n\geq 0$. Then, 
$\mathcal{E}_{\mathrm{inv}}\simeq \mathcal{D}_0\subset \mathcal{D}_1\subset \mathcal{D}_2\subset
\ldots \subset \mathcal{D}_{h}=\mathcal{E}'{}_{\mathrm{inv}}$, $h=\dim\mathfrak{g}$,
is a filtration of the vector space $\mathcal{E}'{}_{\mathrm{inv}}$. 
One can show the following two properties.

Let $\{z_{kx}\}$ be a basis of $CE(\mathfrak{g})_{\mathrm{inv}}$ such that 
$z_{kx}\in CE_k(\mathfrak{g})$ for all $x$. (Such a basis exists
as $l_{\mathfrak{g}i}CE_k(\mathfrak{g})\subset CE_k(\mathfrak{g})$ for all $k$.)

\noindent
$i)$ Let $\mu\in\mathcal{D}_n$ be such that \hphantom{xxxxxxxxxxxxxxxxxx}
\begin{equation}
\delta'\mu=0.
\label{inv27}
\end{equation}
Then, there are $\nu\in \mathcal{D}_{n-1}$, $\alpha_k{}^x\in\mathcal{E}_{\mathrm{inv}}$ with 
$0\leq k\leq n$ such that
\begin{align}
&\mu=\delta'\nu+\sum_{0\leq k\leq n}z_{kx}\otimes \alpha_k{}^x,
\vphantom{\bigg]}
\label{inv28} 
\\
&\delta\alpha_k{}^x=0.
\vphantom{\bigg]}
\label{inv29}  
\end{align}

\noindent
$ii)$ Let $\nu\in \mathcal{D}_{n-1}$, $\alpha_k{}^x\in\mathcal{E}_{\mathrm{inv}}$ with
$0\leq k\leq n$ be such that
\begin{equation}
\delta'\nu+\sum_{0\leq k\leq n}z_{kx}\otimes \alpha_k{}^x=0.
\label{inv30}
\end{equation}
Then, there are $\beta_k{}^x\in\mathcal{E}_{\mathrm{inv}}$ with $0\leq k\leq n$, such that  
\begin{equation}
\alpha_k{}^x=\delta\beta_k{}^x.
\label{inv31}
\end{equation}

By $i$, $ii$, there is a homomorphism $q: H^*(\mathcal{E}'{}_{\mathrm{inv}}, \delta')\rightarrow 
CE(\mathfrak{g})_{\mathrm{inv}}\otimes H^*(\mathcal{E}_{\mathrm{inv}},\delta)$ defined by 
the expression 
\begin{equation}
q([\mu])=\sum_{0\leq k\leq n}z_{kx}\otimes [\alpha_k{}^x],
\label{inv31/2}
\end{equation}
where $\mu$ is expressed as in \eqref{inv28}. 
By \eqref{inv3b}, as $\delta_{\mathfrak{g}}z_{kx}=\frac{1}{2}c^il_{\mathfrak{g}i}z_{kx}=0$, one has
\begin{equation}
\delta'\sum_{0\leq k\leq n}z_{kx}\otimes \gamma_k{}^x
=\sum_{0\leq k\leq n}(-1)^kz_{kx}\otimes \delta\gamma_k{}^x,
\label{inv31/1}
\end{equation}
for $\gamma_k{}^x\in\mathcal{E}_{\mathrm{inv}}$. It follows that $q$ is an isomorphism. 
Thus, if one shows $i$, $ii$, \eqref{inv5} is shown as well.

{\it Proof of $i$. $ii$}. $\mathfrak{g}$ acts on $CE(\mathfrak{g})$ via \eqref{inv3a}
and so, $CE(\mathfrak{g})$  is a representation of $\mathfrak{g}$.
From Lie algebra theory, since $\mathfrak{g}$ is reductive, this representation 
is semisimple. Thus, for any $\mathfrak{g}$--stable subspace $U\subset CE(\mathfrak{g})$
and for any  $\mathfrak{g}$--stable subspace $V\subset U$
there is a $\mathfrak{g}$--stable subspace  $W\subset U$ such that 
$U\simeq V\oplus W$. 
In particular, if $U\subset CE(\mathfrak{g})$ is a $\mathfrak{g}$--stable subspace, then
$U=U_{\mathrm{inv}}\oplus l_{\mathfrak{g}}U$, where 
$U_{\mathrm{inv}}=U\cap(\cap_i\ker l_{\mathfrak{g}i})$ 
and $l_{\mathfrak{g}}U=\lspan_i l_{\mathfrak{g}i}U=U\cap (\lspan_i \im l_{\mathfrak{g}i})$.

Consider $Z_n(CE(\mathfrak{g}))=\ker\delta_{\mathfrak{g}}\cap CE_n(\mathfrak{g})$.
$Z_n(CE(\mathfrak{g}))$ is $\mathfrak{g}$--stable and, therefore, $Z_n(CE(\mathfrak{g}))=
Z_n(CE(\mathfrak{g}))_{\mathrm{inv}}\oplus l_{\mathfrak{g}}Z_n(CE(\mathfrak{g}))$.
Since $\delta_{\mathfrak{g}}=\frac{1}{2}c^il_{\mathfrak{g}i}$, by  \eqref{inv3b}, 
$CE_n(\mathfrak{g})_{\mathrm{inv}}
\subset Z_n(CE(\mathfrak{g}))_{\mathrm{inv}}\subset CE_n(\mathfrak{g})_{\mathrm{inv}}$ and, hence,
$Z_n(CE(\mathfrak{g}))_{\mathrm{inv}}=CE_n(\mathfrak{g})_{\mathrm{inv}}$.
Now, let  $B_n(CE(\mathfrak{g}))=\im\delta_{\mathfrak{g}}\cap CE_n(\mathfrak{g})$.
Since $l_{\mathfrak{g}i}=i_{\mathfrak{g}i}\delta_{\mathfrak{g}}+\delta_{\mathfrak{g}}i_{\mathfrak{g}i}$,
where $i_{\mathfrak{g}i}$ is the degree $-1$ derivation of $CE(\mathfrak{g})$
defined by $i_{\mathfrak{g}i}c^j=\delta^j{}_i$, and $\delta_{\mathfrak{g}}=\frac{1}{2}l_{\mathfrak{g}i}c^i$, 
by  \eqref{inv3b}, 
as $f^j{}_{ji}=0$ for a reductive Lie algebra $\mathfrak{g}$, 
$l_{\mathfrak{g}}Z_n(CE(\mathfrak{g}))\subset B_n(CE(\mathfrak{g}))
\subset l_{\mathfrak{g}}Z_n(CE(\mathfrak{g}))$ and, thus,  
$l_{\mathfrak{g}}Z_n(CE(\mathfrak{g}))=B_n(CE(\mathfrak{g}))$.
In conclusion,
\begin{equation}
Z_n(CE(\mathfrak{g}))= CE_n(\mathfrak{g})_{\mathrm{inv}}\oplus B_n(CE(\mathfrak{g})).
\label{inv32}
\end{equation}
Further, as $Z_n(CE(\mathfrak{g}))$, $CE_n(\mathfrak{g})$ are $\mathfrak{g}$--stable
and $Z_n(CE(\mathfrak{g}))\subset CE_n(\mathfrak{g})$, 
\begin{equation}
CE_n(\mathfrak{g})=Z_n(CE(\mathfrak{g}))\oplus\widetilde{CE}_n(\mathfrak{g}),
\label{inv33}
\end{equation}
for some $\mathfrak{g}$--stable subspace $\widetilde{CE}_n(\mathfrak{g})\subset CE_n(\mathfrak{g})$.
As a consequence, $\delta_{\mathfrak{g}}:\widetilde{CE}_n(\mathfrak{g})
\rightarrow B_{n+1}(CE(\mathfrak{g}))$ is an isomorphism.

From the above discussion, it follows that, for each $n\geq 0$,  there is a basis
$\{z_{nx},r_{nu},s_{ns}\}$ of $CE_n(\mathfrak{g})$ such that $\{z_{nx}\}$, $\{r_{nu}\}$,
$\{s_{ns}\}$ are bases of $CE_n(\mathfrak{g})_{\mathrm{inv}}$, $B_n(CE(\mathfrak{g}))$, 
$\widetilde{CE}_n(\mathfrak{g})$, respectively, with the property that 
\begin{align}
&l_{\mathfrak{g}i}z_{nx}=0,\qquad l_{\mathfrak{g}i}r_{nu}=-A_{ni}{}^v{}_ur_{nv},
\qquad l_{\mathfrak{g}i}s_{ns}=-B_{ni}{}^t{}_ss_{nt},
\vphantom{\bigg]}
\label{inv34} 
\\
&\delta_{\mathfrak{g}}z_{nx}=0,\qquad \delta_{\mathfrak{g}}r_{nu}=0,
\qquad \delta_{\mathfrak{g}}s_{ns}=-Q_n{}^u{}_sr_{n+1u},
\vphantom{\bigg]}
\label{inv35}  
\end{align}
where $A_{ni}$, $B_{ni}$ and $Q_n$ are square matrices with $Q_n$ invertible.
The relation $[l_{\mathfrak{g}i},\delta_{\mathfrak{g}}]=0$ implies further the matrix relation
\begin{equation}
A_{n+1i}Q_n-Q_nB_{ni}=0.
\label{inv36}
\end{equation}

Next, combining \eqref{inv4a}, \eqref{inv4b} and the relation 
$\delta_{\mathfrak{g}}=\frac{1}{2}c^il_{\mathfrak{g}i}$, $\delta'$ can be cast as
\begin{equation}
\delta'=-\delta_{\mathfrak{g}}\otimes 1+1_{\mathfrak{g}}\otimes \delta+c^i\otimes 1\cdot l'{}_i
\label{inv37}
\end{equation}
It follows that, when restricting to $\mathcal{E}'{}_{\mathrm{inv}}$, 
\begin{equation}
\delta'=\delta'{}_1+\delta'{}_2,
\label{inv38}
\end{equation}
where $\delta'{}_1$, $\delta'{}_2$ are given by \hphantom{xxxxxxxxxxxxxxxxxxxxxxxx}
\vspace{-.2cm}
\begin{subequations}
\begin{align}
&\delta'{}_1=-\delta_{\mathfrak{g}}\otimes 1,
\vphantom{\bigg]}
\label{inv39a} 
\\
&\delta'{}_2=1_{\mathfrak{g}}\otimes \delta.
\vphantom{\bigg]}
\label{inv39b} 
\end{align}
\label{inv39}
\end{subequations}
\vspace{-.7cm}

Next, we have the following result. Let $n\geq 0$. If $\mu_n\in \mathcal{C}_n$ is such that 
\begin{equation}
\delta'{}_1\mu_n=0,
\label{inv40}
\end{equation}
then $\mu_n$ is of the special form \hphantom{xxxxxxxxxxxxxxxxxxxxxxxx}
\begin{equation}
\mu_n=\delta'{}_1\nu_{n-1}+z_{nx}\otimes\alpha^x, 
\label{inv41}
\end{equation}
for certain $\nu_{n-1}\in \mathcal{C}_{n-1}$, $\alpha^x\in \mathcal{E}_{\mathrm{inv}}$. 
To see this, we write $\mu_n$ as
\begin{equation}
\mu_n=z_{nx}\otimes\alpha^x+r_{nu}\otimes\beta^u+s_{ns}\otimes \gamma^s,
\label{inv42}
\end{equation}
where $\alpha^x,\beta^u,\gamma^s\in\mathcal{E}$. By \eqref{inv4a}, 
\eqref{inv34}, the condition $l'{}_i\mu_n=0$
implies that
\begin{equation}
l_i\alpha^x=0,\qquad l_i\beta^u=A_{ni}{}^u{}_v\beta^v,
\qquad l_i\gamma^s=B_{ni}{}^s{}_t\gamma^t.
\label{inv43}
\end{equation}
By the first relation \eqref{inv43}, 
$\alpha^x\in \mathcal{E}_{\mathrm{inv}}$. By the 3rd relation \eqref{inv35}, one has 
$r_{nu}=-Q_{n-1}{}^{-1s}{}_u\delta_{\mathfrak{g}}s_{n-1s}$.
Hence, on account of \eqref{inv39a}, one has 
\begin{equation}
r_{nu}\otimes\beta^u=\delta'{}_1\nu_{n-1},
\label{inv45}
\end{equation}
where $\nu_{n-1}$ is given by \hphantom{xxxxxxxxxxxxxxxxxxxxxxxx}
\begin{equation}
\nu_{n-1}=Q_{n-1}{}^{-1s}{}_us_{n-1s}\otimes \beta^u.
\label{inv46}
\end{equation}
Using the 2nd relation \eqref{inv43}, the 3rd relation \eqref{inv34} and 
\eqref{inv36}, one finds that 
$l'{}_i\nu_{n-1}=0$. Hence, $\nu_{n-1}\in\mathcal{C}_{n-1}$.
By \eqref{inv39a}, \eqref{inv35} and the invertibility of the matrix 
$Q_n$ the condition $\delta'{}_1\mu_n=0$ implies that the $\gamma^s$ all vanish,
\eject\noindent
\begin{equation}
\gamma^s=0.
\label{inv48}
\end{equation}
From \eqref{inv42}, \eqref{inv45},  \eqref{inv48}, we get \eqref{inv41}.

The proof of $i$ proceeds by induction on $n$. Let $\mu\in\mathcal{D}_0$ satisfy \eqref{inv27}.
Then, $\mu=1_{\mathfrak{g}}\otimes\alpha$ for some $\alpha\in\mathcal{E}_{\mathrm{inv}}$.
Further, by \eqref{inv4b}, one has $\delta\alpha=0$. Therefore, \eqref{inv28}, \eqref{inv29} hold 
with $\nu=0$. So, $i$ holds for $n=0$.
Suppose now $i$ holds for $n-1$ with $n\geq 1$. Let $\mu\in\mathcal{D}_n$ satisfy \eqref{inv27}.
Write $\mu=\mu_n+\tilde \mu$, where $\mu_n\in \mathcal{C}_n$,
$\tilde\mu\in\mathcal{D}_{n-1}$. Since 
$\delta'{}_1\mathcal{C}_m\subset \mathcal{C}_{m+1}$
$\delta'{}_2\mathcal{C}_m\subset \mathcal{D}_m$ for $m\geq 0$, 
condition \eqref{inv27} implies 
that $\delta'{}_1\mu_n=0$. So, $\mu_n$ satisfies \eqref{inv40} and, so, 
by \eqref{inv41}, there are $\nu_{n-1}\in \mathcal{C}_{n-1}$
and $\alpha_n{}^x\in \mathcal{E}_{\mathrm{inv}}$ such that 
$\mu_n=\delta'{}_1\nu_{n-1}+z_{nx}\otimes\alpha_n{}^x 
=\delta'\nu_{n-1}+z_{nx}\otimes\alpha_n{}^x-\delta'{}_2\nu_{n-1}$.
Setting $\mu^*=\tilde\mu-\delta'{}_2\nu_{n-1}\in \mathcal{D}_{n-1}$, we have then
\begin{equation}
\mu=\delta'\nu_{n-1}+z_{nx}\otimes\alpha_n{}^x +\mu^*.
\label{inv49} 
\end{equation}
Next, since $\delta'\mu=0$ and $\delta'(z_{nx}\otimes\alpha_n{}^x)=(-1)^nz_{nx}\otimes\delta\alpha_n{}^x$
by \eqref{inv38}, \eqref{inv39} and the 1st relation \eqref{inv35},
$(-1)^nz_{nx}\otimes\delta\alpha_n{}^x+\delta'\mu^*=0$.
As $\delta'\mu^*$ has no components of the form $z_{nx}\otimes\gamma^x$, one has 
$\delta\alpha_n{}^x=0$. Thus, $\delta'\mu^*=0$. So, $\mu^*$ satisfies \eqref{inv27}, 
and, so, by the inductive hypothesis, \eqref{inv28}, \eqref{inv29} hold, yielding
\begin{equation}
\mu^*=\delta'\nu^*+\sum_{0\leq k\leq n-1}z_{kx}\otimes \alpha_k{}^x,
\label{inv50} 
\end{equation}
with $\nu^*\in \mathcal{D}_{n-2}$, $\alpha_k{}^x\in\mathcal{E}_{\mathrm{inv}}$ such that
$\delta\alpha_k{}^x=0$ for $0\leq k\leq n-1$. Substituting \eqref{inv50} into 
\eqref{inv49} and setting $\nu=\nu_{n-1}+\nu^*\in\mathcal{D}_{n-1}$, we find that $\mu$ is of the form 
\eqref{inv28} with \eqref{inv29} satisfied. By induction on $n$, $i$ is shown.

The proof of $ii$ also proceeds by induction on $n$.
Let $\nu\in\mathcal{D}_0$, $\alpha_0,\alpha_1{}^x\in\mathcal{E}_{\mathrm{inv}}$ 
satisfy \eqref{inv30}. (Note that $\{z_0{}^x\}=\{1\}$.)
Then, $\nu=-1_{\mathfrak{g}}\otimes\beta$, for some $\beta\in\mathcal{E}_{\mathrm{inv}}$.
Further, by \eqref{inv4b}, one has $\alpha_0=\delta\beta$ and $\alpha_1{}^x=0$. 
Hence, \eqref{inv31} holds. So, $ii$ holds for $n=1$.
Suppose $ii$ holds for $n-2$ with $n\geq 2$. 
Let $\nu\in\mathcal{D}_{n-1}$, 
$\alpha_k{}^x\in\mathcal{E}_{\mathrm{inv}}$ with $0\leq k\leq n$ satisfy \eqref{inv30}.
Write $\nu=\nu_{n-1}+\tilde \nu$, where $\nu_{n-1}\in \mathcal{C}_{n-1}$,
$\tilde\nu\in\mathcal{D}_{n-2}$. Since $\delta'{}_1\mathcal{C}_m\subset \mathcal{C}_{m+1}$
$\delta'{}_2\mathcal{C}_m\subset \mathcal{D}_m$ for $m\geq 0$, 
condition \eqref{inv30} implies that $\delta'{}_1\nu_{n-1}+z_{nx}\otimes \alpha_n{}^x=0$.
As $\delta'{}_1\nu_{n-1}$ has no components of the form $z_{nx}\otimes\gamma^x$, one has 
$\alpha_n{}^x=0$. Hence, $\delta'{}_1\nu_{n-1}=0$. So, $\nu_{n-1}$ satisfies \eqref{inv40} and, so, 
by \eqref{inv41}, there are $\nu_{n-2}\in \mathcal{C}_{n-2}$ and 
$\beta^x\in \mathcal{E}_{\mathrm{inv}}$ such that 
$\nu_{n-1}=\delta'{}_1\nu_{n-2}+z_{n-1x}\otimes\beta^x$. Then, by \eqref{inv38}, \eqref{inv39}, 
$\delta'\nu_{n-1}=\delta'{}_2\nu_{n-1}=(-1)^{n-1}z_{n-1x}\otimes \delta\beta^x-\delta'\delta'{}_2\nu_{n-2}$.
Setting $\nu^*=\tilde\nu-\delta'{}_2\nu_{n-2}\in\mathcal{D}_{n-2}$, we have then
\begin{equation}
\delta'\nu=\delta'\nu^*+(-1)^{n-1}z_{n-1x}\otimes\delta\beta^x.
\label{inv51} 
\end{equation}
Substituting \eqref{inv51} in \eqref{inv30} and recalling that $\alpha_n{}^x=0$, we find 
\begin{equation}
\delta'\nu^*+\sum_{0\leq k\leq n-1}z_{kx}\otimes\alpha^*{}_k{}^x=0,
\label{inv52} 
\end{equation}
where $\alpha^*{}_k{}^x=\alpha_k{}^x+(-1)^{n-1}\delta_{k,n-1}\delta\beta^x\in \mathcal{E}_{\mathrm{inv}}$.
Thus, $\nu^*$, $\alpha^*{}_k{}^x$ satisfy \eqref{inv30}
and, so, by the inductive hypothesis, $\alpha^*{}_k{}^x=\delta\beta^*{}_k{}^x$
for certain $\beta^*{}_k{}^x\in \mathcal{E}_{\mathrm{inv}}$.
Thus, \eqref{inv31} holds. By induction on $n$, $ii$ is shown.
\hfill {\it QED}

\vfill\eject

\section{\normalsize \textcolor{blue}{Lie operations and their equivariant cohomology}}
\label{app:equivcoh}

In this appendix, we recall the basic properties of Lie operations and their cohomology. We 
further provide a self-contained proof of the important cohomology isomorphism
\eqref{oper15}. See refs. \cite{Grueb1,Guillemin1}, for background material. 

A {\it Lie operation} is a quintuplet $(\mathcal{E}, \mathfrak{g},i,l,\delta)$,
where $\mathcal{E}$ is a $\mathbb{Z}$--graded vector space, $\mathfrak{g}$ is a Lie algebra and 
$i:\mathfrak{g}\rightarrow \End_{-1}(\mathcal{E})$, $l:\mathfrak{g}\rightarrow \End_0(\mathcal{E})$
are linear maps and $\delta\in \End_1(\mathcal{E})$ satisfying the commutation relations 
\begin{subequations}
\begin{align}
&[i_i,i_j]=0,
\vphantom{\bigg]}
\label{oper1a}
\\
&[l_i,i_j]=f^k{}_{ij}i_k,
\vphantom{\bigg]}
\label{oper1b}
\\
&[l_i,l_j]=f^k{}_{ij}l_k,
\vphantom{\bigg]}
\label{oper1c}
\\
&[\delta,i_i]=l_i,
\vphantom{\bigg]}
\label{oper1d}
\\
&[\delta,l_i]=0,
\vphantom{\bigg]}
\label{oper1e}
\\
&[\delta,\delta]=0,
\vphantom{\bigg]}
\label{oper1f}
\end{align} 
\label{oper1}
\end{subequations}
\!\!with respect to a chosen basis $\{t_i\}$ of $\mathfrak{g}$.
We note that neither $\mathcal{E}$ is supposed to be an algebra
nor $i_i$, $l_i$, $\delta$ are supposed to be graded derivations.
If $\mathcal{F}\subset\mathcal{E}$ is a subspace, 
$\mathcal{F}_{\mathrm{hor}}=\mathcal{F}\cap(\cap_i\ker i_i)$,
$\mathcal{F}_{\mathrm{inv}}=\mathcal{F}\cap(\cap_i\ker l_i)$ and
$\mathcal{F}_{\mathrm{bas}}=\mathcal{F}\cap(\cap_i(\ker i_i\cap\ker l_i))
=\mathcal{F}_{\mathrm{hor}}\cap\mathcal{F}_{\mathrm{inv}}$  are called the 
{\it horizontal, invariant} and {\it basic} component of 
$\mathcal{F}$, respectively. 

The pairs $(\mathcal{E},\delta)$, $(\mathcal{E}_{\mathrm{bas}},\delta)$ 
are both differential spaces. Their associated cohomologies 
$H^*(\mathcal{E})=H^*(\mathcal{E},\delta)$, 
$H_{\mathrm{bas}}{}^*(\mathcal{E})=H^*(\mathcal{E}_{\mathrm{bas}},\delta)$
are the {\it ordinary} and the {\it basic cohomology} of the Lie operation
$(\mathcal{E}, \mathfrak{g},i,l,\delta)$.

Lie operations can be equipped with connections. 
A {\it connection} of the Lie operation $(\mathcal{E}, \mathfrak{g},i,l,\delta)$
is a linear map $\theta:\mathfrak{g}^\vee\rightarrow \End_1(\mathcal{E})$
satisfying the commutation relations 
\eject\noindent
\begin{subequations}
\begin{align}
&[i_j,\theta^i]=\delta^i{}_j,
\vphantom{\bigg]}
\label{oper4a} 
\\
&[l_j,\theta^i]=-f^i{}_{jk}\theta^k.
\vphantom{\bigg]}
\label{oper4b} 
\end{align}
\label{oper4}
\end{subequations}
\!\!The {\it curvature} of the connection $\theta$ is the linear map 
$\Theta:\mathfrak{g}^\vee\rightarrow \End_2(\mathcal{E})$ defined by
\begin{equation}
\Theta^i=[\delta,\theta^i]+\frac{1}{2}f^i{}_{jk}\theta^j\theta^k.
\label{oper5/1}
\end{equation}
$\Theta$ satisfies the commutation relations 
\begin{subequations}
\begin{align}
&[i_j,\Theta^i]=0,
\vphantom{\bigg]}
\label{oper5a} 
\\
&[l_j,\Theta^i]=-f^i{}_{jk}\Theta^k.
\vphantom{\bigg]}
\label{oper5b} 
\end{align}
\label{oper5}
\end{subequations}
\!\!The Bianchi identities
\begin{subequations}
\begin{align}
&[\delta,\theta^i]=\Theta^i-\frac{1}{2}f^i{}_{jk}\theta^j\theta^k,
\vphantom{\bigg]}
\label{oper6a} 
\\
&[\delta,\Theta^i]=-f^i{}_{jk}\theta^j\Theta^k
\vphantom{\bigg]}
\label{oper6b} 
\end{align}
\label{oper6}
\end{subequations}
\!\!hold.

With the Lie algebra $\mathfrak{g}$, there is associated a canonical Lie operation 
$(W(\mathfrak{g}),\mathfrak{g}$, $i_{\mathfrak{g}},l_{\mathfrak{g}},\delta_{\mathfrak{g}})$,
called {\it the Weil operation}. $W(\mathfrak{g})$ is 
\begin{equation}
W(\mathfrak{g})=\Fun(\mathfrak{g}[1]\oplus \mathfrak{g}[2]),
\label{oper2}
\end{equation}
the algebra of polynomials of the coordinates 
$c^i$, $C^i$ of 
$\mathfrak{g}[1]$, $\mathfrak{g}[2]$ with respect to the basis $\{t_i\}$.
$i_{\mathfrak{g}},l_{\mathfrak{g}},\delta_{\mathfrak{g}}$ are defined by the relations 
\begin{subequations}
\begin{align}
&i_{\mathfrak{g}i}c^j=\delta_i{}^j, \hspace{2cm}
\vphantom{\bigg]}
\label{oper3a}
\\
&i_{\mathfrak{g}i}C^j=0,
\vphantom{\bigg]}
\label{oper3b}
\end{align}
\begin{align}
&l_{\mathfrak{g}i}c^j=-f^j{}_{ik}c^k,
\vphantom{\bigg]}
\label{oper3c}
\\
&l_{\mathfrak{g}i}C^j=-f^j{}_{ik}C^k,
\vphantom{\bigg]}
\label{oper3d}
\\
&\delta_{\mathfrak{g}}c^i=C^i-\frac{1}{2}f^i{}_{jk}c^jc^k.
\vphantom{\bigg]}
\label{oper3e}
\\
&\delta_{\mathfrak{g}}C^i=-f^i{}_{jk}c^jC^k.
\vphantom{\bigg]}
\label{oper3f}
\end{align}
\label{oper3}
\end{subequations}
\!\!Note that (multiplication by) $c^i$ defines a connection of the
Weil operation having 
$C^i$ as its curvature.

For a given Lie operation $(\mathcal{E}, \mathfrak{g},i,l,\delta)$, let us set
\begin{equation}
\mathcal{E}'=\mathcal{E}''=W(\mathfrak{g})\otimes\mathcal{E}.
\label{oper7}
\end{equation}
We can define endomorphisms of $\mathcal{E}'$ by  
\begin{subequations}
\begin{align}
i'{}_i&=i_{\mathfrak{g}i}\otimes 1,
\vphantom{\bigg]}
\label{oper8a} 
\\
l'{}_i&=l_{\mathfrak{g}i}\otimes 1+1_{\mathfrak{g}}\otimes l_i,
\vphantom{\bigg]}
\label{oper8b} 
\\
\delta'&=\delta_{\mathfrak{g}}\otimes 1+1_{\mathfrak{g}}\otimes \delta+c^i\otimes l_i-C^i\otimes i_i.
\vphantom{\bigg]}
\label{oper8c} 
\end{align}
\label{oper8}
\end{subequations}
\!\!Then, $(\mathcal{E}', \mathfrak{g},i',l',\delta')$ is a Lie operation.
By definition, the {\it equivariant cohomology} of the operation 
$(\mathcal{E}, \mathfrak{g},i,l,\delta)$ (in the BRST model) is 
\begin{equation}
H_{\mathrm{equiv}}{}^*(\mathcal{E}):=H_{\mathrm{bas}}{}^*(\mathcal{E}').
\label{oper12}
\end{equation}

When the operation $(\mathcal{E}, \mathfrak{g},i,l,\delta)$ has a connection $\theta$
with curvature $\Theta$, we can define endomorphisms of $\mathcal{E}''$ by  
\begin{subequations}
\begin{align}
i''{}_i&=1_{\mathfrak{g}}\otimes i_i,
\vphantom{\bigg]}
\label{oper9a} 
\\
l''{}_i&=l_{\mathfrak{g}i}\otimes 1+1_{\mathfrak{g}}\otimes l_i,
\vphantom{\bigg]}
\label{oper9b} 
\\
\delta''&=\delta_{\mathfrak{g}}\otimes 1+1_{\mathfrak{g}}\otimes \delta+l_{\mathfrak{g}i}\otimes \theta^i
-i_{\mathfrak{g}i}\otimes \Theta^i.
\vphantom{\bigg]}
\label{oper9c} 
\end{align}
\label{oper9}
\end{subequations}
\!\!Then, $(\mathcal{E}'', \mathfrak{g},i'',l'',\delta'')$  is also a Lie operation. 
A crucial technical result is that the basic cohomologies of $\mathcal{E}$ and $\mathcal{E}''$
are isomorphic
\begin{equation}
H_{\mathrm{bas}}{}^*(\mathcal{E})\simeq H_{\mathrm{bas}}{}^*(\mathcal{E}'').
\label{oper13}
\end{equation}
We shall give a sketch of its proof momentarily. 

The Lie operations $(\mathcal{E}', \mathfrak{g},i',l',\delta')$, 
$(\mathcal{E}'', \mathfrak{g},i'',l'',\delta'')$, which   
we have just constructed are isomorphic, since $i',l',\delta'$ and $i'',l'',\delta''$ are related as 
\begin{subequations}
\begin{align}
i''{}_i&=I^{-1}i'{}_iI,
\vphantom{\bigg]}
\label{oper10a} 
\\
l''{}_i&=I^{-1}l'{}_iI,
\vphantom{\bigg]}
\label{oper10b} 
\\
\delta''&=I^{-1}\delta' I,
\vphantom{\bigg]}
\label{oper10c} 
\end{align}
\label{oper10}
\end{subequations}
where $I\in\Iso_0(\mathcal{E}'',\mathcal{E}')$ is given by 
\begin{equation}
I=\exp(c^i\otimes i_i)\exp(i_{\mathfrak{g}i}\otimes \theta^i).
\label{oper11}
\end{equation}
(The exponential are well defined as the exponential series terminate
after a finite number of terms.)
It follows that \hphantom{xxxxxxxxxxxxxxxxxxx}
\begin{equation}
H_{\mathrm{bas}}{}^*(\mathcal{E}')\simeq H_{\mathrm{bas}}{}^*(\mathcal{E}'').
\label{oper14}
\end{equation}

From \eqref{oper12}, \eqref{oper13}, \eqref{oper14}, we conclude that
\begin{equation}
H_{\mathrm{equiv}}{}^*(\mathcal{E})\simeq H_{\mathrm{bas}}{}^*(\mathcal{E}).
\label{oper15}
\end{equation}
Thus, {\it if the Lie operation $(\mathcal{E}, \mathfrak{g},i,l,\delta)$ admits a connection,
the basic and equivariant cohomologies of $\mathcal{E}$ are equivalent}.
The above fundamental result hinges on the isomorphism \eqref{oper13}, whose proof we 
shall now sketch. 

To begin with, we note that, by \eqref{oper9a}, \eqref{oper9b}, 
\begin{equation}
\mathcal{E}''{}_{\mathrm{bas}}=(W(\mathfrak{g})\otimes\mathcal{E}_{\mathrm{hor}})_{\mathrm{inv}}.
\label{oper20}
\end{equation}
This suggests defining the following subspaces of $\mathcal{E}''{}_{\mathrm{bas}}$ 
\begin{align}
\mathcal{C}_n&=(W(\mathfrak{g})_n\otimes\mathcal{E}_{\mathrm{hor}})_{\mathrm{inv}},
\vphantom{\bigg]}
\label{oper18} 
\\
\mathcal{D}_n&=\bigoplus_{0\leq m\leq n}\mathcal{C}_m,
\vphantom{\bigg]}
\label{oper19} 
\end{align}
where $n\geq 0$. Then, $\mathcal{E}_{\mathrm{bas}}\simeq \mathcal{D}_0\subset 
\mathcal{D}_1\subset \mathcal{D}_2\subset
\ldots \mathcal{E}''{}_{\mathrm{bas}}$ is a filtration of the vector 
space $\mathcal{E}''{}_{\mathrm{bas}}$. One can now show the following two properties.

\noindent
$i)$ Let $\mu\in\mathcal{D}_n$ be such that \hphantom{xxxxxxxxxxxxxxxxxx}
\begin{equation}
\delta''\mu=0.
\label{oper27}
\end{equation}
Then, there are $\nu\in \mathcal{D}_{n-1}$, $\alpha\in\mathcal{E}_{\mathrm{bas}}$ such that
\begin{align}
&\mu=\delta''\nu+1_{\mathfrak{g}}\otimes \alpha,
\vphantom{\bigg]}
\label{oper28} 
\\
&\delta\alpha=0.
\vphantom{\bigg]}
\label{oper29}  
\end{align}

\noindent
$ii)$ Let $\nu\in \mathcal{D}_{n-1}$, $\alpha\in\mathcal{E}_{\mathrm{bas}}$ be such that
\begin{equation}
\delta''\nu+1_{\mathfrak{g}}\otimes \alpha=0.
\label{oper30}
\end{equation}
Then, there is $\beta\in \mathcal{E}_{\mathrm{bas}}$ such that
\begin{equation}
\alpha=\delta\beta.
\label{oper31}
\end{equation}

By $i$, $ii$, there is a homomorphism $q: H^*(\mathcal{E}''{}_{\mathrm{bas}}, \delta'')\rightarrow 
H^*(\mathcal{E}_{\mathrm{bas}},\delta)$ defined by 
\begin{equation}
q([\mu])=[\alpha],
\label{oper31/2}
\end{equation}
where $\mu$ is expressed as in \eqref{oper28}. Since
\begin{equation}
\delta''(1_{\mathfrak{g}}\otimes \gamma)=1_{\mathfrak{g}}\otimes\delta\gamma,
\label{oper31/1}
\end{equation}
for $\gamma\in\mathcal{E}$, by \eqref{oper9c}, $q$ is an isomorphism. 
Thus, if one shows $i$, $ii$, \eqref{oper13} is shown as well.

{\it Proof of $i$, $ii$.} By \eqref{oper9c}, $\delta''$ can be split as 
\begin{equation}
\delta''=\delta''{}_1+\delta''{}_2,
\label{oper16}
\end{equation}
where $\delta''{}_1$, $\delta''{}_2$ are given by 
\begin{subequations}
\begin{align}
&\delta''{}_1=\delta_{\mathfrak{g}}\otimes 1,
\vphantom{\bigg]}
\label{oper17a} 
\\
&\delta''{}_2=1_{\mathfrak{g}}\otimes \delta+l_{\mathfrak{g}i}\otimes \theta^i-i_{\mathfrak{g}i}\otimes \Theta^i.
\vphantom{\bigg]}
\label{oper17b} 
\end{align}
\label{oper17}
\end{subequations}
\vspace{-.5cm}

The following property holds. Let $n\geq 0$. If $\mu_n\in\mathcal{C}_n$ is such that 
\begin{equation}
\delta''{}_1\mu_n=0, 
\label{oper26/1}
\end{equation}
then $\mu_n$ is of the special form
\begin{equation}
\mu_n=\delta''{}_1\nu_{n-1}+\delta_{n,0}1_{\mathfrak{g}}\otimes\alpha,
\label{oper21}
\end{equation}
for certain $\nu_{n-1}\in\mathcal{C}_{n-1}$, $\alpha\in\mathcal{E}_{\mathrm{bas}}$.
To see this, we notice preliminarily that $c^i$, $\tilde C^i:=C^i-\frac{1}{2}f^i{}_{jk}c^jc^k$
are generators of $W(\mathfrak{g})$ such that 
$l_{\mathfrak{g}i}c^j=-f^j{}_{ik}c^k$, $l_{\mathfrak{g}i}\tilde C^j=-f^j{}_{ik}\tilde C^k$ and 
$\delta_{\mathfrak{g}}c^i=\tilde C^i$, 
$\delta_{\mathfrak{g}}\tilde C^i=0$. Now, being $\mu_n\in\mathcal{C}_n$, we have
\begin{equation}
\mu_n=\sum_{p\geq 0,q\geq 0,p+2q=n}c^{i_1}\cdots c^{i_p} \tilde C^{j_1}\cdots \tilde C^{j_q}
\otimes \alpha^{p,q}{}_{i_1\ldots i_p;j_1\dots j_q},
\label{oper22}
\end{equation}
where $\alpha^{p,q}{}_{i_1\ldots i_p;j_1\dots j_q}\in\mathcal{E}$. By \eqref{oper9a}, \eqref{oper9b}, 
the conditions $i''{}_i\mu_n=0$, $l''{}_i\mu_n=0$ imply that 
\begin{subequations}
\begin{align}
&\!i_k\alpha^{p,q}{}_{i_1\ldots i_p;j_1\dots j_q}=0,
\vphantom{\bigg]}
\label{oper23a} 
\\
&\!l_k\alpha^{p,q}{}_{i_1\ldots i_p;j_1\dots j_q}
-\sum_rf^l{}_{ki_r} \alpha^{p,q}{}_{i_1\ldots l \ldots i_p;j_1\dots j_q}
-\sum_sf^l{}_{kj_s} \alpha^{p,q}{}_{i_1\ldots i_p;j_1\dots l \ldots j_q}=0.
\vphantom{\bigg]}
\label{oper23b} 
\end{align}
\label{oper23}
\end{subequations}
\!\!By \eqref{oper17a}, the condition $\delta''{}_1\mu_n=0$ implies further that 
\begin{equation}
\alpha^{p,q}{}_{i_1\ldots \{i_p;j_1\dots j_q\}}=0,\qquad p\geq 1,
\label{oper24}
\end{equation}
where $\{\ldots\}$ stands for complete symmetrization of the enclosed indices. Let us 
\eject\noindent
define
\begin{equation}
\!\!\!\nu_{n-1}=\sum_{p\geq 0,q\geq 1,p+2q=n}\!\!\frac{(-1)^pq}{p+q}\,
c^{i_1}\cdots c^{i_p} c^{i_{p+1}} \tilde C^{j_1}\cdots \tilde C^{j_{q-1}}
\otimes \alpha^{p,q}{}_{i_1\ldots i_p;i_{p+1}j_1\dots j_{q-1}}.
\label{oper25}
\end{equation}
Then, by \eqref{oper9a}, \eqref{oper9b}, \eqref{oper23a}, \eqref{oper23b}, we have 
$i''{}_i\nu_{n-1}=0$, $l''{}_i\nu_{n-1}=0$ so that $\nu_{n-1}\in \mathcal{C}_{n-1}$. Further, by \eqref{oper17a}, 
\eqref{oper24}, 
\begin{equation}
\delta''{}_1\nu_{n-1}=\mu_n-\delta_{n,0}1_{\mathfrak{g}}\otimes\alpha,
\label{oper32}
\end{equation}
where $\alpha=\alpha^{0,0}$. By \eqref{oper23a}, \eqref{oper23b}, $\alpha\in\mathcal{E}_{\mathrm{bas}}$.
\eqref{oper21} follows.

The proof of $i$ proceeds by induction on $n$. Let $\mu\in\mathcal{D}_0$ satisfy \eqref{oper27}.
Then, $\mu=1_{\mathfrak{g}}\otimes\alpha$ for some $\alpha\in\mathcal{E}_{\mathrm{bas}}$.
Using \eqref{oper9c}, one has $\delta\alpha=0$. Hence, \eqref{oper28}, \eqref{oper29} hold 
with $\nu=0$. So, $i$ holds for $n=0$.
Suppose now $i$ holds for $n-1$ with $n\geq 1$. Let $\mu\in\mathcal{D}_n$ satisfy \eqref{oper27}.
Write $\mu=\mu_n+\tilde \mu$, where $\mu_n\in \mathcal{C}_n$,
$\tilde\mu\in\mathcal{D}_{n-1}$. Since 
$\delta''{}_1\mathcal{C}_m\subset \mathcal{C}_{m+1}$
$\delta''{}_2\mathcal{C}_m\subset \mathcal{D}_m$ for $m\geq 0$, 
condition \eqref{oper27} implies that $\delta''{}_1\mu_n=0$. 
So, $\mu_n$ satisfies \eqref{oper26/1} and, so, by \eqref{oper21}, there is $\nu_{n-1}\in \mathcal{C}_{n-1}$
such that $\mu_n=\delta''{}_1\nu_{n-1}=\delta''\nu_{n-1}-\delta''{}_2\nu_{n-1}$.
Thus, $\mu=\delta''\nu_{n-1}+\mu^*$ with $\mu^*=\tilde\mu-\delta''{}_2\nu_{n-1}
\in \mathcal{D}_{n-1}$. As $\delta''\mu=0$, $\delta''\mu^*=0$ as well.
So, $\mu^*$ satisfies \eqref{oper27} and, so, 
by the inductive hypothesis, $\mu^*=\delta''\nu^*+1_{\mathfrak{g}}\otimes \alpha$,
with $\nu^*\in \mathcal{D}_{n-2}$, $\alpha\in\mathcal{E}_{\mathrm{bas}}$ such that
$\delta\alpha=0$.
Hence, $\mu=\delta''\nu+1_{\mathfrak{g}}\otimes \alpha$, where $\nu=\nu_{n-1}
+\nu^*\in\mathcal{D}_{n-1}$. By induction on $n$, $i$ is shown.

The proof of $ii$ also proceeds by induction on $n$.
Let $\nu\in\mathcal{D}_0$, $\alpha\in\mathcal{E}_{\mathrm{bas}}$ satisfy \eqref{oper30}.
Then, $\nu=-1_{\mathfrak{g}}\otimes\beta$ for some $\beta\in\mathcal{E}_{\mathrm{bas}}$.
Using \eqref{oper9c}, one has $\alpha=\delta\beta$. Hence, \eqref{oper31} holds. 
So, $ii$ holds for $n=1$.
Suppose $ii$ holds for $n-2$ with $n\geq 2$. Let $\nu\in\mathcal{D}_{n-1}$, 
$\alpha\in\mathcal{E}_{\mathrm{bas}}$ satisfy \eqref{oper30}.
Write $\nu=\nu_{n-1}+\tilde \nu$, where $\nu_{n-1}\in \mathcal{C}_{n-1}$,
$\tilde\nu\in\mathcal{D}_{n-2}$. Since $\delta''{}_1\mathcal{C}_m\subset \mathcal{C}_{m+1}$
$\delta''{}_2\mathcal{C}_m\subset \mathcal{D}_m$ for $m\geq 0$, 
condition \eqref{oper30} implies that that $\delta''{}_1\nu_{n-1}=0$. 
So, $\nu_{n-1}$ satisfies \eqref{oper26/1} and, so, by \eqref{oper21}, 
there is $\nu_{n-2}\in \mathcal{C}_{n-2}$
such that $\nu_{n-1}=\delta''{}_1\nu_{n-2}=\delta''\nu_{n-2}-\delta''{}_2\nu_{n-2}$.
Thus, $\nu=\delta''\nu_{n-2}+\nu^*$ with $\nu^*=\tilde\nu-\delta''{}_2\nu_{n-2}
\in \mathcal{D}_{n-2}$. By \eqref{oper30}, 
$\delta''\nu^*+1_{\mathfrak{g}}\otimes \alpha=0$. 
So, $\nu^*$ satisfies \eqref{oper30} and, so, 
by the inductive hypothesis, $\alpha=\delta\beta$ for some 
$\beta\in\mathcal{E}_{\mathrm{bas}}$. By induction on $n$, $ii$ is shown. \hfill{\it QED}

{\it Remark} Although to prove the isomorphism \eqref{oper15}
we had to make explicit use at several points 
of a fixed connection $\theta$ of the operation 
$(\mathcal{E}, \mathfrak{g},i,l,\delta)$, the isomorphism can be shown to be independent 
from the choice of $\theta$ and is thus canonical.

\vfill\eject

\section{\normalsize \textcolor{blue}{Superfield formulation of  the N=1 ghost system}}
\label{sec:N=1ghost}

~~~~In this appendix, we show that the $N=1$ ghost system described 
in sect. \ref{sec:BVgauge1} has an elegant superfield formulation.

$N=1$ ghost superfields are elements of the algebra $\mathcal{A}_{\mathfrak{g}|1}((\theta))$, 
where $\theta$ a formal odd variable such that $\partial \theta=-1$. Define the superfields 
\begin{subequations}
\begin{align}
&\mathbb{B}_i=B_i+\theta b_i,
\vphantom{\bigg]}
\label{BVgau113a} 
\\
&\mathbb{c}^i=c^i-\theta C^i.
\vphantom{\bigg]}
\label{BVgau113b} 
\end{align}
\label{BVgau113} 
\end{subequations}
\!\!We have $\mathbb{B}_i\in\mathcal{A}_{\mathfrak{g}|1}((\theta))_{-3}$, 
$\mathbb{c}^i\in\mathcal{A}_{\mathfrak{g}|1}((\theta))_{1}$.
In terms of these, the $\mathfrak{g}$--action \eqref{BVgau110} reads succinctly as
\hphantom{xxxxxxxxxxxxxxxxxxxx}
\begin{subequations}
\begin{align}
&i_{\mathfrak{g}|1i}\mathbb{B}_j=0,
\vphantom{\bigg]}
\label{BVgau114a} 
\\
&i_{\mathfrak{g}|1i}\mathbb{c}^j=\delta_i{}^j
\vphantom{\bigg]}
\label{BVgau114b} 
\\
&l_{\mathfrak{g}|1i}\mathbb{B}_j
=f^k{}_{ij}\mathbb{B}_k,
\vphantom{\bigg]}
\label{BVgau114c} 
\\
&l_{\mathfrak{g}|1i}\mathbb{c}^j
=-f^j{}_{ik}\mathbb{c}^k.
\vphantom{\bigg]}
\label{BVgau114d} 
\end{align}
\label{BVgau114} 
\end{subequations}
\!\!\!The superfield expression of ghost master action \eqref{BVgau111} is
\begin{equation}
S_{\mathfrak{g}|1}=-\int d\theta\Big[\mathbb{B}_i\partial_\theta\mathbb{c}^i
+\frac{1}{2}f^i{}_{jk}\mathbb{B}_i\mathbb{c}^j\mathbb{c}^k\Big].
\label{BVgau115} 
\end{equation} 
\!\!Similarly the quantum BV variations \eqref{BVgau112} read concisely as
\begin{subequations}
\begin{align}
&\delta_{\mathfrak{g}|1}\mathbb{B}_i
=-\partial_\theta\mathbb{B}_i-f^k{}_{ji}\mathbb{B}_k\mathbb{c}^j,
\vphantom{\bigg]}
\label{BVgau116a} 
\\
&\delta_{\mathfrak{g}|1}\mathbb{c}^i
=-\partial_\theta\mathbb{c}^i-\frac{1}{2}f^i{}_{jk}\mathbb{c}^j\mathbb{c}^k.
\vphantom{\bigg]}
\label{BVgau116b} 
\end{align}
\label{BVgau116} 
\end{subequations}
\!\!The superfield formalism was originally worked out in ref. \cite{Horne}. 

\vfill\eject

\section{\normalsize \textcolor{blue}{The modular class of a Lie algebroid}}
\label{app:modular}

~~~~The modular class of a Lie algebroid was first introduced in \cite{Weinstein}.
Let $E$ be a Lie algebroid over the manifold $M$ 
with anchor $\rho_E$ and Lie bracket $[\cdot,\cdot]_E$.
Then, the real line bundle over $M$ 
\begin{equation}
Q_E=\wedge^nT^*M\otimes\wedge^qE,
\label{mod1}
\end{equation}
where $n=\dim M$ and $q=\rank E$, is defined. 
$Q_E$ is called the {\it orientation bundle} of $E$. 

For $\gamma\in\Gamma(Q_E)$, $X\in\Gamma(E)$, we set
\begin{equation}
D_X\gamma=(l_{M\rho_E(X)}\otimes 1_{\wedge^qE}+1_{\wedge^nT^*M}\otimes l_{EX})\gamma,
\label{mod2}
\end{equation}
where $l_{M\rho(X)}$ is the ordinary Lie derivative along the vector field
$\rho_E(X)$ and $l_{EX}$ is defined by
\begin{equation}
l_{EX}(Y_1\wedge\ldots\wedge Y_d)=\sum_{r=1}^dY_1\wedge\ldots\wedge [X,Y_r]_E \wedge\ldots\wedge Y_d,
\qquad Y_r\in\Gamma(E).
\label{mod3}
\end{equation}
Clearly, $D_X\gamma\in\Gamma(Q_E)$.
It can be verified that the map $X\rightarrow D_X$ defines a representation of the Lie
algebroid $E$ in $Q_E$, \hphantom{xxxxxxxxxxxxxxxxxxxx}
\vspace{-.3cm}
\begin{subequations}
\begin{align}
&D_{fX}\gamma=fD_X\gamma,
\vphantom{\bigg]}
\label{mod4}
\\
&D_X(f\gamma)=fD_X\gamma+(\rho_E(X)f)\gamma,
\vphantom{\bigg]}
\label{mod5}
\\
&[D_X,D_Y]\gamma-D_{[X,Y]_E}\gamma=0,
\vphantom{\bigg]}
\label{mod6}
\end{align}
\label{mod4-6}
\end{subequations}
\!\!for $f\in \Fun(M)$ and $X,Y\in\Gamma(E)$, $\gamma\in\Gamma(Q_E)$.

Suppose that the line bundle $Q_E$ is trivial. Then, there is a nowhere vanishing section
$\gamma\in\Gamma(Q_E)$. Further, there is a section $\theta_\gamma\in\Gamma(E^*)$ such that
\begin{equation}
D_X\gamma=\theta_\gamma(X)\gamma,
\label{mod7}
\end{equation}
for $X\in\Gamma(E)$. It can be shown that $\theta_\gamma$ satisfies
\begin{equation}
d_E\theta_\gamma=0.
\label{mod8}
\end{equation}
Further, if one rescales $\gamma$ into $\gamma'=\mathrm{e}^f\gamma$ with $f\in \Fun(M)$, then one has
\begin{equation}
\theta_{\gamma'}=\theta_\gamma+d_Ef.     
\label{mod9}
\end{equation}
Thus, $\theta_\gamma$ is a representative of a well defined cohomology class $\theta_E\in H^1(E)$,
the {\it modular class} of $E$, independent from the choice of $\gamma$. 
If $Q_E$ is not trivial, the modular class of $E$ can still be 
defined as follows. One notes that $Q_E{}^{\otimes 2}$ is trivial and one repeats a similar 
construction with $\gamma$ replaced by a nowhere vanishing section $\nu\in\Gamma(Q_E{}^{\otimes 2})$ 
and $D_X$ replaced by $D_X\otimes 1_{Q_E}+ 1_{Q_E}\otimes D_X$. Then, $\theta_E
=\frac{1}{2}[\theta_\nu]_{H^1(E)}$. 

In general, $E$ is said {\it unimodular} if $\theta_E=0$.

A straightforward calculation shows that, when $Q_E$ is trivial,
$\theta_\gamma$ is given in any local trivialisation of $E$ by 
\begin{equation}
\theta_{\gamma r}=\partial_a\rho_r{}^a+\rho_r{}^a\partial_a\ln\gamma-c^s{}_{sr},
\label{mod10}
\end{equation}
where $\rho_r{}^a$, $c^t{}_{rs}$ are the anchor and structure functions of $E$, respectively.
Upon rescaling $\gamma$ into $\gamma'=\mathrm{e}^f\gamma$ with $f\in \Fun(M)$, one has
\begin{equation}
\theta_{\gamma' r}=\theta_{\gamma r}+\rho_r{}^a\partial_af.
\label{mod11}
\end{equation}
Therefore, it is possible to chose $\gamma$ in such a way that $\theta_\gamma=0$
precisely when $E$ is unimodular.

\vfill\eject


\begin{thebibliography}{99}


\bibitem{BV1}
I.~A.~Batalin and G.~A.~Vilkovisky,
``Gauge algebra and quantization'',
Phys.\ Lett.\ B {\bf 102} (1981) 27.

\bibitem{BV2}
I.~A.~Batalin and G.~A.~Vilkovisky,
``Quantization of gauge theories with linearly dependent generators'',
Phys.\ Rev.\ D {\bf 28} (1983) 2567
(Erratum-ibid.\ D {\bf 30} (1984) 508).

\bibitem{Sam}
J.~Gomis, J.~Paris and S.~Samuel,
``Antibracket, antifields and gauge theory quantization'',
Phys.\ Rept.\  {\bf 259} (1995) 1
[arXiv:hep-th/9412228].

\bibitem{Fiorenza1}
D.~Fiorenza, 
``An introduction to the Batalin-Vilkovisky formalism'',
arXiv: math/0402057.

\bibitem{Schwarz1}
A.~S.~Schwarz,
``Geometry of Batalin-Vilkovisky quantization'',
Commun.\ Math.\ Phys.\  {\bf 155} (1993) 249
[arXiv:hep-th/9205088].

\bibitem{Schwarz2}
A.~S.~Schwarz,
``Semiclassical approximation in Batalin-Vilkovisky formalism'',
Commun.\ Math.\ Phys.\  {\bf 158} (1993) 373
[arXiv:hep-th/9210115].

\bibitem{Zwiebach1}
H.~Hata and B.~Zwiebach,
``Developing the covariant Batalin-Vilkovisky approach to string theory'',
Annals Phys.\  {\bf 229} (1994) 177 [arXiv:hep-th/9301097].

\bibitem{Witten3}
E.~Witten,
``Phases of N = 2 theories in two dimensions'',
Nucl.\ Phys.\  B {\bf 403} (1993) 159
[arXiv:hep-th/9301042].

\bibitem{Spence1}
C.~M.~Hull, G.~Papadopoulos and B.~J.~Spence,
``Gauge symmetries for $(p,q)$ supersymmetric sigma models'',
Nucl.\ Phys.\ B {\bf 363} (1991) 593.

\bibitem{Zucchini6}
R.~Zucchini,
``The Hitchin Model, Poisson-quasi-Nijenhuis Geometry and Symmetry
Reduction'', JHEP {\bf 0710} (2007) 075
[arXiv:0706.1289 [hep-th]].

\bibitem{Zucchini7}
R.~Zucchini,
``Gauging the Poisson sigma model,''
JHEP {\bf 0805} (2008) 018
[arXiv:0801.0655 [hep-th]].

\bibitem{Zucchini8}
R.~Zucchini,
``The Lie algebroid Poisson sigma model'',
JHEP {\bf 0812} (2008) 062
[arXiv:0810.3300 [math-ph]].

\bibitem{AKSZ}
M.~Alexandrov, M.~Kontsevich, A.~Schwartz and O.~Zaboronsky,
``The Geometry of the master equation and topological quantum field theory'',
Int.\ J.\ Mod.\ Phys.\ A {\bf 12} (1997) 1405
[arXiv:hep-th/9502010].

\bibitem{Mnev1}
P.~Mnev,
``Notes on simplicial BF theory'',
arXiv:hep-th/0610326.

\bibitem{Mnev2}
P.~Mnev,
``Discrete BF theory'',
arXiv:0809.1160 [hep-th].

\bibitem{Bonechi1}
F.~Bonechi and M.~Zabzine,
``Poisson sigma model on the sphere'',
Commun.\ Math.\ Phys.\  {\bf 285} (2009) 1033 [arXiv:0706.3164 [hep-th]].

\bibitem{Chevalley1}
C.~Chevalley and S.~Eilenberg, 
``Cohomology theory of Lie groups and Lie algebras'', 
Trans. \ Amer. \ Math. \ Soc. \ {\bf 63} (1948) 85.

\bibitem{Koszul1}
J.~L.~Koszul, 
``Homologie et cohomologie des alg\`ebres de Lie'', 
Bull. \ Soc. \ Math. \ France, {\bf 78} (1950) 65.

\bibitem{Grueb1} W.~Grueb, S.~Halperin and R.~Vanstone,
{\it Connections, Curvature and Cohomology}, vol. III, Academic Press, 
New York (1973).

\bibitem{Mathai1}
V.~Mathai and D.~G.~Quillen,
``Superconnections, Thom classes and equivariant differential forms'',
Topology {\bf 25} (1986) 85.

\bibitem{Moore1}
S.~Cordes, G.~W.~Moore and S.~Ramgoolam,
``Lectures On 2-D Yang-Mills Theory, Equivariant Cohomology And Topological Field Theories'',
Nucl.\ Phys.\ Proc.\ Suppl.\  {\bf 41} (1995) 184
[arXiv:hep-th/9411210].

\bibitem{Guillemin1}
V.~W.~ Guillemin and S.~Sternberg
{\it Supersymmetry And Equivariant De Rham Theory},
Spinger--Verlag Berlin Heidelberg (1999).

\bibitem{Weil1}
A.~Weil, 
``G\'eom\'etrie diff\'erentielle des espaces fibr\'es'', 
(Letters to Chevalley and Koszul), 1949,

\bibitem{HCartan2}
H.~Cartan, 
``La transgression dans un groupe de Lie et dans un fibr\'e principal'',
{\it Colloque de topologie (espaces fibr\'es) (Bruxelles)},  
Li\`ege et Paris (1950) 73.

\bibitem{HCartan1}
H.~Cartan, 
``Notions d'alg\`ebre diff\'erentielle; application aux groupes de Lie et aux 
vari\'et\'es o\`u op\`ere un groupe de Lie'', 
{\it Colloque de topologie (espaces fibr\'es) (Bruxelles)},  
Li\`ege et Paris (1951) 15.

\bibitem{Ouvry1}
S.~Ouvry, R.~Stora and P.~van Baal,
``On the Algebraic Characterization of Witten's Topological Yang-Mills Theory'',
Phys.\ Lett.\  B {\bf 220} (1989) 159.

\bibitem{Kalkman1}
J.~Kalkman,
``BRST model applied to symplectic geometry'',
arXiv:hep-th/9308132.

\bibitem{Dijkgraaf1}
R.~Dijkgraaf and G.~W.~Moore,
``Balanced topological field theories'',
Commun.\ Math.\ Phys.\  {\bf 185} (1997) 411
[arXiv:hep-th/9608169].

\bibitem{Zucchini0}
R.~Zucchini,
``Basic and equivariant cohomology in balanced topological field theory'',
J.\ Geom.\ Phys.\  {\bf 35} (2000) 299
[arXiv:hep-th/9804043].

\bibitem{Vaintrob1}
A.~Yu.~Vaintrob, 
``Lie algebroids and homological vector fields'', 
\ Usp. \ Mat. \ Nauk {\bf 52} (1997) 161 
(English transl.: \ Russ. \ Math. \ Surv. {\bf 52} (1997) 428).

\bibitem{Kubarski}
J.~Kubarski, 
``Invariant cohomology of regular Lie algebroids'', 
International Colloquium on Differential Geometry {\bf VII}, Santiago de Compostela, 1994, 
World Scientific (1995) 137.

\bibitem{Ginzburg}
V.~Ginzburg,
``Equivariant Poisson cohomology and a spectral sequence associated with a
moment map'',  Int.\ J.\ Math.\ {\bf 10} (1999) 977
[arXiv:math.DG/ 9611102]. 

\bibitem{KosmannSchwarzbach1}
Y.~Kosmann-Schwarzbach,
``Derived brackets'',
Lett.\ Math.\ Phys.\  {\bf 69} (2004) 61 [arXiv:math/0312524].

\bibitem{Roytenberg1}
D.~Roytenberg,
``Courant algebroids, derived brackets and even symplectic supermanifolds'', 
Ph. D. thesis, U. C. Berkeley (1999) arXiv:mathDG/ 9910078.

\bibitem{Roytenberg2} 
D.~Roytenberg,
``A note on quasi-Lie bialgebroids and twisted Poisson manifolds'',
Lett.\ Math.\ Phys.\  {\bf 61} (2002) 123 [arXiv:math/0112152].


\bibitem{Weinstein}
S.~Evens, J.-H.~Lu and A.~Weinstein,
``Transverse measures, the modular class, and a cohomology pairing for Lie algebroids'', 
Quart. J. Math. Oxford Ser. (2) {\bf 50} (1999), no. 200, 417 
[arXiv:dg-ga/9610008]. 

\bibitem{Vaisman}
I.~Vaisman, 
``Lectures on the Geometry of Poisson Manifolds'', 
Progress in Mathematics, vol. {\bf 118}, 
Birkhauser Verlag (1994).

\bibitem{Cattaneo1}
A.~S.~Cattaneo and G.~Felder,
``A path integral approach to the Kontsevich quantization formula'',
Commun.\ Math.\ Phys.\  {\bf 212} (2000) 591 
[arXiv:math.qa/9902090].

\bibitem{Horne}
J.~H.~Horne,
``Superspace versions of topological theories'',
Nucl.\ Phys.\  B {\bf 318} (1989) 22.

\end{thebibliography}
\end{document}